\documentclass[preprint,aps,prb]{revtex4-1}

\usepackage[utf8x]{inputenc}
\usepackage{amsmath,amssymb,amsthm}
\usepackage{mathtools}
\usepackage{xspace}
\usepackage[usenames,dvipsnames]{xcolor}
\usepackage[colorlinks,hyperindex,breaklinks]{hyperref}
\usepackage{braket}
\usepackage{hyphenat}
\usepackage{bbold}
\usepackage{enumerate}
\usepackage{tabu,booktabs}
\usepackage{tabularx}
\usepackage[verbose]{placeins}
\usepackage[hang,flushmargin]{footmisc}
\usepackage{comment}

\usepackage[ngerman,british]{babel}

\usepackage{srcltx}

\newcommand{\vecalpha}{\mbox{\boldmath${\alpha}$}}
\newcommand{\vecsigma}{\mbox{\boldmath${\sigma}$}}
\newcommand{\vecnabla}{\mbox{\boldmath${\nabla}$}}
\newcommand{\pbar}{\bar{p}}
\newcommand{\mathbbm}{}

\renewcommand{\vec}[1]{\mathbf{#1}}
\newcommand{\LABEL}[1]{\label{#1}}
\begin{document}
%%\bf
%%\large

\title{Fully relativistic multiple scattering calculations for
general potentials}

\author{H. Ebert, J. Braun, D. K\"odderitzsch and S. Mankovsky}

\affiliation{Department Chemie/Phys.\ Chemie, 
            Ludwig-Maximilians-Universit\"at
           M\"unchen, Germany}

\date{\today}

%\tableofcontents

\begin{abstract}
The formal basis for fully relativistic
 Korringa-Kohn-Rostoker (KKR) or 
multiple scattering 
calculations for the
electronic
 Green function in case of a 
general potential is discussed.
Simple criteria are given to identify situations
 that require  to distinguish between
 right and left hand side solutions to the Dirac equation when
setting up the electronic Green function.
In addition various technical aspects of
an implementation of the relativistic KKR for general local and non-local
 potentials
will be discussed. 
\end{abstract}

%http://www.aip.org/pacs/pacs2010/individuals/pacs2010_regular_edition/index.html

\maketitle

%****************************************************************
%****************************************************************
\section{Introduction}
%****************************************************************
%****************************************************************

Recently there is strong interest in the 
impact of spin-orbit coupling on the 
electronic structure of solids 
and surfaces
as this gives
 rise to many interesting and technically
 important phenomena. 
  In this context 
one may mention the well known
 magneto-crystalline  anisotropy but also the
interesting galvano-magnetic and
 spin transport phenomena.\cite{RMS09,LGK+11,TGFM12,ZCK+14}
Other examples for the central role
of spin-orbit coupling can be found in the
field of spectroscopy as the magneto-optical effects
and the various magneto-dichroic phenomena in
X-ray spectroscopy.\cite{Ebe96b,HE02a,VRW+04}
Finally one may mention the Rashba splitting\cite{APM+08,BMK+14}
of surface states of transition metals
as well as occurrence of the topological surface states
in topological insulators.\cite{HK10,QZ11}

Computational schemes 
used to describe these phenomena or materials, respectively, have
 to account at the same  time for spin-orbit coupling, 
 spin polarization or magnetic ordering as well as the structural
properties of the investigated system 
in a coherent and reliable
way. Among the various available schemes the 
Korringa-Kohn-Rostoker (KKR) or 
multiple scattering method
is especially attractive as it gives direct access
 to the
Green function (GF). Making
 use of the Dyson equation allows for example
 by means of the
corresponding embedding technique to deal with
rather complex systems.\cite{EKM11} Another important field of
application for the KKR-GF method is the 
investigation of disordered systems usually
done in combination with the 
Coherent Potential Approximation (CPA) \cite{Sov67} alloy theory. 

The spin-polarized relativistic (SPR) version 
of the KKR method set up on the basis of the
four-component Dirac formalism that allows
to account for all relativistic effects and spin
magnetism within the framework of L(S)DA
(local (spin) density approximation) on equal
footing was worked out by various authors.\cite{FRA83,SSG84}
Extensions to this approach were made 
to deal with orbital polarization \cite{EB96} as 
well as the presence of a vector potential
 coupling to the total current of the electrons.\cite{EBG97,BMB+12}
Corresponding implementations and 
applications of the KKR method were done in general
making use of the ASA (Atomic Sphere Approximation)
that implies spherical symmetry for the 
potential functions and rotational symmetry for the corresponding vector fields
 coupling to the spin and current of the electrons.
Finally, the so-called full potential (FP) version 
of the SPR-KKR method that removes the mentioned geometrical
restrictions was discussed and 
implemented by various authors.\cite{Tam92,LGG93,WZB+92,HZE+98} 

Compared to the non-relativistic version 
of the KKR method its fully relativistic formulation
leads to a number of technical complications. 
The need to distinguish between 
right (RHS) and left hand side (LHS) solutions to the
Dirac equation was discussed in particular by Tamura 
\cite{Tam92} 
for the case of a general local potential.
Here we extend this work discussing among others
the impact of a non-local site-diagonal potential
and several
practical aspects of corresponding KKR calculations.

%****************************************************************
%****************************************************************
\section{Relativistic Hamiltonian
and Green function for general potentials}
%****************************************************************
%****************************************************************

Starting point of our considerations 
is an effective one-electron 
Hamiltonian $\hat {\cal H} (z)$ that can be 
split into an energy-independent, 
Hermitian part $\hat {\cal H}^{1}   $
 and an energy-dependent, non-Hermitian part:
\begin{eqnarray}
\LABEL{EQ:RGF1-A0}
%
%----------------------------------------------------------------------------------------
\hat {\cal H}(z) & = & \hat {\cal H}^{1}  + \hat \Sigma(z) \\
                 & = & \hat {\cal H}^{0} +\hat {\cal V} + \hat \Sigma(z) 
%
%----------------------------------------------------------------------------------------
\end{eqnarray}
with the Hermitian adjoined operator
\begin{eqnarray}
\LABEL{EQ:RGF1-A1}
%----------------------------------------------------------------------------------------
\hat {\cal H}^{\dagger}(z) & = & \hat {\cal H}^{0\dagger}
    +\hat {\cal V}^{\dagger}   + \hat \Sigma^{\dagger}(z) \\
%
%----------------------------------------------------------------------------------------
\LABEL{EQ:RGF1-A2}
& = &  \hat {\cal H}^{0} +\hat {\cal V} + \hat \Sigma(z^{*}) \; .
\end{eqnarray}
Here $\hat {\cal H}^{0} $ stands for the Hamiltonian of the free-electron
system, $\hat {\cal V} $ for an energy independent Hermitian 
potential and the non-Hermitian
self-energy $\Sigma (z)$ may depend on the energy $z$.
Using  a fully  relativistic formulation based on the four-component 
 Dirac formalism
the real space representation of  $\hat {\cal H}^{1} $ takes the form:\cite{Ros61}
\begin{eqnarray}
%----------------------------------------------------------------------------------------
\LABEL{EQ:RGF1-DEQ-alpha}
\hat {\cal H}^{1} (\vec r)  & =&
- i c \vecalpha \cdot \vecnabla  + \frac{1}{2} \, c^{2} (\beta - 1) + V(\vec r)  \\
%
%----------------------------------------------------------------------------------------
\LABEL{EQ:RGF1-DEQ}
& =& 
i \gamma_{s} \sigma_{r} c \left( \frac{\partial}{\partial r} 
+ \frac{1}{r} - \frac{\beta}{r} \, K \right)
\nonumber \\ && \qquad \qquad \quad
 + \frac{1}{2} \, c^{2} (\beta - 1) + V(\vec r) 
%
%----------------------------------------------------------------------------------------
\end{eqnarray}
 with $\alpha_i$ (see below) and $\beta $ 
 the standard $4 \times 4 $ Dirac matrices,
 $K$ the spin-orbit operator,
 $\sigma_{r} = \vecsigma \cdot \vec r / r $
 the projection of the Pauli matrices 
  and 
\begin{equation}
%
%----------------------------------------------------------------------------------------
\gamma_{s} = \left(
            \begin{array}{cc} 
  0 & -I_{2} \\
-I_{2} &  0
 \end{array}\right)  \;.
%
%----------------------------------------------------------------------------------------
\end{equation}
In Eq.\ (\ref{EQ:RGF1-DEQ})
 atomic Rydberg units ($\hbar = 1$, $e^{2} = 2$, $m = 1/2$) have been used and 
the rest mass energy $c^{2}/2$
 has been subtracted from the energy $z$. 
The potential $V(\vec r)$ is assumed to be local and
 in the most general case it will  be a $4 \times 4$ matrix function according  to:
\begin{eqnarray}
\LABEL{EQ:DEF-POT}
%----------------------------------------------------------------------------------------
 V(\vec r) & = & \bar V(\vec r) + \beta \, \vecsigma \cdot  \vec B(\vec r)  
+ e \vecalpha \cdot \vec A(\vec r)  \\
&=& 
\bar V(\vec r) + 
\left(
            \begin{array}{cc} 
 \vecsigma \cdot \vec B(\vec r) & 0 \\
0 & - \vecsigma \cdot \vec B(\vec r) 
 \end{array}\right) 
\nonumber \\ && \quad \quad
+e \left(\begin{array}{cc} 
 0 & \vecsigma \cdot \vec A(\vec r)  \\
 \vecsigma \cdot \vec A(\vec r) & 0
 \end{array}\right) 
 \nonumber  \\
&=&  
\left( \begin{array}{cc} 
 V^{+}(\vec r)  &  U(\vec r) \\
 U(\vec r)  &  V^{-}(\vec r)
 \end{array}\right) \nonumber
\; .
%----------------------------------------------------------------------------------------
\end{eqnarray}
Here $\bar V(\vec r)$ and  $\beta \, \vecsigma \cdot  \vec B(\vec r)  $
stand  for the 
spin-independent and spin-dependent parts of the potential, respectively,
while the term $e\vecalpha \cdot \vec A(\vec r)$ represents the 
coupling of a vector potential $\vec A(\vec r)$ to the electronic 
current density, with $\vecalpha$ the electronic velocity operator.\cite{Ros61}
Obviously, the auxiliary potential functions $V^{\pm}(\vec r)$ and $ U(\vec r)$ 
are $2 \times 2 $ matrix functions in spin space.\cite{Tam92}

For the real space representation of the 
self energy $ \Sigma(z)$
 one again 
has in general a $4 \times 4$ matrix function  $ \Sigma (\vec r, \vec r \,', z)$
that may be 
written in an analogous way:
%
%----------------------------------------------------------------------------------------
\begin{eqnarray}
\LABEL{EQ:DEF-SELF}
  \Sigma (\vec r, \vec r \,', z) &=& 
     \Sigma^{V} \, (\vec r, \vec r \,',z) + 
\beta \, \vecsigma \, \vec \Sigma^{B} \,
(\vec r, \vec r \,',z)
 \\
 &=& 
 \left(
            \begin{array}{cc} 
 \Sigma^{+}(\vec r, \vec r \,',z)  & 0 \\
0 &  \Sigma^{-}(\vec r, \vec r \,',z) 
 \end{array}\right)
\; , \nonumber
\end{eqnarray}
%----------------------------------------------------------------------------------------
%
where 
we restrict to a spin-dependent self-energy.
A current-dependent one coupling like 
$\vecalpha \cdot \vec \Sigma^{A}$ could be 
introduced as well and treated in analogy to the term 
$\vecsigma \cdot \vec A(\vec r)$ in the local potential.
From Eqs.\  (\ref{EQ:RGF1-A1}) and  (\ref{EQ:RGF1-A2})
one has for  $ \Sigma (\vec r, \vec r \,', z)$ 
the property:
\begin{eqnarray}
\LABEL{EQ:RGF1-SIGMA-ADJOINT}
%----------------------------------------------------------------------------------------
\Sigma^{\dagger} (\vec r, \vec r \,', z) 
 & = &   \Sigma (\vec r \,', \vec r, z^{*})  \;.
%----------------------------------------------------------------------------------------
\end{eqnarray}
In practice it seems to be sufficient for most applications 
to consider a self-energy 
$  \Sigma (\vec r, \vec r \,', z) $ 
%%%%(see Eq.\ \eqref{EQ:DEF-SELF})
that can be represented by an expansion into 
a product of suitable basis  functions:
\begin{eqnarray}
\LABEL{EQ:SELF-EXPANSION}
\Sigma (\vec r, \vec r \,', z)
 &=& 
\sum_{\Lambda \Lambda'} 
 \phi_{\Lambda}(\vec r)\, 
 \Sigma_{\Lambda \Lambda '}(z) 
 \,\phi_{\Lambda '}^{\dagger}(\vec r \,') \;.
\end{eqnarray}
In line with the relativistic representation
the basis  functions $ \phi_{\Lambda}(\vec r)$
are constructed  here as four-component
functions with the index $\Lambda$ 
specifying their spin-angular character \cite{Ros61} (see below).

\medskip

Explicit forms for the local potential $V(\vec{r})$ 
as given by  Eq.\ \eqref{EQ:DEF-POT} can be derived within the 
framework of relativistic density functional theory (DFT).\cite{ED11,Esc96}
Dealing with magnetic solids the relativistic version 
of the local   spin  density approximation (LSDA) to  DFT
is usually adopted.\cite{MV79,RR79,ED11} 
This scheme is derived by applying a Gordon
decomposition of the electronic current density into its
spin and orbital contribution and retaining
only the corresponding spin-dependent part of the 
Hamiltonian.\cite{RC73,MV79}
The term in  Eq.\ \eqref{EQ:DEF-POT} 
involving the vector potential $\vec A(\vec r)$
may be derived within current density functional theory 
(CDFT) \cite{VR87,Die91} 
that in particular accounts for the electronic orbital degrees
of freedom.
Alternatively or in addition, it may
represent the Breit interaction \cite{ED11}  that plays a prominent
role for the magneto-crystalline anisotropy.\cite{Jan88a}

While the non-local self-energy 
$  \Sigma (\vec r, \vec r \,', z) $ 
in Eq.\ \eqref{EQ:DEF-SELF} may stand for example for 
the energy-independent Hartree-Fock potential, it will
in general represent extensions to the standard relativistic
LSDA scheme. Within spectroscopic
investigations, life-time effects are usually represented
by an optical potential corresponding to a local but
complex and energy-dependent potential.\cite{TF86}
Alternatively or in addition, $  \Sigma (\vec r, \vec r \,', z) $ 
may represent correlation effects that are not accounted
for by standard LSDA. Within the rather simple 
L(S)DA+U scheme \cite{AZA91}, the corresponding non-local self-energy
is real and   energy-independent. On the other hand,
the combination of the more sophisticated
dynamical mean field theory (DMFT) 
\cite{GKKR96,Hel07}
with the LSDA
implies a complex  and   energy-dependent   non-local  self-energy.
For both schemes one restricts usually to local correlations
corresponding to a site-diagonal self-energy (see below). 
This restriction is dropped
e.g.\ for cluster variants of the DMFT \cite{GKKR96}
and does not apply to the standard formulation of the 
GW method.\cite{AG98} 
%  author = {F Aryasetiawan and O Gunnarsson - The GW method},

\medskip

The Green function operator $\hat G(z)$ associated with  the
general Hamiltonian $\hat {\cal H} (z)$ in Eqs.\ (\ref{EQ:RGF1-A0}) to (\ref{EQ:RGF1-A2})
is defined 
to be simultaneously the right and left inverse of $\big(z - \hat {\cal H}(z)\big)$:
%
%----------------------------------------------------------------------------------------
\begin{eqnarray}
%
%----------------------------------------------------------------------------------------
\LABEL{EQ:RGF1-B}
\left(z - \hat {\cal H}(z)\right) \, \hat G(z) & = & 1   \\
%
%----------------------------------------------------------------------------------------
\LABEL{EQ:RGF1-C}
 \hat G(z) \, \left(z - \hat {\cal H}(z)\right) & = & 1  
%
%----------------------------------------------------------------------------------------
\end{eqnarray}
implying the relation:
\begin{eqnarray}
\LABEL{EQ:RGF1-GF-ADJOINT}
%----------------------------------------------------------------------------------------
 \hat G^{\dagger}(z) & = & \hat G(z^{*}) \; .
%
%----------------------------------------------------------------------------------------
\end{eqnarray}
In their real space representation  
Eqs.\  (\ref{EQ:RGF1-B}) through (\ref{EQ:RGF1-GF-ADJOINT})
read:
\begin{eqnarray}
%----------------------------------------------------------------------------------------
\LABEL{EQ:RGF1-GFDEF-RHS}
 \left( z\, -  \hat {\cal H}^{1}(\vec r)\right) \, G(\vec r , \vec r \,', z) \hspace{1.5cm} && \nonumber \\
    - \int d^{3} r \,'' \, \Sigma(\vec r , \vec r \,'', z) \,
G(\vec r \,'', \vec r \,', z)  
& = & {\mathbbm{1}}_{4} \, \delta(\vec r\, , \vec r \,') 
\end{eqnarray}
%----------------------------------------------------------------------------------------
\begin{eqnarray}
\LABEL{EQ:RGF1-GFDEF-LHS}
 \delta(\vec r , \vec r \,') \, {\mathbbm{1}}_{4}  & = &  
    G(\vec r , \vec r \,' , z)\left(z - \hat {\cal H}^{1}(\vec r \,')\right) \nonumber \\
&& - \int d^{3} r \,'' \, 
G(\vec r , \vec r \,'' , z)\,  \Sigma(\vec r \,'', \vec r \,', z)
%
%----------------------------------------------------------------------------------------
\end{eqnarray}
\begin{eqnarray}
\LABEL{EQ:RGF1-GFDEF-ADJOINT}
%----------------------------------------------------------------------------------------
G^{\dagger}(\vec r, \vec r \,', z) & = & G(\vec r \,', \vec r , z^{*}) \; .
%
%----------------------------------------------------------------------------------------
\end{eqnarray}
The differential operator contained in $\hat {\cal H}^{1}(\vec r \,')$ 
in Eq.\ \eqref{EQ:RGF1-GFDEF-LHS}
has to be interpreted to act to the left. The more familiar 
right hand side form  can be obtained 
by taking the Hermitian adjoint of this equation and making use of
Eqs.\ \eqref{EQ:RGF1-SIGMA-ADJOINT} and \eqref{EQ:RGF1-GFDEF-ADJOINT}:
\begin{eqnarray}
%----------------------------------------------------------------------------------------
\LABEL{EQ:RGF1-GFDEF-LHS-ADJOINT}
%----------------------------------------------------------------------------------------
 \left( z^{*} -  \hat {\cal H}^{1}(\vec r \,')\right) \,  G(\vec r \,', \vec r , z^{*})
 \hspace{1.5cm} && \nonumber \\ 
- \int d^{3} r \,'' \, 
\Sigma(\vec r \,', \vec r \,'', z^{*})\, G(\vec r \,'', \vec r , z^{*}) 
&  = & \delta(\vec r\, , \vec r \,') \, {\mathbbm{1}}_{4}  \; .
%
%----------------------------------------------------------------------------------------
\end{eqnarray}
Obviously, 
replacing $z^{*}$ by $z$ the 
 original right hand side equation \eqref{EQ:RGF1-GFDEF-RHS}
is recovered.

As shown for the non-relativistic case by various authors,\cite{HFL87,Lay63}
an expression  for the Green function
defined by the  Eqs.\ \eqref{EQ:RGF1-GFDEF-RHS} and  \eqref{EQ:RGF1-GFDEF-LHS}
can be 
  given also for the relativistic case 
in terms of a
  spectral representation:
\begin{eqnarray}
\LABEL{EQ:RGF1-F}
%----------------------------------------------------------------------------------------
G(\vec r, \vec r \,', z) = \sum_{n} \, \frac{\phi_{n}(\vec r, z) \, \psi_{n}^{\dagger}(\vec r \,', z)}{z - E_{n}(z)} \;.
%
%----------------------------------------------------------------------------------------
\end{eqnarray}
Here 
 the so-called right- and left-hand side solutions, 
$\phi_{n}(\vec r , z)$ and $\psi_{n}(\vec r , z)$, respectively,
are  four component functions (bi-spinors) and defined as solutions to the 
following eigenvalue equations 
\begin{eqnarray}
%
%----------------------------------------------------------------------------------------
\LABEL{EQ:EIGEN-RHS}
\left( E_{n}(z) - \hat {\cal H}^{1}(\vec r) \right)\, \phi_{n}(\vec r, z) \hspace{2.1cm} && \nonumber \\
 - \int d^{3} r \, \Sigma(\vec r, \vec r \,',z) \, \phi_{n}(\vec r \,', z) & = & 0 
\\
%
%----------------------------------------------------------------------------------------
\LABEL{EQ:EIGEN-LHS}
\left( E_{n}(z)^{*} - \hat {\cal H}^{1}(\vec r) \right)\, \psi_{n}(\vec r, z)\hspace{2.1cm} && \nonumber \\
 - \int d^{3} r \,' \, \Sigma(\vec r, \vec r \,',z^{*}) \, \psi_{n}(\vec r \,', z) & = & 0
%
%----------------------------------------------------------------------------------------
\end{eqnarray}
that have in general complex
 eigenvalues $E_{n}(z)$.
On the basis of Eqs.\ \eqref{EQ:EIGEN-RHS} and  \eqref{EQ:EIGEN-LHS},
it is straightforward to show that Eq.\ \eqref{EQ:RGF1-F}
 is indeed a solution to 
Eqs.\ \eqref{EQ:RGF1-GFDEF-RHS} and  \eqref{EQ:RGF1-GFDEF-LHS}.
In this context it is interesting to note 
that the homogeneous term
${\mathbbm{1}}_{4} \, \delta(\vec r\, , \vec r \,') $ 
in these equations is ensured to be covered by the closure relation
\begin{eqnarray}
%
%----------------------------------------------------------------------------------------
\sum_{n} \phi_{n}(\vec r, z) \,  \psi_{n}^{\dagger}(\vec r \,', z)  & = & \mathbbm{1}_{4} \, \delta(\vec r, \vec r \,')  \; .
%
%----------------------------------------------------------------------------------------
\end{eqnarray}
In case of a non-vanishing energy dependent self-energy
$ \Sigma(\vec r , \vec r \,', z)$, 
 the set of eigenvalue equations 
  \eqref{EQ:EIGEN-RHS} and  \eqref{EQ:EIGEN-LHS}
 has obviously to be solved for each value of energy $z$.
 For that reason the 
 spectral representation given in 
Eq.\ \eqref{EQ:RGF1-F} may not be very helpful in practice.
Nevertheless, it
 clearly shows that even in case of a non-vanishing 
$ \Sigma(\vec r , \vec r \,', z)$,  a real space
representation of the Green function can in principle  be given.

%****************************************************************
%****************************************************************
\section{Multiple scattering 
or KKR representation of the  Green function}
%****************************************************************
%****************************************************************

The multiple scattering or
KKR-GF formalism aims to supply the 
Green function $G(\vec r, \vec r \,', z)$
for a given energy $z$ without making use of the 
 spectral representation given in 
Eq.\ \eqref{EQ:RGF1-F}. Dealing with an extended system 
as a cluster of atoms or a solid the problem to find the Green
function is subdivided by dealing in a first step with the
scattering from a single potential well associated with an
atom site and treating multiple scattering in a subsequent step.
According to this, the discussion below is restricted  here to a
self-energy $\Sigma(\vec r, \vec r \,',z)$ that is site-diagonal,
i.e.\ $\Sigma(\vec r, \vec r \,',z)=0$ for $\vec r $ or $\vec r \,'$ outside
the regime of the considered potential well. More complex situations
can nevertheless be treated by making use of 
the Dyson equation.\cite{NKH+12}
Furthermore, only
{\em on-the-energy-shell} scattering 
will be considered,
i.e.\ inelastic processes will explicitly 
excluded.

Guided by the  
eigenvalue equations 
  \eqref{EQ:EIGEN-RHS} and  \eqref{EQ:EIGEN-LHS}
 connected with the  
 spectral representation  
Eq.\ \eqref{EQ:RGF1-F} the RHS and LHS solutions to the 
so-called single site Dirac equation will be considered first.
From these the single-site t-matrix and Green function will be
derived. Finally, the multiple scattering will be considered
leading to the Green function of the total system.

%****************************************************************
\subsection{RHS and LHS solutions to the Dirac equation}
%****************************************************************

The RHS solutions $|  \psi (z) \rangle$ 
 to the Dirac equation for a given energy $z$ 
are defined by 
\begin{equation}
%
%----------------------------------------------------------------------------------------
\big(  z     -  \hat {\cal H}(z) \big) \, |  \psi (z) \rangle = 0 \;.
%----------------------------------------------------------------------------------------
%
\end{equation}
With the real space representation of the Hamilton operator
$\hat {\cal H}(z) $
given by  Eqs.\  \eqref{EQ:RGF1-DEQ} and \eqref{EQ:DEF-SELF} this 
corresponds for the wave function
$ \psi_{\nu}(\vec r, z) $ labeled by the index 
${\nu}$ to the equation
%
%----------------------------------------------------------------------------------------
\begin{eqnarray}
\hat {\cal H}^{1} (\vec r) \, \psi_{\nu}(\vec r, z)
 -  \int d^{3} r \,' \, \Sigma(\vec r, \vec r \,',z) \, \psi_{\nu}(\vec r \,', z) &= & \nonumber \\
%----------------------------------------------------------------------------------------
%
\LABEL{EQ:DEQU-RHS}
%----------------------------------------------------------------------------------------
\Bigg[
z -
 i \gamma_{s} \sigma_{r} c \left( \frac{\partial}{\partial r} + \frac{1}{r} - \frac{\beta}{r} \, K \right)   \qquad\qquad && \nonumber \\ 
- V(\vec r) 
- (\beta -1)\, \frac{c^{2}}{2} 
   \Bigg] \psi_{\nu}(\vec r, z) \nonumber\\
%
%----------------------------------------------------------------------------------------
 -  \int d^{3} r \,' \, \Sigma(\vec r, \vec r \,',z) \, \psi_{\nu}(\vec r \,', z) &= &0 \;,
%----------------------------------------------------------------------------------------
\end{eqnarray}
where the general potential $V(\vec r) $ and self-energy 
$\Sigma(\vec r, \vec r \,',z) $
are defined as in Eqs.\ \eqref{EQ:DEF-POT}
and \eqref{EQ:DEF-SELF}.
As multiple scattering is treated in a most suitable way by working
with an angular momentum representation, we make for $\psi_{\nu}(\vec r, z) $ 
the standard ansatz\cite{FRA83,SSG84}
%----------------------------------------------------------------------------------------
\begin{eqnarray}
\LABEL{EQ:ANSATZ-RHS}
\psi_{\nu}(\vec r, z) &=& \sum_{\Lambda}\, \psi_{\Lambda \nu}(\vec r, z) \\
&=&  \sum_{\Lambda} \,\left(
            \begin{array}{c} 
  g_{\Lambda \nu}(r, z)  \chi_{\Lambda}(\hat r) \\
i\, f_{\Lambda \nu}(r, z)   \chi_{-\Lambda}(\hat r)
 \end{array}\right)
%----------------------------------------------------------------------------------------
\end{eqnarray}
with the radial functions $g_{\Lambda \nu}(r, z) $ 
and $f_{\Lambda \nu}(r, z) $ connected 
with 
the large and small, respectively, components of the wave function.
The spin-angular function $\chi_{\Lambda}(\hat r) $
is an eigen function of the spin-orbit operator  $K$ 
%
%----------------------------------------------------------------------------------------
\begin{equation}
\LABEL{EQ:SO-K-AIG-VAL}
K\, \chi_{\Lambda}(\hat r) = - \kappa \,\chi_{\Lambda}(\hat r)
\end{equation}
%----------------------------------------------------------------------------------------
%
with the property
%
%----------------------------------------------------------------------------------------
\begin{equation}
\sigma_{r} \, \chi_{\Lambda}(\hat r) = - \chi_{-\Lambda}(\hat r) \; .
\end{equation}
%----------------------------------------------------------------------------------------
%
Here we used the short-hand notation
 $\Lambda=(\kappa,\mu)$ and
 $-\Lambda=(-\kappa,\mu)$ to give the 
spin-orbit and magnetic quantum numbers $\kappa$ and $\mu$,
respectively.\cite{Ros61}
The index $\nu$ labeling the linearly independent 
solutions  $\psi_{\nu}(\vec r, z) $ 
will be dropped in this section. Later on, it will be replaced by 
a spin-angular index  ($\Lambda'$) 
that reflects the asymptotic behavior of the solution
 $\psi_{\Lambda'}(\vec r, z) $.

Inserting the ansatz 
Eq.\ \eqref{EQ:ANSATZ-RHS} into the Dirac Eq.\ \eqref{EQ:DEQU-RHS}
one is led after some straightforward manipulations to the 
following  set of radial Dirac equations
for the RHS solutions:
\begin{widetext}
\begin{eqnarray}
%
%----------------------------------------------------------------------------------------
\left(
            \begin{array}{cc} 
 z & c\, \left ( \frac{\partial}{\partial \,r} + \frac{-\kappa + 1}{r}  \right)  \\
  c \,\left ( \frac{\partial}{\partial \,r} + \frac{\kappa + 1}{r}  \right) & -(z + c^{2}) 
 \end{array}\right)
\, \left(
            \begin{array}{r} 
  g_{\Lambda}(r,z)   \\
  f_{\Lambda}(r,z)  
 \end{array}\right)  
- \sum_{\Lambda '}\left( \begin{array}{cc} 
 V_{\Lambda \, \Lambda '}^{+}( r) & -i \, U_{\Lambda \, -\Lambda '}( r) \\
i \, U_{-\Lambda \, \Lambda '}( r) & -V_{-\Lambda \, -\Lambda '}^{-}( r)
 \end{array}\right) \, \left(
            \begin{array}{c} 
  g_{\Lambda '}(r,z)   \\
 f_{\Lambda '}(r,z) 
 \end{array}\right)\nonumber\\
%
%----------------------------------------------------------------------------------------
\LABEL{EQ:RAD-DEQU-RHS}
  -  \sum_{\Lambda '} \int  r'^{2} \, d r'  \, \left(
            \begin{array}{c} 
 \Sigma_{\Lambda \, \Lambda '}^{+}( r,  r \,',z)\,\,\,\,\,\,\,  g_{\Lambda '}(r ',z)  \\
- \Sigma_{-\Lambda \, -\Lambda '}^{-}( r,  r \,',z)\,\,\,  f_{\Lambda '}(r ',z) 
 \end{array}\right) & = &0 \; .
%----------------------------------------------------------------------------------------
%
\end{eqnarray}
\end{widetext}
Here we used the 
 matrix element functions connected with  the potential
\begin{eqnarray}
%
%----------------------------------------------------------------------------------------
 V_{\Lambda \, \Lambda '}^{\pm}(r)&  = & \int d \hat r \,\, \chi_{\Lambda}^{\dagger} (\hat r) \, V^{\pm} (\vec r) \, \chi_{\Lambda '}(\hat r) \nonumber\\
%
%----------------------------------------------------------------------------------------
 U_{\Lambda \, \Lambda '}(r)&  = & \int d \hat r \,\chi_{\Lambda}^{\dagger} (\hat r) \,\, U (\vec r) \, \chi_{\Lambda '}(\hat r) \nonumber
\end{eqnarray}
and  the self-energy
%
%----------------------------------------------------------------------------------------
\begin{eqnarray}
 \Sigma_{\Lambda \, \Lambda '}^{\pm}(r, r ', z)  =  \int d \hat r \, \int  d \hat r ' \, 
\chi_{\Lambda}^{\dagger} (\hat r) \,\, \Sigma^{\pm} (\vec r, 
\vec r \, ', z) \, \chi_{\Lambda '}(\hat r \, ') 
\nonumber
\; .
\end{eqnarray}
%----------------------------------------------------------------------------------------
%
%
%
%\subsection{Left-hand side solution}
The LHS solution 
$ \langle \psi^{\times} (z)  | $ corresponding to 
the RHS solution
$ | \psi (z)  \rangle $ 
is defined by the adjoined Dirac equation
\begin{eqnarray}
%
%----------------------------------------------------------------------------------------
\langle \psi^{\times} (z) | \big(z - \hat {\cal H}(z) \big) = 0 
%----------------------------------------------------------------------------------------
%
\end{eqnarray}
with its real space representation  given by 
\begin{eqnarray}
\LABEL{EQ:DEQU-LHS}
%----------------------------------------------------------------------------------------
 \langle \psi^{\times} (z) | \vec r \,' \rangle \, \big(z - \hat {\cal H}^{1}(\vec r \,') \big) 
\hspace{2cm} && \nonumber \\  
 - \int d^{3} r \,''\,\, \langle \psi^{\times} (z) | \vec r \,'' \rangle  \, \Sigma(\vec r \,'', \vec r \,',  z) &= &0 \;.
%----------------------------------------------------------------------------------------
%
\end{eqnarray}
To proceed, we find it more convenient to
switch to the Hermitian adjoined of this equation:
\begin{eqnarray}
\LABEL{EQ:RGF2-A}
%----------------------------------------------------------------------------------------
\big(\hat {\cal H}^{1}(\vec r )  - z^{*}\big) \, \langle \psi^{\times} (z) | \vec r  \rangle^{\dagger} \hspace{2cm} && \nonumber \\  
 + \int d^{3} r \,'\,\, \Sigma^{\dagger}(\vec r , \vec r \,',  z^{*}) \, \langle \psi^{\times} (z) | \vec r \,' \rangle^{\dagger}    &= &0 \; ,
%----------------------------------------------------------------------------------------
%
\end{eqnarray}
where use of the relations
 $\hat {\cal H}^{1}(\vec r ) = \hat {\cal H}^{1\dagger}(\vec r )$ 
 and
  $\Sigma^{\dagger}(\vec r, \vec r \,',  z) 
  = \Sigma(\vec r \,' , \vec r,  z^{*})$
has been made.

Making for $\langle \psi^{\times} (z) | \vec r  \rangle$ 
the ansatz\cite{TF89,Tam92}
\begin{eqnarray}
\LABEL{EQ:ANSATZ-LHS}%----------------------------------------------------------------------------------------
 \langle \psi^{\times} (z) | \vec r \,  \rangle  = \sum_{\Lambda} 
(  g_{\Lambda}^{\times} (r , z) \, \chi_{\Lambda}^{\dagger} (\hat r), \, 
-i f_{\Lambda}^{\times} (r , z) \, \chi_{-\Lambda}^{\dagger} (\hat r)) 
%----------------------------------------------------------------------------------------
%
\end{eqnarray}
one has for its adjoined wave function
\begin{eqnarray}
%
%----------------------------------------------------------------------------------------
 \langle \psi^{\times} (z) | \vec r   \rangle^{\dagger} 
 = \sum_{\Lambda}  \left(
            \begin{array}{r} 
  g_{\Lambda}^{\times *} (r ,z) \,
\chi_{\Lambda} (\hat r) \\
i f_{\Lambda}^{\times *} (r ,z) \, 
\chi_{-\Lambda} (\hat r))  
 \end{array}\right)   \; .
%----------------------------------------------------------------------------------------
%
\end{eqnarray}
Inserting this expression  into Eq.~\eqref{EQ:RGF2-A}
leads to a set of radial  equations that corresponds one-to-one to
Eq.\ \eqref{EQ:RAD-DEQU-RHS}
apart from the replacements
$z \rightarrow z^*$,
$ (g, f) \rightarrow (g^{\times \,*}, f^{\times \,*})$,
and $ \Sigma(\vec r, \vec r \,', z) \rightarrow \Sigma(\vec r, \vec r \,', z^{*})$.
Accordingly, Eq.~\eqref{EQ:RGF2-A} can be rearranged as  Eq.~\eqref{EQ:RAD-DEQU-RHS} to lead to the radial Dirac equations for the LHS solutions
\begin{widetext}
\begin{eqnarray}
%
%----------------------------------------------------------------------------------------
\left(
            \begin{array}{cc} 
 z & c\, \left ( \frac{\partial}{\partial \,r} + \frac{-\kappa + 1}{r}  \right)  \\
  c \,\left ( \frac{\partial}{\partial \,r} + \frac{\kappa + 1}{r}  \right) & -(z + c^{2}) 
 \end{array}\right)
\, \left(
            \begin{array}{r} 
  g_{\Lambda}^{\times }(r,z)   \\
  f_{\Lambda}^{\times }(r,z)  
 \end{array}\right)  
%%%%\qquad && \nonumber \\
%
- \sum_{\Lambda '}\left( \begin{array}{cc} 
 V_{\Lambda ' \, \Lambda}^{+}( r) & i \, U_{-\Lambda' \, \Lambda}( r) \\
-i \, U_{\Lambda' \, -\Lambda}( r) & -V_{-\Lambda' \, -\Lambda}^{-}( r)
 \end{array}\right) \, \left(
            \begin{array}{c} 
  g_{\Lambda '}^{\times }(r,z)   \\
 f_{\Lambda '}^{\times }(r,z) 
 \end{array}\right)\nonumber\\
\qquad && \nonumber \\
%
%----------------------------------------------------------------------------------------
\LABEL{EQ:RAD-DEQU-LHS}
  -  \sum_{\Lambda '} \int  r'^{2} \, d r'  \, \left(
            \begin{array}{c} 
 \Sigma_{\Lambda' \, \Lambda}^{+}( r\,',  r ,z)\,\,\,\,\,\,\,  g_{\Lambda '}^{\times }(r ',z)  \\
- \Sigma_{-\Lambda ' \, -\Lambda}^{-}( r\,',  r ,z)\,\,\,  f_{\Lambda '}^{\times }(r ',z) 
 \end{array}\right) & = & 0 \; .
%----------------------------------------------------------------------------------------
%
\end{eqnarray}
\end{widetext}
Here use of the relations
%
%----------------------------------------------------------------------------------------
\begin{eqnarray}
 V_{\Lambda  \, \Lambda '}^{\pm}(r)^{*} &= & 
 V_{\Lambda' \, \Lambda  }^{\pm}(r)  \\ 
 U_{\Lambda  \, \Lambda '}( r)^{*}  &= & 
 U_{\Lambda' \, \Lambda  }( r) \\
  \Sigma_{\Lambda  \, \Lambda '}^{\pm}( r,  r \,',z^{*})^{*}  &=&
  \Sigma_{\Lambda' \, \Lambda  }^{\pm}( r \,',r,  z    ) 
\end{eqnarray}
%----------------------------------------------------------------------------------------
%
has been made that reflect the Hermiticity of the 
potential (Eq.\ \eqref{EQ:DEF-POT})
as well as the properties of the
 self-energy (Eq.\ \eqref{EQ:DEF-SELF}). 
 Accordingly,
the set of radial Dirac equations \eqref{EQ:RAD-DEQU-RHS} and  \eqref{EQ:RAD-DEQU-LHS} 
for the RHS and LHS, respectively, 
solutions are identical if the following relations hold:
%
%----------------------------------------------------------------------------------------
\begin{eqnarray}
\LABEL{EQ:COND-SYM-V}
 V_{\Lambda  \, \Lambda '}^{\pm}(r) &= & 
 V_{\Lambda' \, \Lambda  }^{\pm}(r)  \\ 
\LABEL{EQ:COND-SYM-U}
   U_{\Lambda  \, \Lambda '}( r)  &= & 
 - U_{\Lambda' \, \Lambda  }( r) \\
\LABEL{EQ:COND-SYM-S}
  \Sigma_{\Lambda  \, \Lambda '}^{\pm}( r,  r \,',z) &=&
  \Sigma_{\Lambda' \, \Lambda  }^{\pm}( r \,',r,  z) \;.
\end{eqnarray}
%----------------------------------------------------------------------------------------
%
Ignoring the self-energy for the moment, the resulting set of radial Dirac equations 
in Eqs.\ \eqref{EQ:RAD-DEQU-RHS}  and \eqref{EQ:RAD-DEQU-LHS}  are completely 
equivalent to those obtained by Tamura \cite{Tam92}. 
Accordingly, the requirements given by him for the RHS and LHS equations being 
the same coincide with Eqs.\ \eqref{EQ:COND-SYM-V} 
and \eqref{EQ:COND-SYM-U}. A more detailed discussion under 
what conditions these relations hold will be given in the next section.
%

%\bigstep

Finally, it should be mentioned that in the context of the
relativistic L(S)DA+U \cite{EPM03} as well as L(S)DA+DMFT \cite{MCP+05}
 Eq.\ \eqref{EQ:RAD-DEQU-RHS}  or \eqref{EQ:RAD-DEQU-LHS}, resp.,
has been dealt with so far  in an approximate way. Because for both schemes
the coupling of the self-energy is usually restricted to the d- or f-electrons
and because the basis functions for the self-energy 
 (see Eq.\ \eqref{EQ:SELF-EXPANSION})
have the same $l$-character, it seems justified to interchange the role 
of the radial functions to be calculated and of 
the basis functions. This 
transfers the radial integro-differential equations into differential
equations as they occur in the case of full potential
type calculations \cite{HZE+98}.
As a consequence, the set up of the corresponding Green function simplifies
in a dramatic way as can be seen from the discussions in section \ref{Sec:Single-site-Green-function}. 

%****************************************************************
\subsection{Relation between the   RHS and LHS solutions }
%****************************************************************

As the RHS and LHS solutions derive from the same Hamiltonian, 
it is obvious that they are not independent from each other. 
In fact, the vector space spanned by the LHS solutions 
is dual to that spanned by the RHS solutions.
In particular,
 Tamura\cite{Tam92}
 could show that the RHS and LHS  solutions are connected via
\begin{eqnarray}
%
%----------------------------------------------------------------------------------------
\LABEL{EQ:TIME-REV-PSI}
\langle \psi^\times(\vec B) | 
= \left( \hat K \, | \psi(-\vec B) \rangle \right)^\dagger 
\end{eqnarray}
by making use of the 
behavior of the Hamiltonian $\hat{\cal H}(\vec B)$ under time reversal
$\hat K$.
This implies 
 for the radial functions the relations\cite{Tam92}
\begin{eqnarray}
%----------------------------------------------------------------------------------------
\LABEL{EQ:TIME-REV-RAD-WF}
\left(
    \begin{array}{r} 
  g_{\kappa \mu}^\times(r,z,\vec B)   \\
  f_{\kappa \mu}^\times(r,z,\vec B)  
 \end{array}\right)   
=
(-1)^{\mu-1/2} S_{\kappa} \,
\left(
    \begin{array}{r} 
  g_{\kappa -\mu}(r,z,-\vec B)   \\
  f_{\kappa -\mu}(r,z,-\vec B)  
 \end{array}
 \right)   \;,
\end{eqnarray}
where  $S_{\kappa}$ stands for the sign of the 
quantum number $\kappa$\cite{Ros61} 
and the vector $\vec B$ represents also the dependence of the wave functions
on the vector potential $\vec A$ as well as the spin dependent part
of the self-energy $\Sigma$ that also reverse sign under time reversal.

Eq.\ \eqref{EQ:TIME-REV-RAD-WF}
 shows that for non-magnetic systems ($\vec B=0$),
 having time reversal symmetry,
  LHS radial functions for $\Lambda=({\kappa,\mu})$
that solve Eq.\  \eqref{EQ:RAD-DEQU-LHS}
 can easily be obtained from the RHS radial functions for 
 $\Lambda=({\kappa,-\mu})$  by multiplying with the phase factor 
 $(-1)^{\mu-1/2} S_{\kappa} $. For magnetic systems ($\vec B\ne0$),
 on the other hand, it may happen that the two
  sets of radial functions have to be calculated individually
(see section \ref{SEC:Practical-aspects}).

  In both cases, however,
the sets of radial differential equations 
for the LHS and RHS solutions,
Eqs.\   \eqref{EQ:RAD-DEQU-RHS} and   \eqref{EQ:RAD-DEQU-LHS}, respectively,
 are identical if
the various potential functions $V_{\Lambda \Lambda '}^{\pm}(r)$,
 $U_{\Lambda \Lambda '}^{\pm}(r)$, and  $\Sigma_{\Lambda \Lambda '}^{\pm}(r,r',z)$
are symmetric
(see Eqs.\ \eqref{EQ:COND-SYM-V} to  \eqref{EQ:COND-SYM-S}).
Accordingly both sets of equations  will be solved by the same
set of linearly independent (unnormalized) radial functions,
i.e.\ these have to be determined only once.
To see under which 
conditions this favorable situation holds, we restrict
 for the moment to the case 
$\Sigma_{\Lambda \Lambda '}^{\pm}(r,r',z)=0$
and expand the real potentials $\bar V(\vec r)$,
 $\vec B(\vec r)$, and  $\vec A(\vec r)$  in terms of 
 real spherical harmonics ${\cal Y}_{L}(\hat r)$ according to:
\begin{eqnarray}
%
%----------------------------------------------------------------------------------------
\LABEL{EQ:EXPAND-VLL}
\bar V(\vec r) &=& \sum_{L}\, V_{L}(r)\, {\cal Y}_{L}(\hat r)\\
%----------------------------------------------------------------------------------------
\LABEL{EQ:EXPAND-BLL}
\vec B(\vec r) &=& \sum_{L}\, \sum_{\lambda}  B_{L}^{\lambda}(r)\,{\cal Y}_{L}(\hat r) \, \hat{e}_\lambda \\
%----------------------------------------------------------------------------------------
\LABEL{EQ:EXPAND-ALL}
\vec A(\vec r) &=& \sum_{L}\, \sum_{\lambda}  A_{L}^{\lambda}(r)\,{\cal Y}_{L}(\hat r) \, \hat{e}_\lambda 
%----------------------------------------------------------------------------------------EXA
%
\end{eqnarray}
 one has
for their matrix elements:
\begin{eqnarray}
%
%----------------------------------------------------------------------------------------
\LABEL{EQ:ME-EXPAND-VLL}
  V_{\Lambda \Lambda '}^{\pm}( r)  &=& \sum_{L} 
\Big( 
\bar V_{L}(r) \, \langle \chi_{\Lambda} | {\cal Y}_{L} |  \chi_{\Lambda '} \rangle \nonumber \\
&& \qquad\quad
\pm
\sum_{\lambda} \, B_{L}^{\lambda}(r) \,
\langle \chi_{\Lambda} | \sigma_{\lambda} \, {\cal Y}_{L} |  \chi_{\Lambda '} \rangle
\Big) \\
%----------------------------------------------------------------------------------------
\LABEL{EQ:ME-EXPAND-ULL}
U_{\Lambda \Lambda '}( r) &=& \sum_{L} \sum_{\lambda} 
 A_{L}^{\lambda}(r) \, \langle \chi_{\Lambda} | \sigma_{\lambda} \, {\cal Y}_{L} |  \chi_{\Lambda '} \rangle \;,
%----------------------------------------------------------------------------------------
%
\end{eqnarray}
with real functions 
$\bar V_{L}(r)$, $ B_{L}^{\lambda}(r)$ and $ A_{L}^{\lambda}(r)$
and $\lambda$ indicating the components of the vector fields
with $\hat{e}_\lambda $ the corresponding unit vector.
For the angular matrix elements
occurring in Eqs.\ \eqref{EQ:ME-EXPAND-VLL} and \eqref{EQ:ME-EXPAND-ULL} 
 one has the property
\begin{eqnarray}
%
%----------------------------------------------------------------------------------------
 \langle \chi_{\Lambda} | {\cal Y}_{L} |  \chi_{\Lambda '} \rangle &=&
 \langle \chi_{\Lambda '} | {\cal Y}_{L} |  \chi_{\Lambda} \rangle^{*} \nonumber \\
%----------------------------------------------------------------------------------------
 \langle \chi_{\Lambda} | \sigma_{\lambda} \, {\cal Y}_{L} |  \chi_{\Lambda '} \rangle &=&
 \langle \chi_{\Lambda '} | \sigma_{\lambda} \, {\cal Y}_{L} |  \chi_{\Lambda} \rangle^{*} \nonumber
%----------------------------------------------------------------------------------------
%
\end{eqnarray}
ensuring the Hermiticity of the corresponding potential terms.

For the special cases $m(L) \geq 0$ and $\lambda = x,$ or $ z$ one finds
in particular
that the angular matrix elements are real, 
implying that  they are symmetric w.r.t.\ the indices $\Lambda$ and  $\Lambda'$.
Having only such terms 
in the expansions in Eqs.\  \eqref{EQ:ME-EXPAND-VLL} 
and \eqref{EQ:ME-EXPAND-ULL} 
also the corresponding potential matrix elements 
are symmetric, i.e.\ the requirement 
specified in Eqs.\  \eqref{EQ:COND-SYM-V} to  \eqref{EQ:COND-SYM-U}
for the LHS and RHS solutions being  identical
are fulfilled. 
The conditions for this to happen are discussed in some
detail in section  \ref{subsec:coupling-scheme}.

\medskip

Finally, 
Eq.\ \eqref{EQ:COND-SYM-S} will in general not hold
for a finite self-energy $\Sigma(\vec r, \vec r \,',z) $
 and accordingly
one has to determine the RHS and LHS solutions on the basis 
Eqs.\   \eqref{EQ:RAD-DEQU-RHS} and   \eqref{EQ:RAD-DEQU-LHS}, 
separately.
Assuming for
$\Sigma(\vec r, \vec r \,',z) $
an expansion as given by Eq.\ \eqref{EQ:SELF-EXPANSION}
with the basis functions  $ \phi_{\Lambda}(\vec r)$
involving real radial functions, the requirement expressed
by Eq.\ \eqref{EQ:COND-SYM-S} reduces to the 
simpler relation:
\begin{eqnarray}
\LABEL{EQ:SELF-EXPANSION-RHS-LHS}
 \Sigma^{\pm}_{\Lambda \Lambda '}(z) = 
 \Sigma^{\pm}_{\Lambda '\Lambda }(z)
\;.
\end{eqnarray}
For a complex, energy-dependent self-energy 
occurring within the  L(S)DA+DMFT scheme 
this relation will in general not be fulfilled.
For the  L(S)DA+U scheme, on the other hand, 
with a real,  energy-independent self-energy 
this relation may hold depending on the symmetry
of the investigated system (see the discussion
above and in section \ref{subsec:coupling-scheme}).
In this case, again one does not have to distinguish 
between the sets of linearly independent 
RHS and LHS solutions to 
Eqs.\  \eqref{EQ:RAD-DEQU-RHS} and   \eqref{EQ:RAD-DEQU-LHS}, 
respectively.

%
%
%****************************************************************
\subsection{Green function for the free electron case}
%****************************************************************

The free electron gas supplies an important reference 
system for the KKR-GF formalism that is used among others in 
connection with the Dyson equation.
As shown by several 
authors \cite{Wei90,Gon92,Tam92,WZB+92,GB99}
the corresponding relativistic free electron Green function $G^{0}(\vec r, \vec r \,', z)$ can
be expressed in terms of the non-relativistic one:
\begin{eqnarray}
%
%----------------------------------------------------------------------------------------
G^{0}(\vec r, \vec r \,', z) = (z- \hat {\cal H}^{0}) \,  G^{0\,{\rm nrel}}(\vec r, \vec r \,', z) \, {\mathbbm{1}}_{4} \nonumber
%----------------------------------------------------------------------------------------
%
\end{eqnarray}
where $\hat {\cal H}^{0}$ is the free-electron Dirac operator
and $G^{0\,{\rm nrel}}(\vec r, \vec r \,', z)$
is given by:\cite{GB99}
\begin{eqnarray}
\LABEL{EQ:GF-0-NONREL}
%----------------------------------------------------------------------------------------
G^{0\,{\rm nrel}}(\vec r, \vec r \,', z) &=& - i\,p \sum_{L} \, j_{L} (\vec r_{<} , z) h_{L}^{+\times }(\vec r_{>},z)  \; .
%----------------------------------------------------------------------------------------
%
\end{eqnarray}
As indicated by the combined angular momentum index $L=(l,m)$
and arguments, 
the spherical Bessel functions     $ j_{l} (pr)$ and
Hankel functions of the first kind $h_{l}^{+}(pr)$ have been 
combined with the complex spherical harmonics   $ Y_{L}(\hat r) $.
In Eq.\ \eqref{EQ:GF-0-NONREL}
the superscript $\times$ indicates 
the LHS side to the free-electron Schr\"odinger equation.
Using complex spherical harmonics this implies 
that the 
complex conjugate has to be taken for $ Y_{L}(\hat r) $.
The arguments $ \vec r_{<} $ and $\vec r_{>}$ 
in Eq.\ \eqref{EQ:GF-0-NONREL} coincide 
with the vectors $\vec r$  and  $\vec r \,'$ depending which
is the shorter or longer one, respectively.

Making use of the eigen functions $\chi_{\Lambda}(\hat r)$
of the spin-orbit operator $K$ 
one finds
\begin{eqnarray}
\LABEL{EQ:GF-FEG}
%----------------------------------------------------------------------------------------
G^{0}(\vec r, \vec r \,', z) &=& - i\,p \sum_{\Lambda} 
\left[ j_{\Lambda} (\vec r , z)\, h_{\Lambda}^{+ \, \times}(\vec r \,',z) \, \theta(r \,' - r) \right .  \nonumber\\
%----------------------------------------------------------------------------------------
%
%----------------------------------------------------------------------------------------
 &&   \left . \quad\quad\; \;\; +  
   \, h_{\Lambda}^{+}(\vec r , z) \, j_{\Lambda}^{\times} (\vec r \,', z)\, \theta(r - r \,') \right ] 
%----------------------------------------------------------------------------------------
%
\end{eqnarray}
where  $\pbar = \zeta p $  is 
 the relativistic momentum 
\begin{eqnarray}
\label{EQ:rel-momentum}
%----------------------------------------------------------------------------------------
p(z) = \left( z (1+z/c^{2})   \right)^{1/2} 
%----------------------------------------------------------------------------------------
%
\end{eqnarray}
scaled by the energy dependent factor
\begin{eqnarray}
\label{EQ:rel-zeta}
%----------------------------------------------------------------------------------------
\zeta(z) = 1+z/c^{2}  
\;.
%----------------------------------------------------------------------------------------
%
\end{eqnarray}
The  relativistic forms of the
 Bessel and (outgoing) first kind Hankel  functions
 are defined accordingly by:
\begin{eqnarray}
%
%----------------------------------------------------------------------------------------
\LABEL{EQ:REG-RHS-j}
j_{\Lambda}(\vec r, z)& = &
% REDEFINED \sqrt{ 1 + z/c^{2} }
\left( \begin{array}{cc} 
& j_{l}(p r) \chi_{\Lambda}(\hat r) \\
\frac{i p c S_{\kappa}}{z + c^{2} } & j_{ \bar l}(p r) \chi_{-\Lambda}(\hat r)
 \end{array}\right)\\
%----------------------------------------------------------------------------------------
%
%
%----------------------------------------------------------------------------------------
\LABEL{EQ:REG-LHS-j}
j_{\Lambda}^{\times}(\vec r, z)& = &
% REDEFINED \sqrt{ 1 + z/c^{2} }
\left( \begin{array}{cc} 
& j_{l}(p r) \chi_{\Lambda}^{\dagger}(\hat r) \\
\frac{- i p c S_{\kappa}}{z + c^{2} } & j_{ \bar l}(p r) \chi_{-\Lambda}^{\dagger}(\hat r)
 \end{array}\right)^{T} \\
%----------------------------------------------------------------------------------------
%
%
%----------------------------------------------------------------------------------------
\LABEL{EQ:REG-RHS-h}
h_{\Lambda}^{+}(\vec r, z)& = &
% REDEFINED \sqrt{ 1 + z/c^{2} }
\left( \begin{array}{cc} 
& h_{l}^{+}(p r) \chi_{\Lambda}(\hat r) \\
\frac{i p c S_{\kappa}}{z + c^{2} } & h_{ \bar l}^{+}(p r) \chi_{-\Lambda}(\hat r)
 \end{array}\right) \\
%----------------------------------------------------------------------------------------
%
%
%----------------------------------------------------------------------------------------
\LABEL{EQ:REG-LHS-h}
h_{\Lambda}^{+ \times}(\vec r, z)& = &
% REDFINED \sqrt{ 1 + z/c^{2} }
\left( \begin{array}{cc} 
& h_{l}^{+}(p r) \chi_{\Lambda}^{\dagger}(\hat r) \\
\frac{- i p c S_{\kappa}}{z + c^{2} } & h_{ \bar l}^{+}(p r) \chi_{-\Lambda}^{\dagger}(\hat r)
 \end{array}\right)^{T}
%----------------------------------------------------------------------------------------
%
\end{eqnarray}
where '$\times $' again
denotes the LHS solution to the Dirac equation,
 $S_{\kappa}=\mbox{\rm sign}(\kappa)$ gives the sign of $\kappa$
and ${ \bar l}= l - S_{\kappa}$.\cite{Ros61}

It should be noted that
the presentation of the free electron Green function in  
Eq.\ \eqref{EQ:GF-FEG} together with the definitions 
in Eqs.\ \eqref{EQ:rel-momentum} to \eqref{EQ:REG-LHS-h}
is not unique.
Alternatively, one may include
a factor $\zeta^{1/2}$ in the definition of the
Bessel and Hankel functions \cite{HZE+98}
or combine the factor $-i\pbar$ with the Hankel functions \cite{WZB+92}
as it is often done.
While all definitions are fully equivalent
they nevertheless 
influence all subsequent expressions and definitions.
With respect to the connection to the non-relativistic 
Green function, the various Lippmann-Schwinger equations and 
matrix elements occurring later-on, the present settings seem to
be most coherent.

%iced that the definition for the 
%relativistic Bessel and first kind Hankel  functions
%differs from that used by Wang et al.\ \cite{WZB+92}.
%Here the energy dependent prefactor $\sqrt{ 1 + z/c^{2} }$
%is applied in a symmetric way to the RHS and LHS solutions
%and the prefactor $-ip$ 
%in Eq.\ \eqref{EQ:GF-FEG}
%is not combined with the Hankel  functions as it is often done.

Finally, in analogy to  Eqs.\  \eqref{EQ:REG-RHS-j} to \eqref{EQ:REG-LHS-h} 
one may introduce in addition the relativistic 
second  kind Hankel  ($h_{\Lambda}^{-}(\vec r, z)$)
and von Neumann  ($n_{\Lambda}(\vec r, z)$)
functions in terms of the standard 
Bessel ($ j_{l} (pr)$),  von Neumann  ($n_{l} (pr)$)
and   Hankel  ($h_{l}^{\pm} (pr)$)  functions
with  $h_{l}^{\pm} (pr) = j_{l} (pr) \pm i n_{l} (pr)$\cite{AS64}.

%****************************************************************
\subsection{Single-site t-matrix
and Lippmann-Schwinger equations   for a general potential}
%****************************************************************

Starting point for the calculation of the single-site Green function
 $G^{n}(\vec r, \vec r \,', z)$ 
associated with a potential well located 
at site $n$
is the Dyson equation
in terms of the single site $t$-matrix operator $\hat t$
that is given in its real space representation by:
\begin{eqnarray}
\LABEL{EQ:SS-DYSON-EQ-T}
%----------------------------------------------------------------------------------------
G^{n}(\vec r, \vec r \,', z)& = & G^{0}(\vec r, \vec r \,', z) + \int d^{3} r\, '' \int d^{3} r \,''' G^{0}(\vec r, \vec r \,'', z)\, 
\nonumber \\
&& \qquad \qquad 
 t(\vec r \,'', \vec r \, ''', z) \,  G^{0}(\vec r\,''' , \vec r \,', z) \; ,
%----------------------------------------------------------------------------------------
%
\end{eqnarray}
where $ G^{0}(\vec r, \vec r \,', z) $
is the free electron Green function.

As $t(\vec r, \vec r \, ', z)$ is restricted to the volume $\Omega$ 
covered by the  single site potential well
that is bound by a sphere of radius $r_{\rm crit}$ the integration can be restricted to 
$r, \,\, r\,' \, < r_{\rm crit}$. 
Making use of the expansion of $G^{0}(\vec r, \vec r \,', z)$
as given by Eq.\ \eqref{EQ:GF-FEG}
 $G(\vec r, \vec r \,', z)$   can be written 
for  $r > r_{\rm crit}$ and $r\, ' > r_{\rm crit}$
 as:
\begin{widetext}
\begin{eqnarray}
\LABEL{EQ:SS-GF-A}
%----------------------------------------------------------------------------------------
G^{n}(\vec r, \vec r \,', z)
%----------------------------------------------------------------------------------------
&=& 
-i \pbar \sum_{\Lambda \Lambda '} \big( j_{\Lambda} (\vec r, z)\, \delta_{\Lambda \Lambda '} - 
i \pbar \, h_{\Lambda}^{+} (\vec r, z) \,t_{\Lambda \Lambda '}(z)\big) \, h_{\Lambda '}^{+ \times} (\vec r \,', z)\, \theta(r ' - r)  \nonumber\\
%----------------------------------------------------------------------------------------
%
%
%----------------------------------------------------------------------------------------
&& \quad \qquad  +  h_{\Lambda}^{+} (\vec r, z)\, \big( j_{\Lambda}^{\times} (\vec r \,', z) \,   \delta_{\Lambda \Lambda '} - i \pbar \, t_{\Lambda \Lambda '}(z)\, h_{\Lambda '}^{+ \times} (\vec r \,', z) \big) \theta(r - r ')  \; ,
%----------------------------------------------------------------------------------------
%
%
\end{eqnarray}
\end{widetext}
where we introduced the formal definition for the single-site $t$-matrix
\begin{eqnarray}
\LABEL{EQ:t-jVj}
%----------------------------------------------------------------------------------------
t_{\Lambda \Lambda '}(z) =  \int d^{3} r \int d^{3} r \,' j_{\Lambda}^{\times}(\vec r,z) \, t(\vec r , \vec r \, ', z)
 \, j_{\Lambda '}(\vec r\,',z) \; .
%----------------------------------------------------------------------------------------
%
%
\end{eqnarray}
Eq.\ \eqref{EQ:SS-GF-A}
suggests to introduce special
 RHS and LHS solutions to the radial Dirac equations
\eqref{EQ:RAD-DEQU-RHS}
 and \eqref{EQ:RAD-DEQU-LHS}, respectively,
 by specifying their asymptotic behavior for $r > r_{\rm crit}$  according to:
\begin{eqnarray}
\LABEL{EQ:RHS-ASYMPT-R}
%----------------------------------------------------------------------------------------
R_{\Lambda}(\vec r, z)& = &\sum_{\Lambda '}  j_{\Lambda} (\vec r, z)\, \delta_{\Lambda \Lambda '} - 
i \pbar \, h_{\Lambda'}^{+} (\vec r, z) \,t_{\Lambda ' \Lambda}(z) \\
%----------------------------------------------------------------------------------------
\LABEL{EQ:RHS-ASYMPT-H}
%----------------------------------------------------------------------------------------
H_{\Lambda}(\vec r, z)& = &h_{\Lambda}^{+} (\vec r, z)  \\
%----------------------------------------------------------------------------------------
%
\LABEL{EQ:LHS-ASYMPT-R}
%----------------------------------------------------------------------------------------
R_{\Lambda}^{\times}(\vec r, z)& = &\sum_{\Lambda '}  j_{\Lambda}^{\times} (\vec r, z)\, \delta_{\Lambda \Lambda '} - 
i \pbar \, t_{\Lambda \Lambda '}(z)\,  h_{\Lambda '}^{+ \times} (\vec r, z) \\
%----------------------------------------------------------------------------------------
\LABEL{EQ:LHS-ASYMPT-H}
%----------------------------------------------------------------------------------------
H_{\Lambda}^{\times}(\vec r, z)& = &h_{\Lambda}^{+ \times} (\vec r, z)  \; ,
%----------------------------------------------------------------------------------------
%
\end{eqnarray}
with the label $\Lambda$ used to specify
 the boundary conditions for these functions
 (see Eq.\ \eqref{EQ:ANSATZ-RHS}).

The functions $R_{\Lambda}(\vec r, z)$ and $R_{\Lambda}^{\times}(\vec r, z)$
in Eqs.\ \eqref{EQ:RHS-ASYMPT-R}  and  \eqref{EQ:LHS-ASYMPT-R},
respectively,
can also be seen as solutions to a
corresponding Lippmann-Schwinger equation. 
For the RHS 
case the two equivalent forms of this equation
in terms of the potential and the t-matrix operator, respectively,
 are given by:
\begin{eqnarray}
%
%----------------------------------------------------------------------------------------
\LABEL{EQ:OP-LSEQ-RHS-V}
| R(z) \rangle   & = & | R^{0}(z) \rangle  + \hat G^{0}(z) \, 
\big(\hat V + \hat \Sigma(z) \big) \, | R(z) \rangle  \\
%----------------------------------------------------------------------------------------
%
%
%----------------------------------------------------------------------------------------
\LABEL{EQ:OP-LSEQ-RHS-T}
  & = & | R^{0}(z) \rangle  + \hat G^{0}(z) \, 
\hat t(z) \, | R^{0}(z) \rangle  \;,
%----------------------------------------------------------------------------------------
%
\end{eqnarray}
with $| R^{0}(z) \rangle $ the solution for the free electron case
as given by Eq.\ \eqref{EQ:REG-RHS-j}.
Adopting again a  real space angular momentum representation
these equations correspond to
\begin{widetext}
\begin{eqnarray}
\LABEL{EQ:SS-GPOT-LSEQU-R-RHS}
%----------------------------------------------------------------------------------------
R_{\Lambda}(\vec r, z)   &= &j_{\Lambda}(\vec r, z)+ \int d^{3} r\, ' \int d^{3} r \,'' \, 
G^{0}(\vec r, \vec r \,' , z) 
% \nonumber\\ && \quad 
\, \big( V(\vec r \,') 
          \delta(\vec r \,', \vec r \,'') +
          \Sigma(\vec r \,', \vec r \,'', z) \big) \, R_{\Lambda}(\vec r \,'', z)   \\
%----------------------------------------------------------------------------------------
%
%
%----------------------------------------------------------------------------------------
\LABEL{EQ:SS-GPOT-LSEQU-R-G0tj-RHS}
 &  = &j_{\Lambda}(\vec r, z) + \int d^{3} r\, ' \int d^{3} r \,'' \, 
G^{0}(\vec r, \vec r \,' , z) 
% \nonumber\\ && \qquad \qquad  
\, t(\vec r \,', \vec r \,'' , z) \, j_{\Lambda}(\vec r \,'', z)  
\; .
%----------------------------------------------------------------------------------------
%
\end{eqnarray}
\end{widetext}
From the boundary conditions reflected by these equations it is obvious 
that the function $R_{\Lambda}(\vec r, z)$ will be regular at the origin 
($r = 0$). On the other hand, Eq.\ \eqref{EQ:SS-DYSON-EQ-T} implies that 
the function $H_{\Lambda}(\vec r, z)$ will in general be irregular at the origin. 
The same applies to the LHS functions 
$R_{\Lambda}^{\times}(\vec r, z)$ and $H_{\Lambda}^{\times}(\vec r, z)$ (see below).

Making use of the explicit expression for the
free-electron Green function given in Eq.\ \eqref{EQ:GF-FEG}
one has for $\vec r$ 
outside the potential regime ($r > r_{\rm crit}$) 
 the asymptotic behavior:
\begin{widetext}
\begin{eqnarray}
%
%----------------------------------------------------------------------------------------
R_{\Lambda}(\vec r, z)  &=& j_{\Lambda}(\vec r, z) -i \pbar \sum_{\Lambda '} \, h_{\Lambda '}^{+}(\vec r, z) \int d^{3} r\, ' \int d^{3} r \,'' \,  j_{\Lambda '}^{\times}(\vec r \,', z) \, 
\big( V(\vec r \,') 
          \delta(\vec r \,', \vec r \,'') +
          \Sigma(\vec r \,', \vec r \,'', z) \big)
 \, R_{\Lambda}(\vec r \,'', z)   \\
%----------------------------------------------------------------------------------------
%----------------------------------------------------------------------------------------
&=& j_{\Lambda}(\vec r, z) -i \pbar \sum_{\Lambda '} \, h_{\Lambda '}^{+}(\vec r, z) \int d^{3} r\, ' \int d^{3} r \,'' \,  j_{\Lambda '}^{\times}(\vec r \,', z) \, t(\vec r \,', \vec r \,'' , z) \, j_{\Lambda}(\vec r \,'', z) 
%----------------------------------------------------------------------------------------
%%
\end{eqnarray}
\end{widetext}
implying for the $t$-matrix the relation:
\begin{eqnarray}
\LABEL{EQ:t-matrix-jVR-RHS}
%----------------------------------------------------------------------------------------
 t_{\Lambda ' \Lambda}(z) & = & \int d^{3} r\, \int d^{3} r \,' \, j_{\Lambda '}^{\times}(\vec r , z) \,
 \nonumber \\
 && \; 
\big( V(\vec r) 
          \delta(\vec r, \vec r \,') +
          \Sigma(\vec r, \vec r \,', z) \big)
 \, R_{\Lambda}(\vec r \,', z)   \; .
% 
%----------------------------------------------------------------------------------------
%
\end{eqnarray}
Dealing with the LHS solutions $\langle R^{\times}(z)| $ to the Dirac equation
one is led to the two equivalent forms 
of the Lippmann-Schwinger equation:
\begin{eqnarray}
%
%----------------------------------------------------------------------------------------
\LABEL{EQ:OP-LSEQ-LHS-V}
\langle R^{\times}(z)| &=& \langle R^{0 \times}(z)| +  
\langle R^{ \times}(z)|\, \big( \hat V + \hat \Sigma(z) \big)\, 
\hat G^{0}(z)  \\
%----------------------------------------------------------------------------------------
\LABEL{EQ:OP-LSEQ-LHS-T}
&=& \langle R^{0 \times}(z)|+  
\langle R^{0 \times}(z)|\, \hat t(z) \, \hat G^{0}(z) 
%----------------------------------------------------------------------------------------
%
\end{eqnarray}
with their real space angular momentum representation given by:
\begin{widetext}
\begin{eqnarray}
\LABEL{EQ:SS-GPOT-LSEQU-R-LHS}
%
%----------------------------------------------------------------------------------------
R_{\Lambda}^{\times}(\vec r, z)  &=& j_{\Lambda}^{\times}(\vec r, z) -i \pbar \sum_{\Lambda '} \,\int d^{3} r\, ' \int d^{3} r \,'' \,  R_{\Lambda}^{\times}(\vec r \,', z) 
\,
\big( V(\vec r \,') 
          \delta(\vec r \,', \vec r \,'') +
          \Sigma(\vec r \,', \vec r \,'', z) \big)
 \, j_{\Lambda '}(\vec r \,'', z)\,   h_{\Lambda '}^{+ \times}(\vec r, z) \\
%----------------------------------------------------------------------------------------
%----------------------------------------------------------------------------------------
\LABEL{EQ:SS-GPOT-LSEQU-R-jtG0-LHS}
&=& j_{\Lambda}^{\times}(\vec r, z) -i \pbar \sum_{\Lambda '} \, \int d^{3} r\, ' \int d^{3} r \,'' \,  j_{\Lambda}^{\times}(\vec r \,', z) \,t(\vec r \,', \vec r \,'' , z) \, j_{\Lambda '}(\vec r \,'', z)\, h_{\Lambda '}^{+ \times}(\vec r, z)
\; .  
%----------------------------------------------------------------------------------------
%%
\end{eqnarray}
\end{widetext}
This implies for the $t$-matrix the alternative and completely equivalent expression
in terms of the LHS solution $\langle R^{\times}(z)|$:
\begin{eqnarray}
\LABEL{EQ:t-matrix-RVj-LHS}
%----------------------------------------------------------------------------------------
 t_{\Lambda \Lambda '}(z)  &=&  \int d^{3} r \int d^{3} r \,' \, R_{\Lambda}^{\times}(\vec r , z) 
 \nonumber \\
&& \;
\big( V(\vec r ) 
          \delta(\vec r , \vec r \,') +
          \Sigma(\vec r , \vec r \,', z) \big)
\,
  j_{\Lambda '}(\vec r \,', z)   \;.
%----------------------------------------------------------------------------------------
%
\end{eqnarray}
To see that Eqs.\ \eqref{EQ:t-matrix-jVR-RHS} 
             and  \eqref{EQ:t-matrix-RVj-LHS}
are indeed equivalent one can insert repeatedly 
Eq.\ \eqref{EQ:OP-LSEQ-RHS-V} and  \eqref{EQ:OP-LSEQ-LHS-V}, respectively.
The resulting series can be replaced by the t-matrix 
that satisfies the relations
\begin{eqnarray}
%
%\LABEL{}
%----------------------------------------------------------------------------------------
 \hat t(z) 
 &=&  
\Big( 1-  \big( \hat V + \hat \Sigma(z) \big) \, \hat G^{0}(z)  \Big)^{-1} \,
 \big( \hat V + \hat \Sigma(z) \big)  \\
 &=&  
 \big( \hat V + \hat \Sigma(z) \big)  \,
\Big( 1-   \hat G^{0}(z) \, \big( \hat V + \hat \Sigma(z) \big) \Big)^{-1}
 \; .
%----------------------------------------------------------------------------------------
\end{eqnarray}
In both cases one is led this way to Eq.\ \eqref{EQ:t-jVj}.
It should be noted that
this manipulation in addition implies the helpful relations:
\begin{eqnarray}
%
%----------------------------------------------------------------------------------------
\LABEL{EQ:VR-tR0-RHS}
\big( \hat V + \hat \Sigma(z) \big) \, | R(z) \rangle   & = & \hat t(z) \,| R^{0}(z) \rangle \\
\LABEL{EQ:RV-R0t-LHS}
\langle R^{\times}(z)| \,\big( \hat V + \hat \Sigma(z) \big)   & = &  \langle R^{0 \times}(z)|\, \hat t(z)
 \; .
%----------------------------------------------------------------------------------------
%
\end{eqnarray}
\medskip

The Lippmann-Schwinger equations  
\eqref{EQ:SS-GPOT-LSEQU-R-RHS} and
\eqref{EQ:SS-GPOT-LSEQU-R-LHS}
for  the RHS and LHS solutions $R_{\Lambda}(\vec r, z)$ 
and $R_{\Lambda}^{\times}(\vec r, z)$, respectively,
together with the expression \eqref{EQ:GF-FEG} 
for the free electron Green function $G^{0}(\vec r, \vec r \,', z)$
imply that these functions 
have   for $r \rightarrow 0 $ the
asymptotic behavior
%
%----------------------------------------------------------------------------------------
\begin{eqnarray}
\LABEL{EQ:ASYPT-0-R-RHS}
  R_{\Lambda}(\vec r, z)           & = & \sum_{\Lambda '} \, j_{\Lambda '}(\vec r, z) \, \alpha^{}_{\Lambda '\Lambda}(z) \\
\LABEL{EQ:ASYPT-0-R-LHS}
     R_{\Lambda}^{\times}(\vec r, z)&  = &\sum_{\Lambda '} \, \alpha^{\times}_{\Lambda\Lambda '}(z)  \, j_{\Lambda '}^{\times}(\vec r, z) \; .
\end{eqnarray}
%----------------------------------------------------------------------------------------
%
Here   the so-called  enhancement factors
have been introduced,
that play an important role for the 
so-called Lloyd formula 
for the integrated density of states.\cite{KB90,TK94,Zel05}
Due to Eqs.\   
\eqref{EQ:SS-GPOT-LSEQU-R-RHS} and
\eqref{EQ:SS-GPOT-LSEQU-R-LHS}
as well as  Eqs.\   
\eqref{EQ:SS-GPOT-LSEQU-R-G0tj-RHS} and
\eqref{EQ:SS-GPOT-LSEQU-R-jtG0-LHS}
or 
alternatively 
Eqs.\  \eqref{EQ:VR-tR0-RHS} and
       \eqref{EQ:RV-R0t-LHS},
respectively,
these quantities are given by:
\begin{widetext}
%----------------------------------------------------------------------------------------
\begin{eqnarray}
\LABEL{ALPHA-h-V-R-RHS}
\alpha^{}_{\Lambda '\Lambda}(z)  
&=& 
\delta_{\Lambda '\Lambda} -i \pbar 
\int d^{3} r \int d^{3} r \,' \,  h_{\Lambda '}^{+\times}(\vec r, z) \, 
\big( V(\vec r ) 
          \delta(\vec r , \vec r \,') +
          \Sigma(\vec r , \vec r \,', z) \big)
 \, R_{\Lambda}(\vec r \,', z) 
\\
%----------------------------------------------------------------------------------------
\LABEL{ALPHA-h-t-j-RHS}
&=& 
\delta_{\Lambda '\Lambda} -i \pbar 
\int d^{3} r \int d^{3} r \,' \,  h_{\Lambda '}^{+\times}(\vec r, z) \, 
          t(\vec r , \vec r \,', z) 
 \, j_{\Lambda}(\vec r \,', z) 
\\
%----------------------------------------------------------------------------------------
%----------------------------------------------------------------------------------------
\LABEL{ALPHA-R-V-h-LHS}
\alpha^{\times}_{\Lambda\Lambda '}(z)  &=& 
\delta_{\Lambda \Lambda'} 
 -i \pbar \int d^{3} r \int d^{3} r \,' \,  R_{\Lambda}^{\times}(\vec r, z) 
\,
\big( V(\vec r ) 
          \delta(\vec r , \vec r \,') +
          \Sigma(\vec r , \vec r \,', z) \big)
 \, h_{\Lambda '}^{+}(\vec r \,', z) \\
%----------------------------------------------------------------------------------------
\LABEL{ALPHA-j-t-h-LHS}
  &=& 
\delta_{\Lambda \Lambda'} 
 -i \pbar \int d^{3} r \int d^{3} r \,' \,  j_{\Lambda}^{\times}(\vec r, z) 
\,
          t(\vec r , \vec r \,', z) 
 \, h_{\Lambda '}^{+}(\vec r \,', z)
\;.
\end{eqnarray}
%----------------------------------------------------------------------------------------
\end{widetext}
The simple behavior given 
 in terms 
of the relativistic spherical Bessel functions
expressed by  Eqs.\ \eqref{EQ:ASYPT-0-R-RHS}
and  \eqref{EQ:ASYPT-0-R-LHS}
clearly shows that $R_{\Lambda}(\vec r, z)$ and  
$R_{\Lambda}^{\times}(\vec r, z)$ 
are indeed  regular solutions at the origin ($r=0$).

As the regular  RHS and LHS solutions $R_{\Lambda}(\vec r, z)$ 
and $R_{\Lambda}^{\times}(\vec r, z)$, respectively,
also their irregular   counterparts 
$H_{\Lambda}(\vec r, z)$ 
and $H_{\Lambda}^{\times}(\vec r, z)$
can be expressed in terms of a corresponding Lippmann-Schwinger
equation.
Taking into account their asymptotic behavior for $r > r_{\rm crit}$
as expressed by Eqs.\ \eqref{EQ:RHS-ASYMPT-H} and \eqref{EQ:LHS-ASYMPT-H},
respectively,
one is led to the expressions:
\begin{widetext}
\begin{eqnarray}
\LABEL{EQ:SS-GPOT-LSEQU-H-RHS}
%----------------------------------------------------------------------------------------
H_{\Lambda}(\vec r, z)   &= &
 \sum_{\Lambda '} h_{\Lambda'}^{+}(\vec r, z) 
\Bigg[ \delta_{\Lambda \Lambda '} -i \pbar   \int d^{3} r\, ' \int d^{3} r \,'' \, 
j_{\Lambda'}(\vec r\, ', z)
\, \big( V(\vec r \,') 
          \delta(\vec r \,', \vec r \,'') +
          \Sigma(\vec r \,', \vec r \,'', z) \big) \, H_{\Lambda}(\vec r \,'', z) \Bigg]  \nonumber \\
  & &\qquad + \int d^{3} r\, ' \int d^{3} r \,'' \, 
G^{0}(\vec r, \vec r \,' , z) 
% \nonumber\\ && \quad 
\, \big( V(\vec r \,') 
          \delta(\vec r \,', \vec r \,'') +
          \Sigma(\vec r \,', \vec r \,'', z) \big) \, H_{\Lambda}(\vec r \,'', z)  \\
%----------------------------------------------------------------------------------------
%
\LABEL{EQ:SS-GPOT-LSEQU-H-LHS}
H_{\Lambda}^{\times}(\vec r, z)  &=& 
 \sum_{\Lambda '}
\Bigg[ \delta_{\Lambda \Lambda '} 
-i \pbar
\,\int d^{3} r\, ' \int d^{3} r \,'' \,  H_{\Lambda}^{\times}(\vec r \,', z) 
\,
\big( V(\vec r \,') 
          \delta(\vec r \,', \vec r \,'') +
          \Sigma(\vec r \,', \vec r \,'', z) \big)
\,  j_{\Lambda'}(\vec r \,'', z)  \Bigg]   h_{\Lambda}^{+\times}(\vec r, z)     \nonumber  \\
  & & \qquad +
\int d^{3} r\, ' \int d^{3} r \,'' \,  H_{\Lambda}^{\times}(\vec r \,', z) 
\,
\big( V(\vec r \,') 
          \delta(\vec r \,', \vec r \,'') +
          \Sigma(\vec r \,', \vec r \,'', z) \big)
\, G^{0}(\vec r \,'', \vec r, z) \;.
%----------------------------------------------------------------------------------------
%
\end{eqnarray}
\end{widetext}
Again these Lippmann-Schwinger equations imply
for the  RHS and LHS solutions $H_{\Lambda}(\vec r, z)$ 
and $H_{\Lambda}^{\times}(\vec r, z)$, respectively,
a simple 
asymptotic behavior  for $r \rightarrow 0 $:
%
%----------------------------------------------------------------------------------------
\begin{eqnarray}
\LABEL{EQ:ASYPT-0-H-RHS}
  H_{\Lambda}(\vec r, z)           & = & \sum_{\Lambda '} \, h_{\Lambda '}^{+}(\vec r, z) \, \beta^{}_{\Lambda '\Lambda}(z) \\
\LABEL{EQ:ASYPT-0-H-LHS}
     H_{\Lambda}^{\times}(\vec r, z)&  = &\sum_{\Lambda '} \, \beta^{\times}_{\Lambda\Lambda '}(z)  \, h_{\Lambda '}^{+\times}(\vec r, z) \; ,
\end{eqnarray}
%----------------------------------------------------------------------------------------
%
with the corresponding enhancement factors
\begin{widetext}
%----------------------------------------------------------------------------------------
\begin{eqnarray}
\LABEL{BETA-j-V-H-RHS}
\beta^{}_{\Lambda '\Lambda}(z)  &=& 
 \delta_{\Lambda \Lambda '} -i \pbar   \int d^{3} r \int d^{3} r \,' \, 
j_{\Lambda'}(\vec r, z)
\, \big( V(\vec r \,') 
          \delta(\vec r , \vec r \,') +
          \Sigma(\vec r , \vec r \,', z) \big) \, H_{\Lambda}(\vec r \,', z) 
\\
%----------------------------------------------------------------------------------------
\LABEL{BETA-H-V-j-LHS}
\beta^{\times}_{\Lambda\Lambda '}(z)  &=& 
 \delta_{\Lambda \Lambda '} 
-i \pbar
\,\int d^{3} r \int d^{3} r \,' \,  H_{\Lambda}^{\times}(\vec r , z) 
\,
\big( V(\vec r ) 
          \delta(\vec r , \vec r \,') +
          \Sigma(\vec r , \vec r \,', z) \big)
\,  j_{\Lambda'}(\vec r \,', z) 
\;.
\end{eqnarray}
%----------------------------------------------------------------------------------------
\end{widetext}
Again, the  simple behavior in terms 
of the relativistic Hankel  functions
clearly shows that $H_{\Lambda}(\vec r, z)$ and  
$H_{\Lambda}^{\times}(\vec r, z)$ 
are indeed  irregular solutions at the origin ($r=0$). 

Finally, it should be emphasized here that the 
 self-energy $\Sigma(\vec r , \vec r \,', z)$
enters the expressions for the various
 enhancement factors 
$ \alpha^{(\times)}_{\Lambda\Lambda '}(z)$,
and  $\beta^{(\times)}_{\Lambda\Lambda '}(z)$,
but does not alter the scaling behavior as such
for the 
corresponding wave functions 
for $r \rightarrow 0 $
as given by Eqs.\ \eqref{EQ:ASYPT-0-R-RHS}, \eqref{EQ:ASYPT-0-R-LHS},
\eqref{EQ:ASYPT-0-H-RHS}, and \eqref{EQ:ASYPT-0-H-LHS},
respectively.

%****************************************************************
\subsection{Relativistic Wronskian}
%****************************************************************

The relativistic form of the
 Wronskian for arbitrary
 RHS and LHS 
 solutions  $\psi_{\nu}$ and 
 $\phi^\times_{\nu'}$ to the 
 corresponding Dirac equations 
\eqref{EQ:DEQU-RHS} and
\eqref{EQ:DEQU-LHS}, respectively, is obtained
by multiplying the radial 
RHS  equation \eqref{EQ:RAD-DEQU-RHS}
from the left with 
 the matrix $A= \Big( \begin{array}{cc} 1 & ~~0 \\[-2ex] 0 &-1  \end{array}\Big)$
and then with
the 
row vector of LHS radial  functions 
$\mbox{\bf \sf x}_{\Lambda''\nu'}^{\times T}(r, z)
=
 \big( g_{\Lambda''\nu'}^{\times}(r,z) \, f_{\Lambda''\nu'}^{\times}(r,z) \big)$. 
Analogously, the radial  
LHS equation \eqref{EQ:RAD-DEQU-LHS}
is also multiplied from the left with the
matrix $A$
and then with the 
row vector of RHS radial  functions 
$\mbox{\bf \sf x}_{\Lambda''\nu}^{T}(r, z)
=
\big( g_{\Lambda''\nu}(r,z) \, f_{\Lambda''\nu}(r,z) \big)$. 
 Both resulting equations are subtracted from each 
other and finally a sum is taken over $\Lambda''$.
Representing the first $2 \times 2$
matrix occurring in 
Eqs.\ \eqref{EQ:RAD-DEQU-RHS}
and \eqref{EQ:RAD-DEQU-LHS}
involving the differential operator $\partial / \partial r$
by the symbol $\mbox{\bf \sf D}_{\kappa''}(r)$
one finds for this term:
%%%%.
%%%%
%%%%
%%%%Dealing with the terms connected with  $\mbox{\bf \sf D}_{\kappa''}(r)$
%%%%one is led to the result:
%
\begin{eqnarray}
%----------------------------------------------------------------------------------------
\LABEL{EQ:WRON-differential}
\sum_{\Lambda''} 
\mbox{\bf \sf x}_{\Lambda''\nu'}^{\times T}(r, z)\,
\mbox{\bf \sf D}_{\kappa''}(r)\,
\mbox{\bf \sf x}_{\Lambda''\nu}^{}(r, z)
\qquad\qquad \nonumber \\ \qquad
- 
\mbox{\bf \sf x}_{\Lambda''\nu}^{T}(r, z)\,
\mbox{\bf \sf D}_{\kappa''}(r)\,
\mbox{\bf \sf x}_{\Lambda''\nu'}^{\times}(r, z)
\qquad\qquad  \nonumber \\
  =
 \frac{\partial}{\partial \,r}
\sum_{\Lambda''}
 \,c  r^2
 \Big(  
g_{\Lambda''\nu'}^{\times}(r, z)\,    f_{\Lambda''\nu}        (r, z)  \quad  \nonumber\\
\hspace{4cm}    -   
g_{\Lambda''\nu}        (r, z)\,    f_{\Lambda''\nu'}^{\times}(r, z) 
\Big)
\; . 
%
%----------------------------------------------------------------------------------------
%%
\end{eqnarray}
%
%where as in the following the index $\nu$
%specifying the asymptotic behaviour of the wave functions
%has been dropped to simplify the notation.

%%the RHS and LHS local potential 
%%%and non-local self-energy terms 
%%%%%%% are represented %%%by $\mbox{\bf \sf V}_{\Lambda''  \Lambda'''}^{(\times)}(r)$
%%%and 
%%%$\mbox{\boldmath{$\Sigma$}}_{\Lambda''  \Lambda'''}^{(\times)}( r,  r \,',z)$, respectively,

Dealing with the  terms connected with the
local potential functions
$\mbox{\bf \sf V}_{\Lambda''  \Lambda'''}^{(\times)}(r) $
(second term in 
Eqs.\ \eqref{EQ:RAD-DEQU-RHS}
and \eqref{EQ:RAD-DEQU-LHS}, respectively)
one finds by making use of their Hermiticity:
\begin{eqnarray}
\LABEL{EQ:WRON-potential}
%----------------------------------------------------------------------------------------
 \sum_{\Lambda''\Lambda'''} 
\mbox{\bf \sf x}_{\Lambda''\nu '}^{\times T}(r, z)\,
\mbox{\bf \sf V}_{\Lambda''  \Lambda'''}^{}(r) \,
\mbox{\bf \sf x}_{\Lambda'''\nu}^{}(r, z)
\quad && \nonumber \\
\;
- 
\mbox{\bf \sf x}_{\Lambda''\nu}^{T}(r, z)\,
\mbox{\bf \sf V}_{\Lambda''  \Lambda'''}^{\times}(r) \,
\mbox{\bf \sf x}_{\Lambda'''\nu'}^{\times}(r, z)
 &= &0 \; .
%
%----------------------------------------------------------------------------------------
%%
\end{eqnarray}
Restricting for the moment to the case of a local potential
Eqs.\ \eqref{EQ:WRON-differential} and
      \eqref{EQ:WRON-potential}
imply for  
 the relativistic Wronskian
of the RHS and LHS functions 
$\psi_{\nu}(\vec r, z)$ and $\phi^\times_{\nu'}(\vec r, z)$, respectively,
 defined by:\cite{Tam82-WRONSKIAN}
\begin{eqnarray}
\LABEL{EQ:WRON-DEF-gf}
%----------------------------------------------------------------------------------------
\left[ \phi^{\times}_{\nu'}  ,   \psi_{\nu}  \right ]& = 
& \sum_{\Lambda'' }c  %%%%%%%%%%%%%%%% DEF r^{2} \,  
\Big(  g_{\Lambda''  \nu'}^{\times}(r, z) \, f_{\Lambda''   \nu }(r, z)
\nonumber \\
&& \quad\quad
-  f_{\Lambda''  \nu' }^{\times}(r, z) \, g_{\Lambda''   \nu }(r, z)  \Big) 
\end{eqnarray}
the simple expression
\begin{eqnarray}
\LABEL{EQ:WRON-gf-const}
%----------------------------------------------------------------------------------------
\left[ \phi^{\times}_{\nu'}  ,   \psi_{\nu}  \right ]
= - \left[ \psi_{\nu}   , \phi^{\times}_{\nu'}   \right]
 = \frac{C}{r^2}\; ,
%
%----------------------------------------------------------------------------------------
%
\end{eqnarray}
where $C$ is a constant.

To fix the constant $C$ in Eq.\ \eqref{EQ:WRON-gf-const} 
one considers the free-electron   solutions  
$\psi_{\nu}(\vec r, z) = j_{\Lambda}(\vec r, z)$ or $h_{\Lambda}^{+}(\vec r, z)$
and  
$\phi^\times_{\nu'}(\vec r, z) = j^\times_{\Lambda'}(\vec r, z)$ or $h_{\Lambda'}^{+\times}(\vec r, z)$, respectively,
given  
in Eqs.\  \eqref{EQ:REG-RHS-j} to \eqref{EQ:REG-LHS-h} for 
which one can identify the label $\nu$ ($\nu'$)
with $\Lambda$ ($\Lambda'$).
Furthermore, these functions have pure spin-angular 
character $\Lambda''$. For that reason there is only
one contribution to the sum in 
Eq.\ \eqref{EQ:WRON-gf-const}
with  $\Lambda''=\Lambda'=\Lambda$.
Making use of the standard Wronskian of 
 the non-relativistic spherical Bessel and Hankel functions \cite{AS64}
that is leading to the relation
\begin{equation}
%
%----------------------------------------------------------------------------------------
S_{\kappa}\Big(j_{l}(x)\, h_{\bar l}^{+}(x) - j_{\bar l}(x)\, h_{l}^{+}(x)\Big)  
= \frac{i}{x^{2}}
%----------------------------------------------------------------------------------------
%
\end{equation}
  one finds
for their relativistic counterparts as defined
by Eqs.\  \eqref{EQ:REG-RHS-j} to \eqref{EQ:REG-LHS-h}:
\begin{eqnarray}
\LABEL{EQ:WRONSKIAN-jj-hh}
\left[j_{\Lambda}^{\times} , j_{\Lambda '} \right] 
&=& \quad \left[h_{\Lambda}^{ \times} , h_{\Lambda '} \right] = 0  \\
\LABEL{EQ:WRONSKIAN-jh-hj}
\left[j_{\Lambda}^{\times} , h_{\Lambda '} \right] 
&=& - \left[h_{\Lambda}^{\times} , j_{\Lambda '} \right] \, = 
w %%% NOTE DEFINITION \frac{i}{pr^2}  \frac{1}{1+E/c^2} 
 \, \delta_{\Lambda \Lambda '}  \; ,
\end{eqnarray}
with
\begin{equation}
\label{rel-w}
%----------------------------------------------------------------------------------------
w = \frac{i}{\zeta pr^2} = \frac{i}{\pbar r^2} 
%----------------------------------------------------------------------------------------
%
\end{equation}
where 
$\zeta(z) = 1+z/c^{2}  $ (see Eq.\ \eqref{EQ:rel-zeta}).

With the asymptotic behavior of the normalized functions 
$R_{\Lambda}(\vec r, z)$, $H_{\Lambda}(\vec r, z)$, 
$R_{\Lambda}^{\times}(\vec r, z)$  and $H_{\Lambda}^{\times}(\vec r, z)$ 
given by Eqs.\ \eqref{EQ:RHS-ASYMPT-R} to \eqref{EQ:LHS-ASYMPT-H} one has
accordingly for $\Sigma (\vec r, \vec r \,', z)=0$:
\begin{eqnarray}
\LABEL{WRONSKIAN-RR-HH}
\left[R_{\Lambda}^{\times} , R_{\Lambda '} \right] 
&=& \quad \left[H_{\Lambda}^{\times} , H_{\Lambda '} \right] = 0  \\
%
%-----------------------------------------------------------------
\LABEL{WRONSKIAN-RH-HR}
\left[R_{\Lambda}^{\times} , H_{\Lambda '} \right] 
&=& - \left[H_{\Lambda}^{\times} , R_{\Lambda '} \right]\, = 
w %%% NOTE DEFINITION \frac{i}{pr^2} \frac{1}{1+E/c^2} 
 \, \delta_{\Lambda \Lambda '}  \;.
\end{eqnarray}
Because of the relation given by Eq.\ \eqref{EQ:WRON-potential}
this holds not only for $r > r_{\rm crit}$, 
but for all $r$. 

 \medskip

As pointed out by  Tamura 
\cite{Tam92}, any general  RHS and LHS solutions,
 $\psi_{\nu}(\vec r,z)$ and 
 $\phi^{\times}_{\nu'}(\vec r,z)$, can be expanded in terms of the 
normalized solutions
$R_{\Lambda}(\vec r,z) $ and 
$H_{\Lambda}(\vec r,z) $, respectively,
\begin{eqnarray}
\label{EQ:28a}
%----------------------------------------------------------------------------------------
\psi_{\nu}(\vec r,z) &= & \sum_{\Lambda} R_{\Lambda}(\vec r,z) \, C_{\Lambda\nu}(z) 
 +
H_{\Lambda}(\vec r,z)\, S_{\Lambda\nu}(z)
\\                    
%----------------------------------------------------------------------------------------
%
\label{EQ:28b}
%----------------------------------------------------------------------------------------
\phi^{\times}_{\nu'}(\vec r,z)& =&  \sum_{\Lambda} C_{\nu'\Lambda}^{\times}(z) \, R_{\Lambda}^{\times}(\vec r,z) +
 S_{\nu'\Lambda}^{\times}(z)\, H_{\Lambda}^{\times}(\vec r,z)                                         
%----------------------------------------------------------------------------------------
%
\end{eqnarray}
with the expansion coefficients given by
 the Wronski relations
\begin{eqnarray}
\label{EQ:29a}
%----------------------------------------------------------------------------------------
 C_{\Lambda \nu}(z) &=& -\frac{1}{w} \, [  H_{\Lambda }^{\times}, \psi_{\nu} ]                
%----------------------------------------------------------------------------------------
%
%\nonumber 
\\
\label{EQ:29b}
%----------------------------------------------------------------------------------------
 S_{\Lambda \nu}(z) &=& \;\;\; \frac{1}{w} \, [  R_{\Lambda }^{\times}, \psi_{\nu} ]         
%----------------------------------------------------------------------------------------
%
%\nonumber 
\\
\label{EQ:29c}
%----------------------------------------------------------------------------------------
 C_{\nu'\Lambda }^{\times}(z)& =& \;\;\; \frac{1}{w} \, [ \phi^{\times}_{\nu'} , H_{\Lambda } ]        
%----------------------------------------------------------------------------------------
%
%\nonumber 
\\
\label{EQ:29d}
%----------------------------------------------------------------------------------------
 S_{\nu'\Lambda }^{\times}(z) &=& -\frac{1}{w} \, [ \phi^{\times}_{\nu'} , R_{\Lambda } ]            
%----------------------------------------------------------------------------------------
\; .
%\nonumber
%
\end{eqnarray}
Because of the asymptotic behavior of $R_{\Lambda }(\vec r,z) $ and 
$H_{\Lambda }(\vec r,z) $ 
(see  Eqs.\ \eqref{EQ:RHS-ASYMPT-R} to \eqref{EQ:LHS-ASYMPT-H})
and the Wronski relations \eqref{EQ:WRONSKIAN-jj-hh} and
                          \eqref{EQ:WRONSKIAN-jh-hj}  for
the spherical functions 
 one has for the Wronski relation 
of the solutions $ \psi_{\nu }(\vec r,z)$ and $ \phi^{\times}_{\nu'}(\vec r,z)$:
\begin{eqnarray}
%
%----------------------------------------------------------------------------------------
[ \phi^{\times}_{\nu'}, \psi_{\nu }] &=&  w\, \sum_{\Lambda } 
  C_{  \nu' \Lambda }^{\times}(z)  \, S_{\Lambda   \nu  }(z) 
- S_{  \nu' \Lambda }^{\times}(z)  \, C_{\Lambda   \nu  }(z)
%----------------------------------------------------------------------------------------
%
\end{eqnarray}
that at the same time is given by Eq.\  \eqref{EQ:WRON-DEF-gf}.
Imposing at an arbitrary point $\vec r$ suitable
values for the small and large
components of  $\phi_{\Lambda}^{\times}(\vec r,z)$ and $\psi_{\Lambda '}(\vec r,z)$,
respectively, 
(indicated by ${\Lambda}$ and ${\Lambda '}$),
Tamura could derive the following additional Wronski
relations of the second kind
for the case of a local potential 
($\Sigma (\vec r, \vec r \,', z)=0$):\cite{Tam92} 
\begin{eqnarray}
\label{EQ:31a}
%----------------------------------------------------------------------------------------
w % NEW DEF \frac{W}{r^{2}}
\, \delta_{\Lambda  \Lambda '} &=& c  \sum_{\Lambda ''} 
  g_{\Lambda ' \Lambda''}^{R \times}(r, z) \, f_{\Lambda  \Lambda ''}^{H}(r, z)     
\nonumber \\ && \qquad
- f_{\Lambda  \Lambda''}^{R \times}(r, z) \,  g_{\Lambda ' \Lambda ''}^{H}(r, z)                     
\\
%----------------------------------------------------------------------------------------
%
\label{EQ:31b}
%----------------------------------------------------------------------------------------
 - w % NEW DEF \frac{W}{r^{2}}
 \, \delta_{\Lambda  \Lambda '} &=& c  \sum_{\Lambda ''} 
  f_{\Lambda ' \Lambda''}^{R \times}(r, z) \, g_{\Lambda  \Lambda ''}^{H}(r, z)  
\nonumber \\ && \qquad
- f_{\Lambda ' \Lambda ''}^{H \times}(r, z) \, g_{\Lambda  \Lambda''}^{R }(r, z)    
\\
%----------------------------------------------------------------------------------------
%
\label{EQ:32a}
%----------------------------------------------------------------------------------------
 0 &=& c  \sum_{\Lambda ''} 
   g_{\Lambda ' \Lambda''}^{R \times} (r, z)  \,g_{\Lambda  \Lambda ''}^{H}(r, z)
\nonumber \\ && \qquad
-  g_{\Lambda ' \Lambda ''}^{H \times}(r, z)  \,g_{\Lambda  \Lambda''}^{R }(r, z)                                
\\
%----------------------------------------------------------------------------------------
%
\label{EQ:32b}
%----------------------------------------------------------------------------------------
 0 &=& c  \sum_{\Lambda ''} 
  f_{\Lambda ' \Lambda''}^{R \times}(r, z) \,   f_{\Lambda  \Lambda ''}^{H}(r, z)  
\nonumber \\ && \qquad
- f_{\Lambda ' \Lambda ''}^{H \times}(r, z) \,  f_{\Lambda  \Lambda''}^{R }(r, z)                                 
%----------------------------------------------------------------------------------------
\;,
\end{eqnarray}
where the superscript   $R$ %and $H$ 
 indicates
 for example 
that 
 the 
small component $ f_{\Lambda ' \Lambda}^{R }(r, z) $  belongs
 to the
 normalized function 
$R_{\Lambda}(\vec r, z)$ .

\medskip

If one considers finally 
for  the arbitrary RHS and LHS functions 
$\psi_{\nu}(\vec r,  z)$ and $\phi^\times_{\nu'}(\vec r,  z)$
 the case of
 a finite self-energy represented by a $2 \times 2$ matrix function 
$\Sigma (\vec r, \vec r \,', z)$ 
(third term in 
Eqs.\ \eqref{EQ:RAD-DEQU-RHS}
and \eqref{EQ:RAD-DEQU-LHS}, respectively)
 one finds for the terms
related to this %%$\Sigma (\vec r, \vec r \,', z)$
the contribution:
\begin{eqnarray}
\label{WRONSKIAN-SELF-ENERGY}
%----------------------------------------------------------------------------------------
 \sum_{\Lambda\Lambda'} 
\mbox{\bf \sf x}_{\Lambda\nu'}^{\times T}(r, z)\,
 \int  r'^{2} \, d r'  \,
\mbox{\boldmath{$\Sigma$}}_{\Lambda  \Lambda '}( r,  r \,',z) \,
\mbox{\bf \sf x}_{\Lambda'\nu}^{}(r', z)
\quad && \nonumber \\
\;
- 
\mbox{\bf \sf x}_{\Lambda\nu}^{T}(r, z)\,
 \int  r'^{2} \, d r'  \,
\mbox{\boldmath{$\Sigma$}}_{\Lambda  \Lambda '}^{\times}( r,  r \,',z)  \,
\mbox{\bf \sf x}_{\Lambda'\nu'}^{\times}(r', z)
 & & 
\LABEL{x-SIGMA-x}
 \; .
%
%----------------------------------------------------------------------------------------
%%
\end{eqnarray}
This expression obviously vanishes only if 
$g_{\Lambda\nu}(r, z) =g_{\Lambda\nu'}^{\times}(r, z)$, 
$f_{\Lambda\nu}(r, z) =f_{\Lambda\nu'}^{\times}(r, z)$ 
and $\Sigma_{\Lambda \Lambda '}^{\pm} (\vec r, \vec r \,', z) 
= \Sigma_{\Lambda ' \Lambda}^{\pm} (\vec r \,', \vec r , z)$. 
This means that the Wronskian of a RHS and LHS solution will 
in general not be given by the simple relation
 in Eq.\ \eqref{WRONSKIAN-RH-HR}.
Due to this,  the  Wronskian relations of second kind
given by Eqs.\  \eqref{EQ:31a} to  \eqref{EQ:32b}
will also not hold
with important consequences for the expression for the single site 
Green function  $G^{n}(\vec r, \vec r \,', z)$ (see below).
%%given in  Eq.\ \eqref{}.

\medskip
Obviously, the expression 
in \eqref{x-SIGMA-x} involving the self-energy 
 $\Sigma(\vec r , \vec r \,', z)$
vanishes for $r > r_{\rm crit}$. For that reason 
the Wronski relation in 
Eq.\ \eqref{WRONSKIAN-RH-HR}
holds for $r> r_{\rm crit}$ even for a non-local potential, i.e.\
in case of $\Sigma(\vec r , \vec r \,', z) \ne 0$
for $r$ and  $r' < r_{\rm crit}$. This property can be exploited when
calculating the t-matrix $t_{\Lambda\Lambda'}$ (see below).
Assuming in addition that  $\Sigma(\vec r , \vec r \,', z) = 0$
 in the limit  $r$ and  $r' \rightarrow 0$,
 Eq.\ \eqref{WRONSKIAN-RH-HR} holds also for this regime.
Expressing the  asymptotic behavior of the wave functions 
$R_{\Lambda}(\vec r, z)$,            $H_{\Lambda}(\vec r, z)$, 
$R_{\Lambda}^{\times}(\vec r, z)$ and $H_{\Lambda}^{\times}(\vec r, z)$
in terms of the enhancement factors
$\alpha^{}_{\Lambda '\Lambda}(z)$,
$\alpha^{\times}_{\Lambda\Lambda '}(z) $,
$\beta^{}_{\Lambda '\Lambda}(z)$, and 
$\beta^{\times}_{\Lambda\Lambda '}(z) $,
as given by 
      Eqs.\ \eqref{EQ:ASYPT-0-R-RHS},      \eqref{EQ:ASYPT-0-R-LHS},
            \eqref{EQ:ASYPT-0-H-RHS}, and  \eqref{EQ:ASYPT-0-H-LHS},
respectively 
one is led for these factors to the relations:
\begin{eqnarray}
%----------------------------------------------------------------------------------------
 \sum_{\Lambda''} 
\alpha^{\times}_{\Lambda\Lambda ''}(z) \, \beta^{}_{\Lambda''\Lambda '}(z) 
&=&  \delta_{\Lambda\Lambda '} \\
 \sum_{\Lambda''} 
\beta^{\times}_{\Lambda\Lambda ''}(z) \, \alpha^{}_{\Lambda''\Lambda '}(z) 
 &=&  \delta_{\Lambda\Lambda '} 
\; .
%
%----------------------------------------------------------------------------------------
%%
\end{eqnarray}
These relations hold in particular for a local potential 
($\Sigma(\vec r , \vec r \,', z) = 0$ for any $\vec r$ and $\vec r \,'$)
and connect the asymptotic behavior of the regular 
and irregular wave functions 
$R_{\Lambda}^{\times}(\vec r, z)$ and  $H_{\Lambda}(\vec r, z)$
as well as 
$H_{\Lambda}^{\times}(\vec r, z)$ and $R_{\Lambda}(\vec r, z)$,
respectively.
\medskip

When dealing with 
matrix elements of the potential 
$ V(\vec r ) 
          \delta(\vec r , \vec r \,') +
          \Sigma(\vec r , \vec r \,', z)$
as occurring in Eqs.\ \eqref{EQ:t-jVj}, 
 \eqref{EQ:t-matrix-jVR-RHS},
 \eqref{ALPHA-h-V-R-RHS},
or  \eqref{BETA-j-V-H-RHS},
that involve a LHS free electron like solution
$ \phi_{\Lambda '}^{\times}(\vec r, z)$ as 
given 
in Eqs.\  \eqref{EQ:REG-RHS-j} to \eqref{EQ:REG-LHS-h},
i.e.\ $ \phi_{\Lambda '}^{\times}(\vec r, z) = j_{\Lambda}^{\times}(\vec r, z)$,
$n_{\Lambda}^{\times}(\vec r, z)$ or
$h_{\Lambda}^{\pm  \times }(\vec r, z)$,
it is in general possible to 
convert the volume integral into a surface integral
that in turn can be expressed by
 a corresponding Wronskian.
This is achieved by expressing
 the integral 
$\int d^{3} r \,' \,  
\big( V(\vec r ) 
          \delta(\vec r , \vec r \,') +
          \Sigma(\vec r , \vec r \,', z) \big)
 \, \psi_{\Lambda}(\vec r \,', z) $
by means of the RHS Dirac equation  \eqref{EQ:DEQU-RHS}
and using 
 the LHS Dirac equation  \eqref{EQ:DEQU-LHS}
for the free electron solution  $ \phi_{\Lambda '}^{\times}(\vec r, z)$ 
($\hat {\cal H}^{1}(\vec r) =  
\hat {\cal H}^{0}(\vec r)$):
%%%%\begin{widetext}
%----------------------------------------------------------------------------------------
\begin{eqnarray}
I^{}_{\Lambda '\Lambda}(z)  
&=&  
\int d^{3} r \int d^{3} r \,' \,  \phi_{\Lambda '}^{\times}(\vec r, z) \, 
\nonumber \\ && \qquad
\big( V(\vec r ) 
          \delta(\vec r , \vec r \,') +
          \Sigma(\vec r , \vec r \,', z) \big)
 \, \psi_{\Lambda}(\vec r \,', z) 
\nonumber
\\
%----------------------------------------------------------------------------------------
%%%%%&=& 
%%%%%\int d^{3} r   \,  \phi_{\Lambda '}^{\times}(\vec r, z) \, 
%%%%%%
%%%%%\big( z + ic \vecalpha \cdot \vecnabla 
%%%%% - \frac{1}{2} \, c^{2} (\beta - 1) \big)
%%%%%%
%%%%% \, \psi_{\Lambda}(\vec r , z) 
%%%%%\nonumber \\
%%%%%%----------------------------------------------------------------------------------------
%%%%%&=& 
%%%%% i c
%%%%%\int d^{3} r  \, 
%%%%%\vecnabla \big(
%%%%% \phi_{\Lambda '}^{\times}(\vec r, z) \, 
%%%%%%
%%%%%\vecalpha
%%%%%%
%%%%% \, \psi_{\Lambda}(\vec r , z) 
%%%%% \big)
%%%%%\nonumber \\
%%%%%%----------------------------------------------------------------------------------------
%%%%%&=& 
%%%%% i c
%%%%%\int   d  \vec a  \,  
%%%%% \phi_{\Lambda '}^{\times}(\vec r, z) \, 
%%%%%%
%%%%%\vecalpha
%%%%%%
%%%%% \, \psi_{\Lambda}(\vec r , z) 
%%%%%%%
%%%%%\nonumber \\
%%%%%%----------------------------------------------------------------------------------------
%
%
&=& 
 i c
\Bigg\{
\int_{r=r_{\rm crit}}   d  \hat r  \, r^2 \,  
 \phi_{\Lambda '}^{\times}(\vec r, z) \, 
\alpha_r
 \, \psi_{\Lambda}(\vec r , z)  
\nonumber \\
&& \quad 
- \lim_{r\rightarrow 0}
\int   d  \hat r  \, r^2 \,  
 \phi_{\Lambda '}^{\times}(\vec r, z) \, 
\alpha_r
 \, \psi_{\Lambda}(\vec r , z)  \Bigg\}
\label{EQ:INT-V-VOL-SURF}
 \\
%----------------------------------------------------------------------------------------
%
\label{EQ:INT-V-VOL-WRON-RHS}
&=& 
  c
\Bigg\{
 r_{\rm crit}^2 \,
\Big(
   g_{\Lambda '}^{\phi \times}(r, z) \, f_{\Lambda '\Lambda}^{\psi}(r, z)
\nonumber \\ && \qquad\qquad
-  f_{\Lambda '}^{\phi \times}(r, z) \, g_{\Lambda '\Lambda}^{\psi}(r, z)
\Big)_{r=r_{\rm crit}}
\nonumber \\
&& \quad 
- \lim_{r\rightarrow 0}
 r^2 \,
\Big(
   g_{\Lambda '}^{\phi \times}(r, z) \, f_{\Lambda '\Lambda}^{\psi}(r, z)
\nonumber \\ && \qquad\qquad\quad
-  f_{\Lambda '}^{\phi \times}(r, z) \, g_{\Lambda '\Lambda}^{\psi}(r, z)
\Big)
  \Bigg\}
 \\
%----------------------------------------------------------------------------------------
%
\label{EQ:INT-V-VOL-WRONSKIAN-RHS}
&=& 
 r_{\rm crit}^2 \,
\Big[ \phi_{\Lambda '}^{\times} , \psi_{\Lambda} \Big]_{r=r_{\rm crit}}
\nonumber \\
&& \quad 
- \lim_{r\rightarrow 0}
 r^2 \,
\Big[ \phi_{\Lambda '}^{\times} , \psi_{\Lambda} \Big]_{r}
\; ,
\end{eqnarray}
%----------------------------------------------------------------------------------------
with $g_{\Lambda '}^{\phi\times}(r, z) = g_{\Lambda '' \Lambda'}^{\phi\times}(r, z)\,\delta_{\Lambda '' \Lambda'}$
and  $f_{\Lambda '}^{\phi\times}(r, z) = f_{\Lambda '' \Lambda'}^{\phi\times}(r, z)\,\delta_{\Lambda '' \Lambda'}$.
In an analogous way one finds the relation:
%----------------------------------------------------------------------------------------
\begin{eqnarray}
I^{\times}_{\Lambda\Lambda '}(z)  
&=&  
\int d^{3} r \int d^{3} r \,' \,  \psi_{\Lambda}^{\times}(\vec r, z) \, 
\nonumber \\ && \qquad
\big( V(\vec r ) 
          \delta(\vec r , \vec r \,') +
          \Sigma(\vec r , \vec r \,', z) \big)
 \, \phi_{\Lambda '}(\vec r \,', z) 
\nonumber
\\
%----------------------------------------------------------------------------------------
%
\label{EQ:INT-V-VOL-WRON-LHS}
&=& 
  c
\Bigg\{
 r_{\rm crit}^2 \,
\Big(
   g_{\Lambda '\Lambda}^{\psi\times}(r, z) \, f_{\Lambda '}^{\phi}(r, z)
\nonumber \\ && \qquad\qquad
-  f_{\Lambda '\Lambda}^{\psi\times}(r, z) \, g_{\Lambda '}^{\phi}(r, z)
\Big)_{r=r_{\rm crit}}
\nonumber \\
&& \quad 
- \lim_{r\rightarrow 0}
 r^2 \,
\Big(
   g_{\Lambda '\Lambda}^{\psi\times}(r, z) \, f_{\Lambda '}^{\phi}(r, z)
\nonumber \\ && \qquad\qquad\quad
-  f_{\Lambda '\Lambda}^{\psi\times}(r, z) \, g_{\Lambda '}^{\phi}(r, z)
\Big)
  \Bigg\}
 \\
%----------------------------------------------------------------------------------------
%
\label{EQ:INT-V-VOL-WRONSKIAN-LHS}
&=& 
 r_{\rm crit}^2 \,
\Big[ \psi_{\Lambda}^{\times} , \phi_{\Lambda '} \Big]_{r=r_{\rm crit}}
\nonumber \\
&& \quad 
- \lim_{r\rightarrow 0}
 r^2 \,
\Big[ \psi_{\Lambda}^{\times} , \phi_{\Lambda '} \Big]_{r}
\; .
\end{eqnarray}
%----------------------------------------------------------------------------------------
%%%%\end{widetext}
It should be mentioned that converting the volume
integrals over an atomic cell (in general a polyhedron) by
means of Gauss theorem one is led to a complicated integral over
the surface of the cell.\cite{WZB+92}
As one has 
$ V(\vec r ) 
          \delta(\vec r , \vec r \,') +
          \Sigma(\vec r , \vec r \,', z) =0 $
          for $\vec r$ or $\vec r \,'$ outside 
          the cell the volume integral can be performed over a sphere 
          of radius $r_{\rm crit}$  that circumscribes the cell.
          Accordingly, in this case the surface normal
          is always parallel to $\hat r$ leading to a very simple
          surface integral.
       The same applies to the surface integral over the sphere 
       with $r\rightarrow 0$.

Applying Eq.\ \eqref{EQ:INT-V-VOL-WRON-RHS}
when dealing with the t-matrix $ t_{\Lambda ' \Lambda}(z)$
via  Eq.\ \eqref{EQ:t-matrix-jVR-RHS} 
one has 
$ \phi_{\Lambda}^{\times}(\vec r, z)=
     j_{\Lambda}^{\times}(\vec r, z)$ and 
$\psi_{\Lambda}(\vec r , z)= 
 R_{\Lambda}(\vec r , z)         $.
In this case the second term in  Eq.\ \eqref{EQ:INT-V-VOL-WRON-RHS}
does not contribute.
Making use of the asymptotic behavior
of $R_{\Lambda}(\vec r , z)         $ for $r>r_{\rm crit}$
as given by   Eq.\ \eqref{EQ:RHS-ASYMPT-R}
one is led to the identity $ t_{\Lambda ' \Lambda}(z)= t_{\Lambda ' \Lambda}(z)$.
This obviously confirms
 the consistency of the various transformations leading 
to  Eq.\ \eqref{EQ:INT-V-VOL-WRON-RHS}. In a similar way one can 
deal with the  enhancement factor $\alpha^{}_{\Lambda '\Lambda}(z)  $
as given by  Eq.\ \eqref{ALPHA-h-V-R-RHS}.
In this case one has 
$ \phi_{\Lambda}^{\times}(\vec r, z)=
   h_{\Lambda '}^{+\times}(\vec r, z)$ and 
$\psi_{\Lambda}(\vec r , z)= 
 R_{\Lambda}(\vec r , z)$.
Expressing the  asymptotic behavior
of $R_{\Lambda}(\vec r , z)         $ for $r\rightarrow 0$
by means of  Eq.\ \eqref{EQ:ASYPT-0-R-RHS}
one gets  $\alpha^{}_{\Lambda '\Lambda}(z)
=  \alpha^{}_{\Lambda '\Lambda}(z)  $ confirming once more the 
coherence of the various expressions.
The same is true when dealing with Eqs.\ 
\eqref{ALPHA-R-V-h-LHS},
\eqref{BETA-j-V-H-RHS} and 
\eqref{BETA-H-V-j-LHS}
for 
  $\alpha^{\times}_{\Lambda '\Lambda}(z)$,
  $\beta_{\Lambda '\Lambda}(z)$, and 
  $\beta^{\times}_{\Lambda '\Lambda}(z)$,
respectively, using  Eqs.\ 
\eqref{EQ:INT-V-VOL-WRON-RHS} and
\eqref{EQ:INT-V-VOL-WRON-LHS}.
However, it should be stressed that these 
equations are nevertheless quite helpful 
(see for example section \ref{SEC:Practical-aspects}).

%****************************************************************
\subsection{Single-site Green function for a general potential}
%****************************************************************

\label{Sec:Single-site-Green-function}

Inserting the regular and irregular solutions to the RHS and LHS 
Dirac equations, $R(\vec r, z)$,  $H(\vec r, z)$, $R^{\times}(\vec r, z)$ and $H^{\times}(\vec r, z)$, respectively, 
specified by their asymptotic behavior $r > r_{\rm crit}$ given by
Eqs.\ \eqref{EQ:RHS-ASYMPT-R} to \eqref{EQ:LHS-ASYMPT-H}
one can write the expression
for the  single site  Green function $G^{n}(\vec r, \vec r \,', z)$ 
 in Eq.\ \eqref{EQ:SS-GF-A}  in a 
compact way:
\begin{eqnarray}
\LABEL{EQ:SS-GF-RH-OUT}
%----------------------------------------------------------------------------------------
G^{n}(\vec r, \vec r \,', z)= -i \pbar \sum_{\Lambda} \, R_{\Lambda}(\vec r, z) \, H_{\Lambda}^{\times}(\vec r\,', z)\, \theta(r '- r)  \nonumber\\
%----------------------------------------------------------------------------------------
%
%
%----------------------------------------------------------------------------------------
+  H_{\Lambda}(\vec r, z) \,  R_{\Lambda}^{\times}(\vec r\,', z) \,\theta(r - r ') 
%----------------------------------------------------------------------------------------
\;.
\end{eqnarray}
Eq.~\eqref{EQ:SS-GF-RH-OUT}  holds by construction
for  $r > r_{\rm crit}$ and $r\, ' > r_{\rm crit}$, i.e.\ 
it is a solution to the
defining equations
\eqref{EQ:RGF1-GFDEF-LHS} and \eqref{EQ:RGF1-GFDEF-ADJOINT}
for the regime  
with $V(\vec r) = 0$ and $\Sigma(\vec r, \vec r \,', z) = 0$.

For $r < r_{\rm crit}$ and $r\,' > r_{\rm crit}$ or $r > r_{\rm crit}$ and $r\,' < r_{\rm crit}$ 
Eq.~\eqref{EQ:SS-GF-RH-OUT} gives still the proper solution 
to Eqs.\ \eqref{EQ:RGF1-GFDEF-RHS} and \eqref{EQ:RGF1-GFDEF-LHS} 
as the inhomogeneous term, i.e.\ the $\delta$-function does not show up 
and the functions 
$R_{\Lambda}(\vec r, z)$,            $H_{\Lambda}(\vec r, z)$, 
$R_{\Lambda}^{\times}(\vec r, z)$ and $H_{\Lambda}^{\times}(\vec r, z)$ 
solve the corresponding RHS and LHS Dirac equations \eqref{EQ:DEQU-RHS} 
and \eqref{EQ:RGF2-A}, respectively. 
In addition,  the Green function adopts this way the 
required regular 
asymptotic behavior for  $r \rightarrow 0$.
For $r < r_{\rm crit}$ and $r\,' < r_{\rm crit}$, 
however,  the last property is no more sufficient as the inhomogeneous term 
has to be recovered when inserting Eq.~\eqref{EQ:SS-GF-RH-OUT} 
into Eq.\ \eqref{EQ:RGF1-GFDEF-RHS} or \eqref{EQ:RGF1-GFDEF-LHS}.
A  proof that 
this additional requirement is
indeed satisfied by the product representation 
for $ G^{n}(\vec r, \vec r \,', z)$
in  Eq.~\eqref{EQ:SS-GF-RH-OUT}
has been given by 
 Tamura \cite{Tam92}  for the case of a
local
potential $V(\vec r)$; i.e.\ $\Sigma(\vec r, \vec r\,', z) = 0$.
This proof relies on the fact that
the RHS and LHS solutions to the Dirac equation satisfy the 
Wronski relation of second kind given by
 Eq.\ \eqref{EQ:31a}. This in turn is ensured by the
validity of the Wronski relation of the first kind
in Eqs.\ \eqref{WRONSKIAN-RR-HH} and \eqref{WRONSKIAN-RH-HR} that hold for any local
potential. For a non-local self-energy  $\Sigma(\vec r, \vec r\,', z) $
present, however, the Wronski relation of first kind
given by Eq.\ \eqref{EQ:WRON-gf-const} does not hold
anymore because of the non-vanishing term in
Eq.\ \eqref{WRONSKIAN-SELF-ENERGY}. As a consequence one has to conclude that the
product ansatz for the Green function
given in Eq.\ \eqref{EQ:SS-GF-RH-OUT} is no more acceptable for a
non-local self-energy.
The same
conclusion can be drawn
from an 
alternative proof that Eq.\ \eqref{EQ:SS-GF-RH-OUT} is acceptable
for a local potential
given in   appendix \ref{appendix-A}.

As suggested by Eq.\ \eqref{EQ:SS-DYSON-EQ-T}  an {\em on-the-energy-shell }
representation of the Green function
can nevertheless be given within the framework of scattering theory
also for a finite non-local self-energy   $\Sigma(\vec r, \vec r\,', z) $   without
making use of the spectral representation
(see Eq.\ \eqref{EQ:RGF1-F}). Starting again from Eq.\ \eqref{EQ:SS-DYSON-EQ-T}
without restrictions concerning
$ \vec r$  and $ \vec r \, '$ one is led to an
expansion of the
Green function in terms of the
Bessel and Hankel functions
\begin{eqnarray}
\LABEL{GF-j-h-t-matrix-exp}
%----------------------------------------------------------------------------------------
G(\vec r, \vec r\, ')& = & 
\sum\limits_{\Lambda  \Lambda '}j_{\Lambda }(\vec r)\,
G_{\Lambda \Lambda '}^{jj}(r,r')\,
j_{\Lambda '}^{\times}(\vec r\, ') \nonumber\\
%----------------------------------------------------------------------------------------
&&\quad+ j_{\Lambda }(\vec r)\,
G_{\Lambda \Lambda '}^{jh}(r,r') \,
h_{\Lambda '}^{\times}(\vec r\, ') \nonumber\\
%----------------------------------------------------------------------------------------
&&\quad+ h_{\Lambda }(\vec r)\,
G_{\Lambda \Lambda '}^{hj}(r,r')\,
j_{\Lambda '}^{\times}(\vec r\, ') \nonumber\\
%----------------------------------------------------------------------------------------
&&\quad +h_{\Lambda }(\vec r)\,
G_{\Lambda \Lambda '}^{hh}(r,r')\,
h_{\Lambda '}^{\times}(\vec r\, ')
%----------------------------------------------------------------------------------------
\;,
%----------------------------------------------------------------------------------------
\end{eqnarray}
with the  expansion coefficient functions $G_{\Lambda \Lambda '}^{\alpha\beta}(r,r') $
($\alpha \,(\beta) = j,\,h$)
given as integrals involving the real space representation
$t(\vec r, \vec r \, ', z)$ of the t-operator
(see right hand side of  Eqs.\ \eqref{CC-hth} to \eqref{SS-jtj}
for explicit expressions).
While Eq.\ \eqref{GF-j-h-t-matrix-exp} clearly shows that
a scattering representation of the Green function
can indeed be given, it is not very helpful for practical 
applications as it requires an explicit expression for $t(\vec r, \vec r \, ', z)$.
A useful expression for the 
 single site Green function  $ G^{n}(\vec r, \vec r \,', z)$
for the case   $\Sigma (\vec r, \vec r \,', z)\ne0$
can nevertheless be 
deduced by making use of 
 the Dyson equation. 
  Denoting $G^{1}(z)$ the 
Green function operator connected with the 
 Hamiltonian ${\cal H}^{1}$ that 
contains only the local potential one has:
\begin{eqnarray}
\LABEL{DYSON-SS-G1SIG}
\hat G^{n}(z)&=& \hat G^{1}(z) + 
\hat G^{1}(z)\, 
\hat \Sigma(z)\, \hat G^{n}(z)
\; .
\end{eqnarray}
This equation can be solved by a series expansion, i.e.\
by inserting the equation repeatedly into itself leading to:
\begin{eqnarray}
\LABEL{DYSON-SS-G1SIG-EXPAND}
\hat G^{n}(z)&=& 
\hat G^{1}(z) + 
\hat G^{1}(z)\, 
\hat \Sigma(z)\, \hat G^{1}(z) 
\nonumber \\
&& \;
+
\hat G^{1}(z)\, 
\hat \Sigma(z)\, \hat G^{1}(z) 
\hat \Sigma(z)\, \hat G^{1}(z) 
+ ...
\nonumber
\\
%---------------------------------------------------------------------------
\LABEL{DYSON-SS-G-G1-Dt-G1}
&=&
\hat G^{1}(z) + 
\hat G^{1}(z)\, 
\Delta \hat t(z)\, \hat G^{1}(z) 
\; .
\end{eqnarray}
with 
\begin{eqnarray}
\Delta \hat t(z) &=&
\hat \Sigma(z) +
\hat \Sigma(z)
\, \hat G^{1}(z) \hat \Sigma(z)
\nonumber \\
&& \;
+
\hat \Sigma(z)
\, \hat G^{1}(z) \hat \Sigma(z)
\, \hat G^{1}(z) \hat \Sigma(z)
\nonumber \\
&& \;
+
\hat \Sigma(z)\, \hat G^{1}(z) 
\hat \Sigma(z)\, \hat G^{1}(z) 
\hat \Sigma(z)\, \hat G^{1}(z) 
\hat \Sigma(z)
+ ...
\nonumber
\\
%---------------------------------------------------------------------------
&=&
\hat \Sigma(z)
+
\hat \Sigma(z)
\, \hat G^{1}(z) \, 
\Delta \hat t(z)\nonumber
\\
%---------------------------------------------------------------------------
&=&
\LABEL{EQ:Delta-t-SIG-G1-SIG}
\big(
1- \hat \Sigma(z) \,\hat G^{1}(z)
\big)^{-1} \,
  \hat \Sigma(z)
\\
%---------------------------------------------------------------------------
&=&
  \hat \Sigma(z) \,
\big(
1- \hat G^{1}(z)  \,\hat \Sigma(z)
\big)^{-1} 
\; ,
\end{eqnarray}
 where $\Delta \hat t(z) $ represents 
the correction to the single site t-matrix operator
due to the non-local self-energy $ \hat \Sigma(z)$.

The real space representation of
 Eq.\ \eqref{DYSON-SS-G1SIG}
 can be reformulated as follows:
\begin{widetext}
\begin{eqnarray}
\LABEL{EQ:SS-GF-RS-G1-SIG-G}
G^{n}(\vec r, \vec r\,' , z)&=& G^{1}(\vec r, \vec r\,' , z) + 
\int d^{3} r\,'' \int d^{3} r\,''' 
\,
%\nonumber \\
%&& 
G^{1}(\vec r, \vec r\,'' , z)\, 
\Sigma(\vec r\,'', \vec r\,''' , z)\, G^{n}(\vec r\,''', \vec r\,' , z)
\\
%---------------------------------------------------------------------------
&=&
G^{1}(\vec r, \vec r\,' , z)\, 
-i \pbar \sum_{\Lambda}
\Bigg[
 R_{\Lambda}^{1}(\vec r, z) 
\int_{r}^{r_{\rm crit}} d^{3} r\,'' \int d^{3} r\,''' 
H_{\Lambda}^{1\times}(\vec r\,'', z)\,
\Sigma(\vec r\,'', \vec r\,''' , z)\, G^{n}(\vec r\,''', \vec r\,' , z)
\nonumber \\
&& \qquad\qquad\qquad\qquad\qquad
+
H_{\Lambda}^{1}(\vec r, z) 
\int_{0}^{r} d^{3} r\,'' \int d^{3} r\,''' 
R_{\Lambda}^{1\times}(\vec r\,'', z)\,
\Sigma(\vec r\,'', \vec r\,''' , z)\, G^{n}(\vec r\,''', \vec r\,' , z)
\Bigg]
 \\
%---------------------------------------------------------------------------
&=&
\LABEL{EQ:SS-GF-RAHB-RHS}
G^{1}(\vec r, \vec r\,' , z)\, 
+ \sum_{\Lambda}
 R_{\Lambda}^{1}(\vec r, z) 
\,
A_{\Lambda}^{1\times}(r,\vec r\,' , z)
%\nonumber \\
%%
%&& \qquad\qquad\qquad\qquad
%
+
H_{\Lambda}^{1}(\vec r, z) 
\, 
B_{\Lambda}^{1\times}(r,\vec r\,', z)
%
%----------------------------------------------------------------------------------------
\; .
\end{eqnarray}
\end{widetext}
with the auxiliary functions 
\begin{eqnarray}
%
%---------------------------------------------------------------------------
A_{\Lambda}^{1\times}(r,\vec r\,' , z) &=&
-i \pbar
\int_{r}^{r_{\rm crit}} d^{3} r\,'' \int d^{3} r\,'''   \, 
H_{\Lambda}^{1\times}(\vec r\,'', z)\,
\nonumber \\ && \qquad  
\Sigma(\vec r\,'', \vec r\,''' , z)\, G^{n}(\vec r\,''', \vec r\,' , z)
 \\
%---------------------------------------------------------------------------
B_{\Lambda}^{1\times}(r,\vec r\,', z)
&=&
-i \pbar
\int_{0}^{r} d^{3} r\,'' \int d^{3} r\,'''  \,
R_{\Lambda}^{1\times}(\vec r\,'', z)\,
\nonumber \\ && \qquad  
\Sigma(\vec r\,'', \vec r\,''' , z)\, G^{n}(\vec r\,''', \vec r\,' , z)
%
%----------------------------------------------------------------------------------------
\; .
\end{eqnarray}
Here we used the explicit expression for $G^{1}(\vec r, \vec r \,', z)$
\begin{eqnarray}
\LABEL{EQ:SS-GF1-RH}
%----------------------------------------------------------------------------------------
G^{1}(\vec r, \vec r \,', z)= -i \pbar \sum_{\Lambda} \, R_{\Lambda}^{1}(\vec r, z) \, H_{\Lambda}^{1\times}(\vec r\,', z)\, \theta(r '- r)  \nonumber\\
%----------------------------------------------------------------------------------------
%
%
%----------------------------------------------------------------------------------------
+  H_{\Lambda}^{1}(\vec r, z) \,  R_{\Lambda}^{1\times}(\vec r\,', z) \,\theta(r - r ') 
%-------------------------------------------------------------------------------------
\end{eqnarray}
in terms of the regular and irregular wave functions
$R_{\Lambda}^{1}(\vec r, z)$, 
$H_{\Lambda}^{1}(\vec r, z)$, 
$R_{\Lambda}^{1\times}(\vec r, z)$ 
and $H_{\Lambda}^{1\times}(\vec r, z)$ 
connected with the local Hamiltonian ${\cal H}^{1}(\vec r)$
in correspondence to Eq.~\eqref{EQ:SS-GF-RH-OUT}
that holds for any $\vec r$ and $\vec r\,'$.

For  $r > r_{\rm crit}$ and $r\, ' > r_{\rm crit}$
one may use for 
$G^{n}(\vec r, \vec r \,', z)$
the representation given 
in Eq.\ \eqref{EQ:SS-GF-RH-OUT}
in terms of the regular and irregular wave functions
$R_{\Lambda}(\vec r, z)$, 
$H_{\Lambda}(\vec r, z)$, 
$R_{\Lambda}^{\times}(\vec r, z)$ 
and $H_{\Lambda}^{\times}(\vec r, z)$ 
connected with the full Hamiltonian ${\cal H}(\vec r,\vec r \, ',z)$.
With this 
one finds immediately an explicit
expression for the angular momentum representation
of 
$\Delta \hat t(z)$:
\begin{eqnarray}
\LABEL{EQ:Delta-t-R1-SIG-R}
%---------------------------------------------------------------------------
\Delta t_{\Lambda\Lambda'}(z)
&=&
\int d^{3} r \int d^{3} r\,' \nonumber \\
&& \quad 
R_{\Lambda}^{1\times}(\vec r, z)\,
\Sigma(\vec r, \vec r\,' , z)\,
R_{\Lambda'}(\vec r\,', z)
%
%----------------------------------------------------------------------------------------
\; .
\end{eqnarray}
Accordingly,  the full t-matrix
$ t_{\Lambda\Lambda'}(z)$ can be obtained from 
\begin{eqnarray}
%
%---------------------------------------------------------------------------
 t_{\Lambda\Lambda'}(z)
&=&
 t_{\Lambda\Lambda'}^{1}(z) +
\Delta t_{\Lambda\Lambda'}(z)
%
%----------------------------------------------------------------------------------------
\; ,
\end{eqnarray}
with $t_{\Lambda\Lambda'}^{1}(z) $
the t-matrix corresponding to the local Hamiltonian ${\cal H}^{1}(\vec r)$.
These relations fully coincide with those one is led to
if one calculates the regular function
$R_{\Lambda}(\vec r, z)$
connected with the full Hamiltonian ${\cal H}(\vec r,\vec r \, ',z)$
via the Lippmann-Schwinger equation 
with the reference system referring to the local Hamiltonian
 ${\cal H}^{1}(\vec r)$ with its associated solutions
$R_{\Lambda}^{1}(\vec r, z)$, 
$H_{\Lambda}^{1}(\vec r, z)$, 
$R_{\Lambda}^{1\times}(\vec r, z)$ 
and $H_{\Lambda}^{1\times}(\vec r, z)$
(see section \ref{SEC:BORN}).
Furthermore, one may note that Eq.\ \eqref{EQ:Delta-t-R1-SIG-R}
is obviously  the counterpart to Eq.\ \eqref{EQ:t-matrix-jVR-RHS}
for the full t-matrix.

Obviously, the representation of the 
single site 
Green function 
$G^{n}(\vec r, \vec r \,', z)$
given by Eq.\ \eqref{EQ:SS-GF-RAHB-RHS} is not unique.
 Starting from the alternative 
form of the Dyson equation, 
\begin{eqnarray}
\LABEL{EQ:DYSON-Gz-Gz-Sig-G1z}
\hat G(z) = \hat G^{1}(z) + \hat G(z) \, \hat \Sigma(z) \, \hat G^{1}(z)
\; ,
\end{eqnarray}
 one 
would get the corresponding expressions:
\begin{eqnarray}
G^{n}(\vec r, \vec r\,' , z)&=&
\LABEL{EQ:SS-GF-ARBH-LHS}
G^{1}(\vec r, \vec r\,' , z)\,
\nonumber \\ && 
+ \sum_{\Lambda}
A_{\Lambda}^{1}(\vec r, r',z)
\,
 R_{\Lambda}^{1\times}(\vec r\,' , z) 
%\nonumber \\
%%
%&& \qquad\qquad\qquad\qquad
%
\nonumber \\ && \qquad 
+
B_{\Lambda}^{1}(\vec r, r', z)
\, 
H_{\Lambda}^{1\times}(\vec r\,' , z) 
%
%----------------------------------------------------------------------------------------
\; .
\end{eqnarray}
with  
\begin{eqnarray}
%
%---------------------------------------------------------------------------
A_{\Lambda}^{1}(\vec r,  r',z) &=&
-i \pbar \int_{r'}^{r_{\rm crit}} d^{3} r\,'' \int d^{3} r\,''' 
G^{n} (\vec r, \vec r \, '' , z)\,
\nonumber \\ && \qquad \quad 
\Sigma(\vec r\,'', \vec r\,''' , z)\,
H_{\Lambda}^{1}(\vec r\,''', z)
 \\
%---------------------------------------------------------------------------
B_{\Lambda}^{1}(\vec r,  r',z)
&=&
-i \pbar \int_{0}^{r'} d^{3} r\,'' \int d^{3} r\,'''  \,
G^{n}(\vec r, \vec r\,'' , z)\,
\nonumber \\ && \qquad  \quad 
\Sigma(\vec r\,'', \vec r\,''' , z)\, 
R_{\Lambda}^{1\times}(\vec r\,''', z)
%
%----------------------------------------------------------------------------------------
\\
%---------------------------------------------------------------------------
\LABEL{EQ:Delta-t-R-SIG-R1}
\Delta t_{\Lambda\Lambda'}(z)
&=&
\int d^{3} r \int d^{3} r\,'
%% \nonumber \\
%%&& \quad 
\nonumber \\ && \qquad 
R_{\Lambda}^{\times}(\vec r, z)\,
\Sigma(\vec r, \vec r\,' , z)\,
R_{\Lambda}^{1}(\vec r\,', z)
%
%----------------------------------------------------------------------------------------
\; ,
\end{eqnarray}
where again  one may note that Eq.\ \eqref{EQ:Delta-t-R-SIG-R1}
is obviously  the counterpart to Eq.\ \eqref{EQ:t-matrix-RVj-LHS}
for the full t-matrix.

As 
Eq.\ \eqref{DYSON-SS-G-G1-Dt-G1}
indicates, 
one can get a more symmetric representation
for  the 
single site 
Green function 
$G^{n}(\vec r, \vec r \,', z)$
in terms of the solutions
$R_{\Lambda}^{1}(\vec r, z)$, 
$H_{\Lambda}^{1}(\vec r, z)$, 
$R_{\Lambda}^{1\times}(\vec r, z)$ 
and 
$H_{\Lambda}^{1\times}(\vec r, z)$
associated with 
the local Hamiltonian
 ${\cal H}^{1}(\vec r)$
 than given by Eqs.\
\eqref{EQ:SS-GF-RAHB-RHS}
and 
\eqref{EQ:SS-GF-ARBH-LHS}.
Performing a series expansion w.r.t.\ the self-energy
one is led to:
\begin{eqnarray}
G^{n}(\vec r, \vec r\,' , z)&=&
\LABEL{EQ:SS-GF-CDEF-LHS}
\quad \sum_{\Lambda\Lambda'}
 R_{\Lambda}^{1}(\vec r , z)  \,
G_{\Lambda \Lambda '}^{RR}(r,r',z)
\,
 R_{\Lambda'}^{1\times}(\vec r\,' , z) 
\nonumber \\ &&
%%
%&& \qquad\qquad\qquad\qquad
%
+ \sum_{\Lambda\Lambda'}
 R_{\Lambda}^{1}(\vec r , z)  \,
G_{\Lambda \Lambda '}^{RH}(r,r',z) \, 
H_{\Lambda'}^{1\times}(\vec r\,' , z) 
\nonumber \\
%----------------------------------------------------------------------------------------
&& 
+ \sum_{\Lambda\Lambda'}
  H_{\Lambda}^{1}(\vec r , z)  \,
G_{\Lambda \Lambda '}^{HR}(r,r',z)
\,
 R_{\Lambda'}^{1\times}(\vec r\,' , z) 
\nonumber \\ &&
%%
%&& \qquad\qquad\qquad\qquad
%
+\sum_{\Lambda\Lambda'}
 H_{\Lambda}^{1}(\vec r , z)  \,
G_{\Lambda \Lambda '}^{HH}(r,r',z) \, 
H_{\Lambda'}^{1\times}(\vec r\,' , z) 
\; .
\end{eqnarray}
with  
\begin{widetext}
\begin{eqnarray}
\LABEL{EQ:SS-GF-EXP-COEFF-RR}
%---------------------------------------------------------------------------
G_{\Lambda \Lambda '}^{RR}(r,  r',z) &=&
(-i \pbar)^2 
\int_{r}^{r_{\rm crit}} d^{3}r_{1} \int_{r'}^{r_{\rm crit}} d^{3}r_{2} \,
 H_{\Lambda}^{1\times}( \vec r_{1} , z)\,
\Sigma(\vec r_{1}, \vec r_{2} , z)\,
H_{\Lambda'}^{1}(\vec r_{2}, z) 
\nonumber \\
&&+ (-i \pbar)^2 
\int_{r}^{r_{\rm crit}} d^{3}r_{1} 
\int d^{3}r_{2} 
\int d^{3}r_{3} 
\int_{r'}^{r_{\rm crit}} d^{3}r_{4} \,
 H_{\Lambda}^{1\times}( \vec r_{1} , z)\,
 \nonumber \\
 && \qquad\qquad\qquad\qquad\qquad\qquad\qquad
\Sigma(\vec r_{1}, \vec r_{2} , z)\,
G^{1}( \vec r_{2} ,\vec r_{3} , z)\,
\Sigma(\vec r_{3}, \vec r_{4} , z)\,
H_{\Lambda'}^{1}(\vec r_{4}, z) + ...
 \\
\LABEL{EQ:SS-GF-EXP-COEFF-RH}
%---------------------------------------------------------------------------
G_{\Lambda \Lambda '}^{RH}(r,  r',z)
&=&
(-i \pbar)^2 \int_{r}^{r_{\rm crit}} d^{3}r_{2} \int_{0}^{r'} d^{3}r_{1}  \,
H_{\Lambda}^{1}(\vec r_{1} , z)
\,
\Sigma(\vec r_{1}, \vec r_{2} , z)\, 
R_{\Lambda'}^{1\times}(\vec r_{2}, z) + ...
 -i \pbar \, \delta_{\Lambda \Lambda '}\,  \theta(r ' - r) 
%
%----------------------------------------------------------------------------------------
%
\\
\LABEL{EQ:SS-GF-EXP-COEFF-HR}
%---------------------------------------------------------------------------
G_{\Lambda \Lambda '}^{HR}(r,  r',z) &=&
(-i \pbar)^2 \int_{0}^{r} d^{3}r_{1} \int_{r'}^{r_{\rm crit}} d^{3}r_{2} \,
R_{\Lambda}^{1\times}( \vec r_{1} , z)\,
\Sigma(\vec r_{1}, \vec r_{2} , z)\,
H_{\Lambda'}^{1}(\vec r_{2}, z) + ...
 -i \pbar \, \delta_{\Lambda \Lambda '}\,  \theta(r - r ') 
 \\
\LABEL{EQ:SS-GF-EXP-COEFF-HH}
%---------------------------------------------------------------------------
G_{\Lambda \Lambda '}^{HH}(r,  r',z)
&=&
(-i \pbar)^2 \int_{0}^{r} d^{3}r_{2} \int_{0}^{r'} d^{3}r_{1}  \,
R_{\Lambda}^{1}(\vec r_{1} , z)
\,
\Sigma(\vec r_{1}, \vec r_{2} , z)\, 
R_{\Lambda'}^{1\times}(\vec r_{2}, z)  + ...
%
%----------------------------------------------------------------------------------------
\; ,
\end{eqnarray}
\end{widetext}
where the terms involving $\delta_{\Lambda \Lambda '}$
represent the contributions connected 
with $G^{1}(\vec r, \vec r \,', z)$ (see Eq.\  \eqref{EQ:SS-GF1-RH}).

Eq.\  \eqref{EQ:SS-GF-CDEF-LHS} together with 
Eqs.\ \eqref{EQ:SS-GF-EXP-COEFF-RR}
 to   \eqref{EQ:SS-GF-EXP-COEFF-HH})
obviously provides an explicit expression for the Green function 
in terms of the solutions
$R_{\Lambda}^{1}(\vec r, z)$, 
$H_{\Lambda}^{1}(\vec r, z)$, 
$R_{\Lambda}^{1\times}(\vec r, z)$ 
and $H_{\Lambda}^{1\times}(\vec r, z)$
 associated with ${\cal H}^{1}(\vec r)$.
As these in turn can be expanded w.r.t.\ the Bessel and Hankel
functions (see the corresponding expressions in
Eqs.\ \eqref{PHI-j-h-RHS}, \eqref{PHI-j-h-LHS},  \eqref{F-j-h-RHS} and \eqref{F-j-h-LHS})
Eq.\ \eqref{EQ:SS-GF-CDEF-LHS}
could be expressed as well in terms of the Bessel
and Hankel functions. The new
expansion coefficient functions  obviously correspond
to those used in Eq.\ \eqref{GF-j-h-t-matrix-exp}.

\medskip

While Eqs.\ \eqref{EQ:SS-GF-RAHB-RHS},   \eqref{EQ:SS-GF-ARBH-LHS}  and 
\eqref{EQ:SS-GF-CDEF-LHS}
may be solved iteratively or by summing up the 
series expansion, respectively,
one notes that the single site Green function is  not
obtained as a simple product representation as in 
Eq.~\eqref{EQ:SS-GF-RH-OUT} that completely
decouples the dependency on $\vec r$ and $\vec r \,'$.
 In fact, Zeller\cite{Zel15}
has given arguments that this convenient representation 
should not exist for a self-energy that has no restrictions
concerning its dependency on $\vec r$ and $\vec r \,'$.

\medskip

In the following the special case of a
self-energy 
$  \Sigma (\vec r, \vec r \,', z) $ 
is considered that is 
represented by an expansion into 
a product of  basis functions
$ \phi_{\Lambda}(\vec r)$
according to Eq.\ \eqref{EQ:SELF-EXPANSION}.
Inserting this expression into Eq.\ \eqref{EQ:SS-GF-RS-G1-SIG-G}
for the single site Green function
$G(\vec r, \vec r\,' , z)$ 
one gets
together with Eq.\ \eqref{EQ:SS-GF1-RH}
for $G^{1}(\vec r, \vec r \,', z)$:
\begin{widetext}
%
%=============================================================================
\begin{eqnarray}
\LABEL{EQ:SS-GF-RS-G1-SIG-FACT-G}
G^{n}(\vec r, \vec r\,' , z)&=&
%---------------------------------------------------------------------------
G^{1}(\vec r, \vec r\,' , z)\, 
+
\sum_{\Lambda\Lambda'}
\Bigg[
\int d^{3} r\,''  \,
G^{1}(\vec r, \vec r\,'' , z)\, 
 \phi_{\Lambda}(\vec r\,'')
 \Bigg] 
 \,
 \Sigma_{\Lambda \Lambda '}(z)  
\Bigg[
 \int d^{3} r\,'''\,
\phi_{\Lambda '}^{\dagger}(\vec r \,''') \,
 G^{n}(\vec r\,''', \vec r\,' , z)
 \Bigg] 
 \\
%=============================================================================
 &=&
 \LABEL{EQ:SS-GF-G1-S-SIG-S}
G^{1}(\vec r, \vec r\,', z)\, 
%%\nonumber \\
%%&&
+ \sum_{\Lambda\Lambda'}
 \rho_{\Lambda}^{1}(\vec r , z)
 \,
 \Sigma_{\Lambda \Lambda '}(z)  
\, \rho_{\Lambda '}^{\times}(\vec r \,', z)
\end{eqnarray}
%=============================================================================
%
with the auxiliary function
%
%=============================================================================
\begin{eqnarray}
\LABEL{EQ:S1-DEF-RHS}
\rho_{\Lambda}^{1}(\vec r , z) 
&=&
\int d^{3} r\,'  \,
G^{1}(\vec r, \vec r\,' , z)\,  \phi_{\Lambda}(\vec r\,')
\\
%=============================================================================
 &=&
 -i \pbar
\sum_{\Lambda_1} 
\Bigg[
R_{\Lambda_1}^{1}(\vec r, z) 
\int_{r}^{r_{\rm crit}} d^{3} r\,'  \,
H_{\Lambda_1}^{1\times}(\vec r\,', z)\, 
 \phi_{\Lambda}(\vec r\,'')
+ 
H_{\Lambda_1}^{1}(\vec r, z) 
\int_{0}^{r} d^{3} r\,'  \,
R_{\Lambda_1}^{1\times}(\vec r\,', z)\, 
 \phi_{\Lambda}(\vec r\,')
\Bigg] 
\\
%=============================================================================
 &=&
\LABEL{EQ:S1-EXPANSION-RHS}
\sum_{\Lambda_1} 
\Bigg[
R_{\Lambda_1}^{1}(\vec r, z) \,
C_{\Lambda_1\Lambda}^{1}(r, z)
+ 
H_{\Lambda_1}^{1}(\vec r, z) \,
S_{\Lambda_1\Lambda}^{1}(r, z)
\Bigg] \; .
\end{eqnarray}
%=============================================================================
%%%  \\
%%%%=============================================================================
%%%\nonumber \\
%=============================================================================
%%% &=&
%%%G^{1}(\vec r, \vec r\,', z)\, 
%%%-i \pbar  \sum_{\Lambda\Lambda'}
%%%\Big[
%%%\bar R_{\Lambda}^{1}(\vec r, z) 
%%%+ \bar H_{\Lambda}^{1}(\vec r, z) 
%%%\Big]
%%% \,
%%% \Sigma_{\Lambda \Lambda '}(z)  
%%%\, S_{\Lambda '}^{\times}(\vec r \,', z)
%%%\\
%=============================================================================
%%%\; ,
%%%%
%%%end{eqnarray}
%%%%=============================================================================
%%%%
%%%where we used the renormalized
%%%regular and irregular functions:
%%%%
%%%%=============================================================================
%%%\begin{eqnarray}
%%%%=============================================================================
%%%\bar R_{\Lambda}^{1}(\vec r, z) 
%%% &=&
%%%\sum_{\Lambda'} 
%%%R_{\Lambda'}^{1}(\vec r, z) 
%%%\int_{r}^{r_{\rm crit}} d^{3} r\,' 
%%%H_{\Lambda'}^{1\times}(\vec r\,', z)\, 
%%% \phi_{\Lambda}(\vec r\,')
%%% \\
%%%%=============================================================================
%%% \bar H_{\Lambda}^{1}(\vec r, z) 
%%% &=&
%%%\sum_{\Lambda'} 
%%%H_{\Lambda'}^{1}(\vec r, z) 
%%%\int_{0}^{r} d^{3} r\,' 
%%%R_{\Lambda'}^{1\times}(\vec r\,', z)\, 
%%% \phi_{\Lambda}(\vec r\,')
%%%%=============================================================================
%%%\; ,
%%%%
%
that can be calculated directly for a self-energy
as given by Eq.\ \eqref{EQ:SELF-EXPANSION}.
The remaining function  $\rho_{\Lambda}^{\times}(\vec r , z)$
can be obtained iteratively from the following expression:
%
%=============================================================================
\begin{eqnarray}
\LABEL{EQ:S-DEF-LHS}
\rho_{\Lambda }^{\times}(\vec r , z) 
&=&
 \int d^{3} r\,'
\,
\phi_{\Lambda }^{\dagger}(\vec r \, ') \,
 G^{n}(\vec r\,', \vec r , z)
  \\
%=============================================================================
&=&
 \int d^{3} r\,'
\, \phi_{\Lambda }^{\dagger}(\vec r \,') \,
 G^{1}(\vec r\,', \vec r , z)
%\nonumber \\
%&&
+ 
 \sum_{\Lambda_2}
 \sum_{\Lambda_1}
\Bigg[
 \int d^{3} r\,'
 \,
\phi_{\Lambda }^{\dagger}(\vec r \,') \, 
\rho_{\Lambda_1}(\vec r\,', z) 
\Bigg]
 \,
 \Sigma_{\Lambda_1 \Lambda_2}(z)  
\, \rho_{\Lambda_2}^{\times}(\vec r, z)
  \\
%=============================================================================
&=&
\rho_{\Lambda}^{1\times}(\vec r, z) 
%\nonumber \\
%&&
+
 \sum_{\Lambda_1\Lambda_2}
\langle 
\phi_{\Lambda } 
|  
\rho_{\Lambda_1}(z)
\rangle
 \,
 \Sigma_{\Lambda_1 \Lambda_2}(z)  
\, \rho_{\Lambda_2}^{\times}(\vec r, z)
\; ,
\end{eqnarray}
%=============================================================================
%
\end{widetext}
where $\rho_{\Lambda}^{1\times}(\vec r, z) $
is defined in analogy to Eqs.\ \eqref{EQ:S1-DEF-RHS} and \eqref{EQ:S-DEF-LHS}.

As in the case of Eq.\ \eqref{EQ:SS-GF-RAHB-RHS}
 the representation for the single site Green function
in Eq.\ \eqref{EQ:SS-GF-G1-S-SIG-S}
 is obviously not unique. Starting from the alternative 
form of the Dyson equation given by
Eq.\ \eqref{EQ:DYSON-Gz-Gz-Sig-G1z}
 one would fix 
$\rho_{\Lambda}^{1\times}(\vec r, z)$
 but would have to determine  $\rho_{\Lambda}(\vec r, z)$, 
 in analogy to Eq.\ \eqref{EQ:SS-GF-ARBH-LHS},
where 
$\rho_{\Lambda}^{1\times}(\vec r, z)$
and $\rho_{\Lambda}(\vec r, z)$
are
again
 defined in analogy to Eqs.\ \eqref{EQ:S1-DEF-RHS} and \eqref{EQ:S-DEF-LHS}.

To get to a symmetric form for the single
site Green function 
$G(\vec r, \vec r\,' , z)$ 
one may start from Eq.\ \eqref{EQ:SS-GF-RS-G1-SIG-G}.
Making use of the expansion for the self-energy 
$  \Sigma (\vec r, \vec r \,', z) $ 
given by Eq.\ \eqref{EQ:SELF-EXPANSION}
and 
inserting Eqs.\ \eqref{EQ:SS-GF-RS-G1-SIG-G} again and again into itself
one is led to the following 
series expansion:
\begin{widetext}
%
%=============================================================================
\begin{eqnarray}
G^{n}(\vec r, \vec r\,' , z)&=&
G^{1}(\vec r, \vec r\,' , z)\, 
+
\sum_{\Lambda\Lambda'}
 \rho_{\Lambda}^{1}(\vec r , z)
 \,
 \Sigma_{\Lambda \Lambda '}(z)  
\, \rho_{\Lambda '}^{\times}(\vec r \,', z) 
\nonumber \\
%---------------------------------------------------------------------------
&=& 
G^{1}(\vec r, \vec r\,' , z)\, 
+
\sum_{\Lambda\Lambda'}
 \rho_{\Lambda}^{1}(\vec r , z)
 \,
 \Sigma_{\Lambda \Lambda '}(z)  
\, \rho_{\Lambda '}^{1\times}(\vec r \,', z) 
\nonumber \\
&& \qquad
+
\sum_{\Lambda\Lambda'}
\sum_{\Lambda_1\Lambda_2}
 \rho_{\Lambda}^{1}(\vec r , z)
 \,
 \Sigma_{\Lambda \Lambda_1}(z)  \,
\tilde G^1_{\Lambda_1 \Lambda_2}(z)
 \,
 \Sigma_{\Lambda_2 \Lambda'}(z)  
\, \rho_{\Lambda '}^{1\times}(\vec r \,', z)  
+ ...
\nonumber 
\\
%---------------------------------------------------------------------------
&=& 
G^{1}(\vec r, \vec r\,' , z)\, 
+
\sum_{\Lambda\Lambda'}
 \rho_{\Lambda}^{1}(\vec r , z)
 \,
\Big[ 
\underline \Sigma(z)  
+
\underline \Sigma(z)  
\, \tilde{\underline{G}}^1(z) \,
\underline \Sigma(z)  
+
\underline \Sigma(z)  
\, \tilde{\underline{G}}^1(z) 
\,
\underline \Sigma(z)  
\, \tilde{\underline{G}}^1(z) 
\,
\underline \Sigma(z)  
+ ...
\Big]_{\Lambda \Lambda '}
\, \rho_{\Lambda '}^{1\times}(\vec r \,', z) 
\nonumber \\
\nonumber \\
%---------------------------------------------------------------------------
&=& 
\LABEL{EQ:SS-GF-G1-S1-Gamma-S1}
G^{1}(\vec r, \vec r\,' , z)\, 
+
\sum_{\Lambda\Lambda'}
 \rho_{\Lambda}^{1}(\vec r , z)
 \,
\Gamma_{\Lambda \Lambda '}(z)
\, \rho_{\Lambda '}^{1\times}(\vec r \,', z) \; ,
\end {eqnarray}
%=============================================================================
%
where the underline indicates matrices w.r.t.\ the spin-angular index
$\Lambda$.
The auxiliary Green function matrix  $ \tilde{\underline{G}}^1(z)$
is defined by the  projection 
of the single site Green function $G^{1}(\vec r, \vec r\,' , z)$
on to the basis functions $\phi_{\Lambda}(\vec r )$
%
%=============================================================================
\begin{eqnarray}
 \tilde{G}^1_{\Lambda \Lambda '}(z) &= &
\int d^{3} r  
\int d^{3} r\,' 
\phi_{\Lambda}^{\dagger}(\vec r) \
G^{1}(\vec r, \vec r\,' , z)\, 
 \phi_{\Lambda '}(\vec r \,') \; ,
\end {eqnarray}
%=============================================================================
while  the matrix 
  $\underline \Gamma(z)$ is given by
%
%=============================================================================
\begin{eqnarray}
\underline \Gamma(z)&=& 
\underline \Sigma(z)  
+
\underline \Sigma(z)  
\, \tilde{\underline{G}}^1(z) \,
\underline \Sigma(z)  
+
\underline \Sigma(z)  
\, \tilde{\underline{G}}^1(z) 
\,
\underline \Sigma(z)  
\, \tilde{\underline{G}}^1(z) 
\,
\underline \Sigma(z)  
+ ...
\nonumber \\
%---------------------------------------------------------------------------
&=&
\underline \Sigma(z)  
+
\underline \Sigma(z)  
\, \tilde{\underline{G}}^1(z) \,
\Big[
\underline \Sigma(z)  
+
\underline \Sigma(z)  
\, \tilde{\underline{G}}^1(z) \,
\underline \Sigma(z)  
+
\underline \Sigma(z)  
\, \tilde{\underline{G}}^1(z) 
\,
\underline \Sigma(z)  
\, \tilde{\underline{G}}^1(z) 
\,
\underline \Sigma(z)  
+ ...
\Big]
\nonumber \\
%---------------------------------------------------------------------------
&=&
\underline \Sigma(z)  
+
\underline \Sigma(z)  
\, \tilde{\underline{G}}^1(z) \,
\underline \Gamma(z)
\nonumber \\
%---------------------------------------------------------------------------
&=&
\LABEL{EQ:SS-GF-Gamma-renorm-SIG}
\Big(
\underline 1
+
\underline \Sigma(z)  
\, \tilde{\underline{G}}^1(z) 
\Big)^{-1} \,
\underline  \Sigma(z)  \; .
\end {eqnarray}
%=============================================================================
%
\end{widetext}
Using now the expansion of $\rho_{\Lambda}^{1}(\vec r , z)$ 
in terms of $R_{\Lambda_1}^{1}(\vec r, z)$ and
$H_{\Lambda_1}^{1}(\vec r, z)$
as given in Eq.\ \eqref{EQ:S1-EXPANSION-RHS}
together with its counterpart for  $\rho_{\Lambda}^{1\times}(\vec r, z) $
one can give the expansion coefficient functions
in Eq.\ \eqref{EQ:SS-GF-CDEF-LHS} explicitly as:
\begin{widetext}
\begin{eqnarray}
\LABEL{CC-R-SIG-R}
%----------------------------------------------------------------------------------------
G_{\Lambda \Lambda '}^{RR}(r,r') &=&
(-i \pbar)^2
\sum_{\Lambda ''\Lambda '''}  
C_{\Lambda \Lambda ''}^{1}(r,z) \, \Gamma_{\Lambda '' \Lambda '''}(z) \, C_{\Lambda ''' \Lambda '}^{1\times}(r ',z) 
\\
%
%----------------------------------------------------------------------------------------
\LABEL{CS-R-SIG-H}
G_{\Lambda \Lambda '}^{RH}(r,r') &=&
(-i \pbar)^2
\sum_{\Lambda ''\Lambda '''}  
C_{\Lambda \Lambda ''}^{1}(r,z) \, \Gamma_{\Lambda '' \Lambda '''}(z) \, S_{\Lambda ''' \Lambda '}^{1\times}(r ',z) 
% \nonumber \\ & &
 -i \pbar \, \delta_{\Lambda \Lambda '}\,  \theta(r ' - r) 
\\
%----------------------------------------------------------------------------------------
%
\LABEL{SC-H-SIG-R}
G_{\Lambda \Lambda '}^{HR}(r,r') &=&
(-i \pbar)^2
\sum_{\Lambda ''\Lambda '''}  
S_{\Lambda \Lambda ''}^{1}(r,z) \, \Gamma_{\Lambda '' \Lambda '''}(z) \, C_{\Lambda ''' \Lambda '}^{1\times}(r ',z) 
% \nonumber \\ & &  
-i \pbar \, \delta_{\Lambda \Lambda '}\,  \theta(r - r ') 
\\ 
%----------------------------------------------------------------------------------------
%
\LABEL{SS-H-SIG-H}
G_{\Lambda \Lambda '}^{HH}(r,r') &=&
(-i \pbar)^2
\sum_{\Lambda ''\Lambda '''}  
S_{\Lambda \Lambda ''}^{1}(r,z) \, \Gamma_{\Lambda '' \Lambda '''}(z) \, S_{\Lambda ''' \Lambda '}^{1\times}(r ',z) 
%%----------------------------------------------------------------------------------------
\;.
\end{eqnarray}
\end{widetext}
Comparing Eq.\  \eqref{EQ:SS-GF-Gamma-renorm-SIG}
with  Eq.\  \eqref{EQ:Delta-t-SIG-G1-SIG}
one can see that the renormalized self-energy
matrix  $\underline \Gamma(z)$
essentially corresponds to the change in the t-matrix
$\Delta \underline t(z)$ caused by the 
self-energy matrix w.r.t.\ that for the  local Hamiltonian.

Furthermore, one should mention that 
Eqs.\  \eqref{EQ:SS-GF-G1-S-SIG-S}
and \eqref{EQ:SS-GF-G1-S1-Gamma-S1}
 clearly show
that in case of a product representation of the self-energy
(see Eq.\ \eqref{EQ:SELF-EXPANSION}), a product 
representation of the Green function can be given as well. In fact in this 
case the arguments \cite{Zel15} given against a product representation as 
in Eq.~\eqref{EQ:SS-GF-RH-OUT} don't apply anymore.

\medskip

Finally, it should be noted that instead of working with the set of functions  
$R_{\Lambda}(\vec r, z)$, $H_{\Lambda}(\vec r, z)$,
 $R_{\Lambda}^{\times}(\vec r, z)$ and $H_{\Lambda}^{\times}(\vec r, z)$ 
specified by Eqs.~\eqref{EQ:RHS-ASYMPT-R} to \eqref{EQ:LHS-ASYMPT-H} 
it is sometimes more convenient to work with an alternative set of regular 
($Z_{\Lambda}(\vec r, z)$, $Z_{\Lambda}^{\times}(\vec r, z)$)
 and irregular  ($J_{\Lambda}(\vec r, z)$, $J_{\Lambda}^{\times}(\vec r, z)$) 
functions that are related to the original ones by the relations:
\begin{eqnarray}
\LABEL{EQ:RHS-ASYMPT-RZ}
%----------------------------------------------------------------------------------------
R_{\Lambda}(\vec r, z)& = &\sum_{\Lambda '}  Z_{\Lambda'} (\vec r, z)
 \,t_{\Lambda' \Lambda }(z) \\
%----------------------------------------------------------------------------------------
\LABEL{EQ:RHS-ASYMPT-HJ}
%----------------------------------------------------------------------------------------
-i \pbar H_{\Lambda}(\vec r, z)& = & Z_{\Lambda}(\vec r, z)
- \sum_{\Lambda '} J_{\Lambda'}(\vec r, z) \,
\,m_{\Lambda' \Lambda }(z)  \\
%----------------------------------------------------------------------------------------
%
\LABEL{EQ:LHS-ASYMPT-RZ}
%----------------------------------------------------------------------------------------
R_{\Lambda}^{\times}(\vec r, z)& = &\sum_{\Lambda '} 
 t_{\Lambda\Lambda' }(z) \, Z_{\Lambda '}^{\times} (\vec r, z) \\
%----------------------------------------------------------------------------------------
\LABEL{EQ:LHS-ASYMPT-HJ}
%----------------------------------------------------------------------------------------
-i \pbar H_{\Lambda}^{\times}(\vec r, z)& = &
 Z_{\Lambda '}^{\times} (\vec r, z) 
- \sum_{\Lambda '} 
m_{\Lambda \Lambda '}(z) \, J_{\Lambda '}^{\times} (\vec r, z) 
%----------------------------------------------------------------------------------------
%
\end{eqnarray}
where the matrix $\underline{m}(z)$ is the inverse of the single site $t$-matrix, i.e. 
$m_{\Lambda \Lambda '}(z) 
= \Big( \underline{t}^{-1}(z) \Big)_{\Lambda \Lambda '}$. 
The alternative set of functions obviously have the asymptotic behavior 
for $r > r_{\rm crit}$:
\begin{eqnarray}
\LABEL{EQ:RHS-ASYMPT-Z}
%----------------------------------------------------------------------------------------
Z_{\Lambda}(\vec r, z)& = &\sum_{\Lambda '}  
j_{\Lambda'} (\vec r, z)\,m_{\Lambda' \Lambda }(z) - 
i \pbar \, h_{\Lambda}^{+} (\vec r, z)   \\
%----------------------------------------------------------------------------------------
\LABEL{EQ:RHS-ASYMPT-J}
%----------------------------------------------------------------------------------------
J_{\Lambda}(\vec r, z)& = &j_{\Lambda} (\vec r, z)  \\
%----------------------------------------------------------------------------------------
%
\LABEL{EQ:LHS-ASYMPT-Z}
%----------------------------------------------------------------------------------------
Z_{\Lambda}^{\times}(\vec r, z)& = &\sum_{\Lambda '}  
m_{\Lambda \Lambda '}(z)\,j_{\Lambda'}^{\times} (\vec r, z) - 
i \pbar \,   h_{\Lambda}^{+ \times} (\vec r, z) \\
%----------------------------------------------------------------------------------------
\LABEL{EQ:LHS-ASYMPT-J}
%----------------------------------------------------------------------------------------
J_{\Lambda}^{\times}(\vec r, z)& = &j_{\Lambda }^{\times} (\vec r, z)  \; .
%----------------------------------------------------------------------------------------
%
\end{eqnarray}
Starting from Eq.~\eqref{EQ:SS-GF-RH-OUT} and using the relations \eqref{EQ:RHS-ASYMPT-RZ}
 - \eqref{EQ:LHS-ASYMPT-HJ} the single site Green function 
$G^{n}(\vec r, \vec r\,', z)$ can now be written as 
\begin{eqnarray}
%
%----------------------------------------------------------------------------------------
G^{n}(\vec r, \vec r \,', z) & = & \sum\limits_{\Lambda \Lambda '}\, Z_{\Lambda} (\vec r, z)\, 
t_{\Lambda \Lambda '}(z) \, Z_{\Lambda '}^{\times} (\vec r \,', z) \nonumber\\
%----------------------------------------------------------------------------------------
&& - \sum\limits_{\Lambda }\, \big( Z_{\Lambda} (\vec r, z)\, J_{\Lambda }^{\times} (\vec r \,', z) \, 
\theta(r ' - r)   \nonumber\\
\LABEL{EQ:GF-SS-OAK}
%----------------------------------------------------------------------------------------
&&\, \qquad + \, J_{\Lambda} (\vec r, z)\,  Z_{\Lambda }^{\times} (\vec r \,', z) \, 
\theta(r - r ') \big) 
%----------------------------------------------------------------------------------------
\end{eqnarray}
Here it is interesting to note that Eq.\ \eqref{EQ:GF-SS-OAK} that is based on the 
so-called Oak-Ridge-Bristol convention, i.e. normalization of the wave function 
according to Eqs.~\eqref{EQ:RHS-ASYMPT-Z}  to \eqref{EQ:LHS-ASYMPT-J}, 
holds for the case of an arbitrary $t$-matrix. Originally, its non-relativistic 
counterpart was derived by assuming explicitly a symmetric $t$-matrix 
\cite{FS80}. If one coherently distinguishes between the RHS and LHS solutions 
and their associated Lippmann-Schwinger equations, this restriction 
is obviously not necessary.

It should be emphasized once more that Eq.~\eqref{EQ:GF-SS-OAK} as  Eq.~\eqref{EQ:SS-GF-RH-OUT}
holds for any $\vec r$ and $\vec r\,'$ in case of a local Hamiltonian $\hat{\cal H}^{1}$.
In case of a non-local self-energy involved in $\hat{\cal H}$ both equations hold 
only if at least $r$ or $r\,'$ is larger than $r_{\rm crit}$.

Finally, it should be noted that the 
normalization of the regular and irregular wave functions
according to 
Eqs.\ \eqref{EQ:RHS-ASYMPT-R} to  \eqref{EQ:LHS-ASYMPT-H} 
or alternatively according to 
Eqs.\ \eqref{EQ:RHS-ASYMPT-Z} to  \eqref{EQ:LHS-ASYMPT-J} 
are not the only possible ones.
Other sets of functions and with this other 
representations of the Green function can be obtained by
imposing a suitable asymptotic behavior for the regular functions
and constructing the irregular functions accordingly.\cite{BGZ92,WZB+92}
See also appendix \ref{appendix-A} concerning this.

%****************************************************************
\subsection{Green function for extended systems}
%****************************************************************
The multiple scattering or KKR
formalism allows to obtain the Green function
of an extended system by introducing the corresponding (total)
t-matrix operator $\hat T(z)$ and its decomposition into the 
site ($i,j$) resolved scattering path operator $ \hat \tau^{ij}(z)$:\cite{FS80}
\begin{eqnarray}%----------------------------------------------------------------------------------------
\LABEL{DYSON-G0TG0}
\hat G(z) & = & \hat G^{0}(z) +  \hat G^{0}(z)\, \hat T(z) \, \hat G^{0}(z) \\
%----------------------------------------------------------------------------------------
\hat T(z) &=& \sum_{i, j} \hat \tau^{ij}(z)   
\end{eqnarray}
with the equations of motion
\begin{eqnarray}
%----------------------------------------------------------------------------------------
 \hat \tau^{ij}(z) &=& \hat t^{i}(z)\, \delta_{ij} + \hat t^{i}(z)\, \hat G^{0}(z)\,
\sum_{k \neq i}\, \hat \tau^{kj}(z) \\ 
%----------------------------------------------------------------------------------------
  &=&\hat  t^{i}(z)\, \delta_{ij} + \sum_{k \neq j}\, \hat \tau^{ik}(z)\, \hat G^{0}(z)\, \hat t^{j}(z) \;  
%----------------------------------------------------------------------------------------
\end{eqnarray}
and the free electron Green operator $\hat G^{0}(z)$.
In terms of the single-site Green function for site 
$n$ 
%%%%%($\hat G^{ss}(z)=\hat G^{n}(z)$) 
Eq.\ \eqref{DYSON-G0TG0}
can be rewritten as:\cite{FS80}
\begin{eqnarray}
%
%%%%\LABEL{EQ:GFTS-GF-OPER}
%----------------------------------------------------------------------------------------
\hat G(z) & = & \hat G^{n}(z) +  \hat G^{n}(z)\, \hat T_{nn}(z) \, \hat G^{n}(z)
%----------------------------------------------------------------------------------------
\end{eqnarray}
with the auxiliary operator 
\begin{eqnarray}
%
%----------------------------------------------------------------------------------------
 \hat T_{nn}(z) & = & \sum_{{i \neq n \atop j \neq m}} \hat \tau^{ij}(z) \; .
%----------------------------------------------------------------------------------------
\end{eqnarray}
To get the real space representation of this expression one makes use of 
the single site Green function 
connected with site $n$ according to Eqs.\  \eqref{EQ:SS-GF-RH-OUT},
\eqref{EQ:RHS-ASYMPT-H}
and \eqref{EQ:LHS-ASYMPT-H}:
\begin{eqnarray}
%
%----------------------------------------------------------------------------------------
G^{n}(\vec r, \vec r \,', z) & = & - i \pbar \sum\limits_{\Lambda }\, R_{\Lambda} (\vec r, z)\, 
h_{\Lambda}^{+ \times} (\vec r\,' , z)\, \theta(r ' - r)   \nonumber\\
%----------------------------------------------------------------------------------------
&&\;\;\; \qquad + \, h_{\Lambda}^{+} (\vec r, z)\,  R_{\Lambda }^{\times} (\vec r \,', z) \, 
\theta(r - r ') 
%----------------------------------------------------------------------------------------
\end{eqnarray}
that is valid for either $r'$ or $r$ being outside the potential well 
centered at site $n$. 
This leads to:
\begin{eqnarray}
%
%----------------------------------------------------------------------------------------
G(\vec r, \vec r \,', z)  & = & G^{n}(\vec r, \vec r \,', z) +  
(- i \pbar)^{2} \int d r\,'' \int d r\,''' 
\nonumber\\ 
%----------------------------------------------------------------------------------------
&& 
\quad \sum\limits_{\Lambda} R_{\Lambda} (\vec r, z)\, 
h_{\Lambda}^{+ \times} (\vec r\,'' , z)\,  
%\nonumber\\ 
%&&
\sum\limits_{{i \neq n \atop j \neq n}} \tau^{ij}(\vec r\,'', \vec r \,''', z)   \nonumber\\
%----------------------------------------------------------------------------------------
&& \hspace{0.5cm} \sum\limits_{\Lambda '} 
h_{\Lambda '}^{+} (\vec r\,''', z)\,  R_{\Lambda '}^{\times} (\vec r \,', z) \;.
%----------------------------------------------------------------------------------------
\end{eqnarray}
Taking into account that $ \tau^{ij}(\vec r, \vec r \,', z)$ 
is zero for $\vec r$ ($\vec r \,'$) 
being outside the atomic cell $i$ ($j$) and re-expanding 
the Hankel functions $h_{\Lambda}^{+ \times} (\vec r\,'' , z)$
 and $h_{\Lambda '}^{+} (\vec r\,''', z)$
  around the sites $i$ and $j$,
respectively,
making use of the 
so-called real space KKR structure constants $G_{\Lambda \Lambda '}^{0\, n m}(z)$
 \cite{Wei90,Gon92,GB99}
one gets finally for the site-diagonal Green function:
\begin{eqnarray}
\LABEL{EQ:GF-tot-G}
%----------------------------------------------------------------------------------------
G(\vec r, \vec r \,', z) & = & G^{n}(\vec r, \vec r \,', z) +  \sum\limits_{\Lambda \Lambda '} 
R_{\Lambda} (\vec r, z)\, G_{\Lambda \Lambda '}^{nn}(z) R_{\Lambda '}^{\times} (\vec r \,', z) \\ 
%----------------------------------------------------------------------------------------
\LABEL{EQ:GF-tot-TAU}
%----------------------------------------------------------------------------------------
 & =&  G^{n}(\vec r, \vec r \,', z) 
+ \sum\limits_{\Lambda \Lambda '} Z_{\Lambda} (\vec r, z)
\nonumber \\
&& 
\qquad \quad \big( \tau_{\Lambda \Lambda '}^{nn}(z)- t_{\Lambda \Lambda '}^{n}(z)\big)
 Z_{\Lambda '}^{\times} (\vec r \,', z) \;,
%----------------------------------------------------------------------------------------
\end{eqnarray}
with $\vec r $ and $\vec r \,'$ lying both within the cell $n$.
In the first expression we used 
the so-called structural Green function matrix 
 $\underline{G}^{nm}(z)$ that is connected to the scattering path operator 
matrix by the expression:\cite{ZDU+95}
\begin{eqnarray}
%
%----------------------------------------------------------------------------------------
 \underline\tau^{nm}(z) & = &  
\underline t^{n}(z)\, \underline G^{nm}(z) \, \underline t^{m}(z)
+  \underline t^{n}(z)\,\delta_{nm} \; . 
%----------------------------------------------------------------------------------------
%----------------------------------------------------------------------------------------
\end{eqnarray}
The second term in Eqs.~\eqref{EQ:GF-tot-G} and \eqref{EQ:GF-tot-TAU} represents 
the so-called back scattering contribution to
 the Green function, that is given 
in terms of the regular RHS and LHS solutions 
to the full Hamiltonian $\hat {\cal H}(z)$
that may contain a non-local self-energy. 
This means, that in contrast to the
single site Green function $G^{n}(\vec r, \vec r \,', z) $, the conventional
 expression for the back scattering
Green function \cite{FS80} is not affected by the presence of a 
non-local but site-diagonal self-energy.

An expression for the so-called site-off-diagonal 
 Green function $G(\vec r_n, \vec r_m, z)$ 
 connected with sites $n$ and $m$
 is derived in analogy to the site-diagonal one  \cite{FS80}
leading to:
\begin{eqnarray}
\LABEL{EQ:GF-tot-G-nm}
%----------------------------------------------------------------------------------------
G(\vec r_n, \vec r_m, z) & = & \sum\limits_{\Lambda \Lambda '} 
R_{\Lambda}^{n} (\vec r_n, z)\, G_{\Lambda \Lambda '}^{nm}(z)\, R_{\Lambda '}^{\times m} (\vec r_m, z) \\ 
%----------------------------------------------------------------------------------------
\LABEL{EQ:GF-tot-TAU-nm}
%----------------------------------------------------------------------------------------
 & =&  \sum\limits_{\Lambda \Lambda '} Z_{\Lambda}^{n} (\vec r_n, z)
\, \tau_{\Lambda \Lambda '}^{nm}(z) \,
 Z_{\Lambda '}^{\times m} (\vec r_m, z) \;,
%----------------------------------------------------------------------------------------
\end{eqnarray}
where cell centered coordinates $\vec r_n$ and $\vec r_m $ have been used.
Again, the Green function is expressed in terms of
 the regular RHS and LHS solutions 
to the full Hamiltonian $\hat {\cal H}(z)$
that may contain a non-local self-energy. 

For the sake of completeness we give the equation of motion
for the scattering path operator in its angular momentum
representation used in practice:
\begin{eqnarray}
\LABEL{EQ:SPO-EOM-A}
%----------------------------------------------------------------------------------------
 \underline{ \tau}^{ij}(z) &=& \underline{ t}^{i}(z)\, \delta_{ij} 
+ \underline{ t}^{i}(z)\, \underline{ G}^{0}(z)\,
\sum_{k \neq i}\, \underline{ \tau}^{kj}(z) \\ 
%----------------------------------------------------------------------------------------
\LABEL{EQ:SPO-EOM-B}
  &=&\underline{  t}^{i}(z)\, \delta_{ij} + \sum_{k \neq j}\, 
\underline{ \tau}^{ik}(z)\, \underline{ G}^{0}(z)\, \underline{ t}^{j}(z) \;  .
%----------------------------------------------------------------------------------------
\end{eqnarray}
For a finite system these equations can be solved by a simple matrix inversion. 
For a periodic system a solution is obtained by Fourier transformation.
For other, more complex geometries, corresponding techniques are available
to solve the multiple scattering problem \cite{Wei90,Gon92,GB99,EKM11}.

%****************************************************************
%****************************************************************
\section{Practical aspects}
%****************************************************************
%****************************************************************

\label{SEC:Practical-aspects}

%****************************************************************
\subsection{Computer codes 
to deal simultaneously
 with the RHS and LHS radial equations}
%****************************************************************
From the  form of the radial Dirac equations
 \eqref{EQ:RAD-DEQU-RHS} and
   \eqref{EQ:RAD-DEQU-LHS}
 for the RHS and LHS, respectively, solutions 
it is obvious that the two sets of solutions can be obtained with
one and the same radial
differential equation solver. Dealing with the LHS equation
one has just to do the replacement
%
%----------------------------------------------------------------------------------------
\begin{eqnarray}
\LABEL{EQ:REPLACE-V}
 V_{\Lambda  \, \Lambda '}^{\pm}(r) &\rightarrow& 
 V_{\Lambda' \, \Lambda  }^{\pm}(r)  \\ 
\LABEL{EQ:REPLACE-U}
   U_{\Lambda  \, \Lambda '}( r)  &\rightarrow & 
 - U_{\Lambda' \, \Lambda  }( r) \\
\LABEL{EQ:REPLACE-S}
  \Sigma_{\Lambda  \, \Lambda '}^{\pm}( r,  r \,',z) &\rightarrow&
  \Sigma_{\Lambda' \, \Lambda  }^{\pm}( r \,',r,  z) \;.
\end{eqnarray}
Dealing with the regular solutions one has to  impose the proper boundary conditions
according to Eqs.\ \eqref{EQ:RHS-ASYMPT-R}
and \eqref{EQ:LHS-ASYMPT-R}, respectively.
To have the same form for these equations for the RHS and LHS solutions,
one may introduce for the sake of convenience the auxiliary 
LHS t-matrix
$t_{\Lambda\Lambda'}^{\times}(z) $ by the definition
%
%----------------------------------------------------------------------------------------
\begin{eqnarray}
\LABEL{EQ:t-LHS}
t_{\Lambda\Lambda'}^{\times} (z) = t_{\Lambda'\Lambda} (z) \; ,
\end{eqnarray}
%----------------------------------------------------------------------------------------
%
that has no real physical meaning.
Using 
$t_{\Lambda\Lambda'}^{\times} (z) $ within 
Eq.\ \eqref{EQ:LHS-ASYMPT-R}
one gets the same form as 
Eq.\ \eqref{EQ:RHS-ASYMPT-R}.
 As a consequence making the replacement 
for the potential matrix element functions 
one can use one and the same computer routine
dealing with the RHS and LHS solutions including
the normalization of the 
wave function.
In this case, Eq.\ \eqref{EQ:t-LHS} can be used to check the 
consistency of the numerical results.
However, it should be stressed that the 
LHS t-matrix $t_{\Lambda\Lambda'}^{\times} (z) $
has no other meaning and
for that reason does not show up otherwise
as it can be clearly seen  in particular 
from Eqs.\ \eqref{EQ:t-matrix-jVR-RHS},
\eqref{EQ:t-matrix-RVj-LHS},
\eqref{EQ:SPO-EOM-A}, and
\eqref{EQ:SPO-EOM-B}.

%****************************************************************
\subsection{Direct solution of the radial Dirac 
equations and calculation of the single site t-matrix}
%****************************************************************

In the following some schemes to deal with the RHS and LHS radial Dirac equations
given by Eqs.~\eqref{EQ:RAD-DEQU-RHS} and \eqref{EQ:RAD-DEQU-LHS} are
briefly  discussed (for alternative schemes 
and numerical aspects see also Refs.\ \onlinecite{Kordt12,GAK+15}).
In the beginning we restrict to the case of a local potential, i.e. 
$\Sigma(\vec r, \vec r\,', z) = 0$. In this case one has to deal with sets 
of couples differential equations, that can be handled by standard techniques.

Starting the direct integration \cite{PFTV86}
 of Eq.~\eqref{EQ:RAD-DEQU-RHS} or \eqref{EQ:RAD-DEQU-LHS} 
 imposing regular boundary 
conditions at $r = 0$ \cite{FRA83,SSG84}
one may get the unnormalized regular wave functions:
\begin{eqnarray}
%
%----------------------------------------------------------------------------------------
\phi_{\Lambda}(\vec r, z) = \sum_{\Lambda '} \, \phi_{\Lambda ' \Lambda} (\vec r, z)\;. \nonumber
%----------------------------------------------------------------------------------------
%
\end{eqnarray}
Using the auxiliary quantities
\begin{eqnarray}
%
%----------------------------------------------------------------------------------------
a_{\Lambda ' \Lambda}(z)& = - &i \pbar r^{2}_{\rm crit} \, \left [ h_{\Lambda '}^{-}, \phi_{\Lambda ' \Lambda} \right ]_{r=r_{\rm crit}} \nonumber \\
%----------------------------------------------------------------------------------------
b_{\Lambda ' \Lambda}(z)& =  &i \pbar r^{2}_{\rm crit}\, \left [ h_{\Lambda '}^{+}, \phi_{\Lambda ' \Lambda} \right ]_{r=r_{\rm crit}} \nonumber 
%----------------------------------------------------------------------------------------
%
\end{eqnarray}
one can express the $t$-matrix as \cite{EG88}:
\begin{eqnarray}
\LABEL{EQ:t-matrix-EQ88}
%----------------------------------------------------------------------------------------
\underline{t}(z) = \frac{i}{2 \pbar} \, \bigg(\underline{a}(z) \, \Big( {\underline{b}}(z)\Big)^{-1} - \underline{1} \bigg)\;. 
%----------------------------------------------------------------------------------------
%
\end{eqnarray}

Using an alternative complete and linearly independent set of arbitrarily 
normalized functions $\psi_{\Lambda}$ defined by
\begin{eqnarray}
%
%----------------------------------------------------------------------------------------
\phi_{\Lambda}(\vec r, z) = \sum_{\Lambda '} \, \psi_{\Lambda '} (\vec r, z) \, \gamma_{\Lambda ' \Lambda} (z) \nonumber
%----------------------------------------------------------------------------------------
%
\end{eqnarray}
one can straightforwardly show that:
\begin{eqnarray}
%
%----------------------------------------------------------------------------------------
\underline{t}\{\phi\}(z) & = &\frac{i}{2 \pbar} \, \bigg( {\underline{a}}\{\phi\}(z) \, 
\Big( {\underline{b}}\{\phi\}(z)\Big)^{-1} - \underline{1} \bigg) \nonumber \\
%----------------------------------------------------------------------------------------
%
& = &\frac{i}{2 \pbar} \, \bigg( {\underline{a}}\{\psi\}(z) \, 
\Big( {\underline{b}}\{\psi\}(z)\Big)^{-1} - \underline{1} \bigg) \nonumber
%----------------------------------------------------------------------------------------
%
\end{eqnarray}
with
\begin{eqnarray}
%
%----------------------------------------------------------------------------------------
{\underline{a}}\{\phi\}(z) & = &  {\underline{a}}\{\psi\}(z)\,\underline{\gamma}(z) \nonumber \\
%----------------------------------------------------------------------------------------
%
{\underline{b}}\{\phi\}(z) & = &  {\underline{b}}\{\psi\}(z)\, \underline{\gamma}(z)\; . \nonumber
%----------------------------------------------------------------------------------------
%
\end{eqnarray}
This implies that Eq.~\eqref{EQ:t-matrix-EQ88} leads to the proper $t$-matrix as long as 
it is derived from a complete set of solutions being regular at the origin.

%****************************************************************
\subsection{Solution of the radial Dirac 
equations using the variable-phase approach}
%****************************************************************

\LABEL{SEC:variable-phase}

As an alternative to the direct integration of Eqs.~\eqref{EQ:RAD-DEQU-RHS} 
and \eqref{EQ:RAD-DEQU-LHS} one may extend the variable-phase approach of 
Calogero  \cite{Cal67} in an appropriate way.

Starting with the Lippmann-Schwinger equation \eqref{EQ:SS-GPOT-LSEQU-R-RHS}
\begin{eqnarray}
%
%----------------------------------------------------------------------------------------
R_{\Lambda}(\vec r, z) = j_{\Lambda}(\vec r, z) + \int\limits_{0}^{r_{\rm crit}} d^{3} r' \, G^{0}(\vec r, \vec r\, ',z)\, 
V(\vec r \, ') \, R_{\Lambda}(\vec r\, ',z) \nonumber
%----------------------------------------------------------------------------------------
%
\end{eqnarray}
one may introduce a set of auxiliary functions $\tilde {R}_{\Lambda}$
by the definition
\begin{eqnarray}
%
%----------------------------------------------------------------------------------------
R_{\Lambda}(\vec r, z) = \sum_{\Lambda '} \, \tilde R_{\Lambda '}(\vec r, z)\,  A_{\Lambda ' \Lambda}(z) \; .\nonumber
%----------------------------------------------------------------------------------------
%
\end{eqnarray}
Setting for the unknown expansion coefficients
\begin{eqnarray}
\LABEL{EQ:PHAS-A}
%
%----------------------------------------------------------------------------------------
 A_{\Lambda ' \Lambda}(z)& =& \delta_{\Lambda ' \Lambda} - i \pbar \, \int\limits_{0}^{r_{\rm crit}} d^{3} r'
 \, h_{\Lambda '}^{+\times}(\vec r\, ',z) 
\,  V(\vec r \, ') \, 
\nonumber \\
&& \qquad \qquad
\sum_{\Lambda ''} \, \tilde R_{\Lambda ''}(\vec r\, ',z)\,  A_{\Lambda '' \Lambda}(z)
%----------------------------------------------------------------------------------------
%
\end{eqnarray}
one gets a Volterra type of integral equations:
\begin{eqnarray}
%
%----------------------------------------------------------------------------------------
\tilde R_{\Lambda}(\vec r, z) &=& j_{\Lambda}(\vec r, z) 
\nonumber \\
&& \; 
+ \pbar \, \sum_{\Lambda '} \int\limits_{0}^{r} d^{3} r' \, 
N_{\Lambda '}(\vec r, \vec r\, ',z)\, 
 V(\vec r \, ') \, \tilde R_{\Lambda}(\vec r\, ',z) \nonumber
%----------------------------------------------------------------------------------------
%
\end{eqnarray}
with the auxiliary functions:
\begin{eqnarray}
%
%----------------------------------------------------------------------------------------
N_{\Lambda }(\vec r, \vec r\, ',z) = n_{\Lambda }(\vec r, z)\, j_{\Lambda }^{\times}(\vec r\, ', z) -
                                    j_{\Lambda }(\vec r, z)\, n_{\Lambda }^{\times}(\vec r\, ', z) 
\nonumber \;,
%----------------------------------------------------------------------------------------
%
\end{eqnarray}
where $ n_{\Lambda }(\vec r, z)$
and  $ n_{\Lambda }^{\times}(\vec r, z)$ 
are  relativistic von Neumann functions 
defined in analogy to Eqs.~\eqref{EQ:REG-RHS-j} through \eqref{EQ:REG-LHS-h}.
Introducing in addition the auxiliary functions
\begin{eqnarray}
%
%----------------------------------------------------------------------------------------
 C_{\Lambda ' \Lambda}(r, z)& = \delta_{\Lambda ' \Lambda}& -\, \pbar \, \int\limits_{0}^{r} d^{3} r' \, n^{\times}_{\Lambda '}(\vec r\, ', z) 
\,  V(\vec r \, ') \,  \tilde R_{\Lambda}(\vec r\, ',z) \nonumber \\
%----------------------------------------------------------------------------------------
 S_{\Lambda ' \Lambda}(r, z)& = & -\, \pbar \, \int\limits_{0}^{r} d^{3} r' \, j^{\times}_{\Lambda '}(\vec r\, ', z) 
\,  V(\vec r \, ') \,  \tilde R_{\Lambda}(\vec r\, ',z) \nonumber
%----------------------------------------------------------------------------------------
%
\end{eqnarray}
one finds the relation
\begin{eqnarray}
%
%----------------------------------------------------------------------------------------
\tilde R_{\Lambda}(\vec r, z) =  \sum_{\Lambda '} 
       j_{\Lambda '}(\vec r,z)\, C_{\Lambda ' \Lambda}(r, z)
     - n_{\Lambda '}(\vec r,z)\, S_{\Lambda ' \Lambda}(r, z)  \nonumber \;.
%----------------------------------------------------------------------------------------
%
\end{eqnarray}
With the definition of $A$ given by Eq.~\eqref{EQ:PHAS-A} one has:
\begin{eqnarray}
%
%----------------------------------------------------------------------------------------
 A_{\Lambda ' \Lambda}^{-1}(z)& = &\delta_{\Lambda ' \Lambda} + i \pbar \, \int\limits_{0}^{r_{\rm crit}} d^{3} r'
 \, h_{\Lambda '}^{+\times}(\vec r\, ',z) 
\, 
\nonumber \\
&& \qquad \qquad\qquad
 V(\vec r \, ') \, \sum_{\Lambda ''} \, \tilde R_{\Lambda ''}(\vec r\, ',z) \nonumber \\
%----------------------------------------------------------------------------------------
%
 & = &  C_{\Lambda ' \Lambda}(r, z) - i \,  S_{\Lambda ' \Lambda}(r, z)\hspace{1.4cm} 
\text{for}\hspace{0.3cm} r \geq r_{\rm crit} \nonumber 
%----------------------------------------------------------------------------------------
%
\end{eqnarray}
or in matrix notation
\begin{eqnarray}
%
%----------------------------------------------------------------------------------------
%
 \underline A(z) & = & \big( {\underline {C}}(r_{\rm crit}, z) - 
i \,  {\underline {S}}(r_{\rm crit}, z) \big)^{-1} \;. \nonumber 
%----------------------------------------------------------------------------------------
%
%
\end{eqnarray}
From the asymptotic form of the Lippmann-Schwinger equation one finally finds 
for the $t$-matrix:
\begin{eqnarray}
%
%----------------------------------------------------------------------------------------
%
 {\underline {t}}(z) & = & - \frac{1}{ \pbar } \, {\underline {S}}(r_{\rm crit},z)\, 
\big( {\underline {C}}(r_{\rm crit},z) - i\, {\underline {S}}(r_{\rm crit}, z) \big)^{-1}\;. \nonumber 
%----------------------------------------------------------------------------------------
%
%
\end{eqnarray}
It can be shown straightforwardly that this expression is completely equivalent 
to that given by Eq.~\eqref{EQ:t-matrix-EQ88}.

Here it should be noted that Calogero extended the variable-phase approach to deal 
with non-local potentials \cite{Cal67}. However, he considered only 
the non-relativistic case with no coupling of angular momentum channels. 
In fact the treatment of a non-spherical non-local potential by using 
the variable-phase approach is not straightforward.

%****************************************************************
\subsection{Solution of the radial Dirac 
equations via Born series expansion}
%****************************************************************
\LABEL{SEC:BORN}

The Lippmann-Schwinger equation for the RHS and LHS solutions can be solved alternatively 
by means of a Born series as was demonstrated 
for the non-relativistic \cite{Dri90,OA05} 
as well as relativistic \cite{HZE+98} case. Here we give the extension of this approach 
to the case of a finite self-energy. To make use of this scheme it is convenient to 
define the intermediate reference (r) 
system by restricting to a local spherical symmetric 
scalar potential and rotational symmetric
 vector fields (see Eqs.\ \eqref{ASA-V} to \eqref{ASA-A}).
This means that the remaining 
local potential functions $\Delta V(\vec r)$ and the self-energy 
are seen as an additional potential for the single-site problem.
 Having solved the radial 
equations for the regular and irregular solutions 
$R_{\Lambda}^{r}(r,z)$ and  $H_{\Lambda}^{r}(r,z)$ for the reference system, its single-site Green function is given by
\begin{eqnarray}
%
%----------------------------------------------------------------------------------------
G^{r}(\vec r, \vec r \,', z) & = & - i \pbar \sum\limits_{\Lambda }\, R_{\Lambda}^{r} (\vec r, z)\, 
H_{\Lambda}^{r \times} (\vec r\,' , z)\, \theta(r ' - r)   \nonumber\\
%----------------------------------------------------------------------------------------
&&\;\;\; \qquad + \, H_{\Lambda}^{r} (\vec r, z)\,  R_{\Lambda }^{r \times} (\vec r \,', z) \, 
\theta(r - r ') \nonumber \;,
%----------------------------------------------------------------------------------------
\end{eqnarray}
where one doesn't have to distinguish the radial functions for the RHS and LHS solutions, 
i.e.\ $g^{r}_{\Lambda ' \Lambda}(r,z) \equiv g_{\Lambda ' \Lambda}^{r\times}(r,z)$, etc. (see below). 
Using the auxiliary 
radial functions $P^{r}_{\Lambda ' \Lambda}(r,z) = r g^{r}_{\Lambda ' \Lambda}(r,z)$ 
and $Q^{r}_{\Lambda ' \Lambda}(r,z) = c r f^{r}_{\Lambda ' \Lambda}(r,z)$ 
one finds for the regular RHS solution the following 
radial Lippmann-Schwinger equation:
\begin{widetext}
\begin{eqnarray}
%
%----------------------------------------------------------------------------------------
\left ( \begin{array}{c} P_{\Lambda ' \Lambda}(r, z) \\ 
Q_{\Lambda ' \Lambda}(r, z) \end{array} \right )
 = \left ( \begin{array}{c} P_{\Lambda ' \Lambda}^{r}(r, z) \\ 
Q_{\Lambda ' \Lambda}^{r}(r, z) \end{array} \right ) 
+ \sum\limits_{\Lambda ''} \left\{ 
\left ( \begin{array}{c} P_{\Lambda ' \Lambda ''}^{r}(r, z) \\ 
Q_{\Lambda ' \Lambda ''}^{r}(r, z) \end{array} \right ) \, A_{\Lambda '' \Lambda}(r, z)
+ \left ( \begin{array}{c} \tilde {P}_{\Lambda ' \Lambda ''}^{r}(r, z) \\ 
\tilde {Q}_{\Lambda ' \Lambda ''}^{r}(r, z) \end{array} \right ) \, 
B_{\Lambda '' \Lambda}(r, z) \right\} \nonumber
%----------------------------------------------------------------------------------------
%
\end{eqnarray}
%%%%%%%%%%\end{widetext}
%
where $\tilde P_{\Lambda ' \Lambda}^{r}(r, z)$ and
 $\tilde Q_{\Lambda ' \Lambda}^{r}(r, z)$
denote the radial functions connected with the 
 large and small components  of the irregular solution 
$H_{\Lambda}^{r \times}(\vec r, z)$ and the phase functional matrices:
%
%%%%%%%%\begin{widetext}
\begin{eqnarray}
%
%----------------------------------------------------------------------------------------
%----------------------------------------------------------------------------------------
A_{\Lambda ' \Lambda}(r,z) &= &
-i \pbar\int\limits_r^{r_{\rm crit}}dr'\!\!  \sum\limits_{\Lambda'' \Lambda'''}
\bigg\{ \tilde{P}_{\Lambda''\Lambda}^{r}(r\,',z)
\Delta V_{\Lambda''\Lambda'''}^{+}(r\,')
P_{\Lambda'''\Lambda}(r\,',z)
+ \frac{1}{c^2}
\tilde{Q}_{\Lambda''\Lambda}^{r}(r\,',z)
\Delta V_{\Lambda''\Lambda'''}^{-}(r\,')
Q_{\Lambda'''\Lambda}(r\,',z)  \nonumber \\
%----------------------------------------------------------------------------------------
&& \qquad  \qquad \qquad\quad
+ \tilde {P}_{\Lambda '' \Lambda_{G}}^{r}( r \,')  
\int d r '' \, \Sigma_{\Lambda '' \Lambda '''}^{+}(r \,', r\,'') \, 
 P_{\Lambda ''' \Lambda}(r \,'')  
\nonumber \\
&& \qquad  \qquad \qquad\quad
+ \frac{1}{c^{2}}\, \tilde {Q}_{\Lambda '' \Lambda_{G}}^{r}( r \,') 
\int d r '' \,\Sigma_{-\Lambda '' -\Lambda '''}^{-}(r \,', r\,'')  \, 
 Q_{\Lambda ''' \Lambda}(r \,'') 
\bigg\}
\nonumber \\
%
%----------------------------------------------------------------------------------------
%----------------------------------------------------------------------------------------
B_{\Lambda ' \Lambda}(r,z) &= &
-i \pbar \int\limits_0^rdr'\!\! \sum\limits_{\Lambda'' \Lambda'''}
\bigg\{ P_{\Lambda''\Lambda '}^{r}(r\,',z)
\Delta V_{\Lambda''\Lambda'''}^{+}(r\,')
P_{\Lambda'''\Lambda}(r\,',z)
%\right.  \nonumber \\
%&&\hspace{1.5cm}
+ \frac{1}{c^2}
Q_{\Lambda''\Lambda '}^{r}(r\,',z)
\Delta V_{\Lambda''\Lambda'''}^{-}(r\,')
%\left. 
Q_{\Lambda'''\Lambda}(r\,',z)
\nonumber \\
&& \qquad  \qquad \qquad\quad
+ {P}_{\Lambda '' \Lambda_{G}}^{r}( r \,')  \int d r '' \, \Sigma_{\Lambda '' \Lambda '''}^{+}(r \,', r\,'') \, 
 P_{\Lambda ''' \Lambda}(r \,'')
\nonumber \\
&& \qquad  \qquad \qquad\quad
+  \frac{1}{c^{2}}\,
\, {Q}_{\Lambda '' \Lambda_{G}}^{r}( r \,') \int d r '' \,\Sigma_{-\Lambda '' -\Lambda '''}^{-}(r \,', r\,'')  \, 
 Q_{\Lambda ''' \Lambda}(r \,'') 
\bigg\} \nonumber 
%
%----------------------------------------------------------------------------------------
%
\end{eqnarray}
\end{widetext}
have been introduced.\\

Comparing the asymptotic behavior of the regular solutions $R_{\Lambda}^{r}$
for the reference system to that for the system including $\Delta V$ and 
$\Sigma$ one is led to the simple expression for the $t$-matrix
\begin{eqnarray}
%
%----------------------------------------------------------------------------------------
t_{\Lambda \Lambda '}(z) = t_{\Lambda \Lambda '}^{r}(z) +
 \Delta t_{\Lambda \Lambda '}(z)
%----------------------------------------------------------------------------------------
%
\end{eqnarray}
with
\begin{eqnarray}
%
%----------------------------------------------------------------------------------------
-i \pbar\, \Delta t_{\Lambda \Lambda '}(z) = B_{\Lambda \Lambda '}(r_{\rm crit}, z) 
%----------------------------------------------------------------------------------------
%
\end{eqnarray}
where $t_{\Lambda \Lambda '}^{r}$ is the $t$-matrix of the reference system.

%****************************************************************
\subsection{Wave function coupling scheme and full-potential picking rules}
%****************************************************************

\label{subsec:coupling-scheme}

Making use of a muffin-tin or ASA geometry implies a spherical symmetric
spin-independent potential and rotational symmetric vector 
fields according to:
\begin{eqnarray}
%
%----------------------------------------------------------------------------------------
\LABEL{ASA-V}
\bar V(\vec r) &=& \bar V(r)  \\
%----------------------------------------------------------------------------------------
\LABEL{ASA-B}
\vec B(\vec r) &=& \hat B\, B(r)  \\
%----------------------------------------------------------------------------------------
\LABEL{ASA-A}
\vec A(\vec r) &=& \hat A\, A(r) \; .
%----------------------------------------------------------------------------------------
%
\end{eqnarray}
Having the vectors  $\hat B$ and $\hat A$ 
along $\hat z$ or $\hat x$ the potential matrix functions are symmetric
implying that one has not to distinguish between RHS and LHS 
radial solutions. 
This favorable situation can always be achieved by working in a 
local frame with $\hat B || \hat A || \hat z'$ 
or $|| \hat x'$. Choosing $\hat z'$ has the great advantage
 that one has the most simple coupling scheme 
for the wave functions
\begin{eqnarray}
%
%----------------------------------------------------------------------------------------
\phi_{\Lambda}(\vec r, z) = \sum_{\Lambda '} \, \phi_{\Lambda ' \Lambda}(\vec r, z)
\hspace{0.5cm} \phi = R,\, H,\, Z \text{  or  } J \nonumber  \;.
%----------------------------------------------------------------------------------------
%
\end{eqnarray}
Allowing only coupling with $\Delta \, l = 0$ 
one has at most 2 terms in the summation $\sum_{\Lambda '}$ \cite{FRA83,SSG84}
(see also Eq.\ \eqref{EQ:RAD-DEQU-RHS}).

In case of a full-potential type  calculation
the non-vanishing terms $\bar V_{L}(r)$, $B_{L}^{\lambda}(r)$ and $A_{L}^{\lambda}(r)$ 
occurring in Eqs.\ \eqref{EQ:EXPAND-VLL} to  \eqref{EQ:EXPAND-ALL}
reflect the local symmetry of an atomic site.
Accordingly, the corresponding sets can be 
determined  by application of  the so-called picking rules 
that depend on the various symmetry operations being  present in the local point group.\cite{Kur77}
Table  \ref{TAB-PICK-RULE-1} gives for the various possible 
symmetry elements the 
resulting restriction for the non-vanishing expansion terms with angular 
character $L=(l,m)$.
%
%
%%\begin{widetext}
%FFFFFFFFFFFFFFFFFFFFFFFFFFFFFFFFFFFFFFFFFFFFFFFFFFFFFFFFFFFFFFFFFFFFFFFFFFFFFFFFFFFFFFFFFFFFFFF
\begin{table}[hbt!]
  \begin{center}
  \begin{tabular}{|ll|l|}
  \hline
   Symmetry element &  & Picking 
                                 rule for $L=(lm)$ \\
   \hline
   Inversion center &  & $(2\lambda,\pm\mu)$ \\
%----------------------------------------------------------------------------------
   $n$-fold rotation $\parallel\hat{z}$ &  & $(l,\pm n\mu)$ \\
%----------------------------------------------------------------------------------
   $n$-fold rotation-inversion $\parallel \hat{z}$ & &\\
  \hspace{0.6cm}    $n:$ even & &
                                                        $(2\lambda,\pm n\mu)$; 
\\
&&
                                        $(2\lambda+1,\pm n(\mu+\frac{1}{2}))$\\
%----------------------------------------------------------------------------------
    \hspace{0.6cm}        $n:$ odd  & &
                                        $(2\lambda,\pm n\mu)$\\
%----------------------------------------------------------------------------------
   $2$-fold axis $\parallel\, \hat{y}$ &  & $(2\lambda,+\mu)$; 
                                                                      $(2\lambda+1,-\mu)$ \\
%----------------------------------------------------------------------------------
   $2$-fold axis $\parallel\, \hat{x}$ & & $(l,l-2\mu)$; $(l,-(l-2\mu+1))$\\
%----------------------------------------------------------------------------------
   Symmetry plane && \\
   \hspace{0.6cm}   $\perp \, \hat z$ &  & $(l,\pm(l-2\mu))$ \\
%----------------------------------------------------------------------------------
   \hspace{0.6cm}   $\perp \, \hat y$ &  & $(l,+\mu)$ \\
%----------------------------------------------------------------------------------
   \hspace{0.6cm}   $\perp \, \hat x$ &  & $(l,+2\mu)$; $(l,-2\mu-1)$ \\
   \hline
  \end{tabular} 
  \end{center} 
  \caption{ Picking  rule for $L=(lm)$
for the expansion of a scalar function in terms of
  real spherical harmonics ${\cal Y}_{lm}\left(\hat{r}\right)$ 
    imposed by a  symmetry element. The  
            parameters $\lambda$ and $\mu$ specifying the allowed
            quantum numbers $(lm)$ in column 2 are integer
            numbers $\geq\ 0$ with $\mu$ taking all values
            for which $|m(\mu)| \leq l$ holds.\cite{Kur77}}
  \LABEL{TAB-PICK-RULE-1}
\end{table}
%FFFFFFFFFFFFFFFFFFFFFFFFFFFFFFFFFFFFFFFFFFFFFFFFFFFFFFFFFFFFFFFFFFFFFFFFFFFFFFFFFFFFFFFFFFFFFFFFF
%%\end{widetext}
%
From this table one can see that 
for the presence of a mirror plane perpendicular to $\hat y$
all expansion terms with  $m(L) <  0$ have to vanish.
Thus, if there is a local mirror plane $\sigma$ one can work with
a local frame of reference for which   $\sigma \perp  \hat y'$.
As a consequence of this choice one has again the situation
that all potential matrix elements are symmetric and for that 
reason one has not to distinguish between RHS and LHS
solutions. 
This is illustrated by Fig.\ \ref{FIG-hcp-unit-cell},
that gives  the unit cell of a hcp solid 
 for two different
choices of the coordinate system.
%
%FFFFFFFFFFFFFFFFFFFFFFFFFFFFFFFFFFFFFFFFFFFFFFFFFFFFFFFFFFFFFFFFFFFFFFFFFFFFFFFFFFFFFFFFFFFFFFF
\begin{figure}[hbt!]
\begin{center}
\includegraphics[scale=0.3,clip]{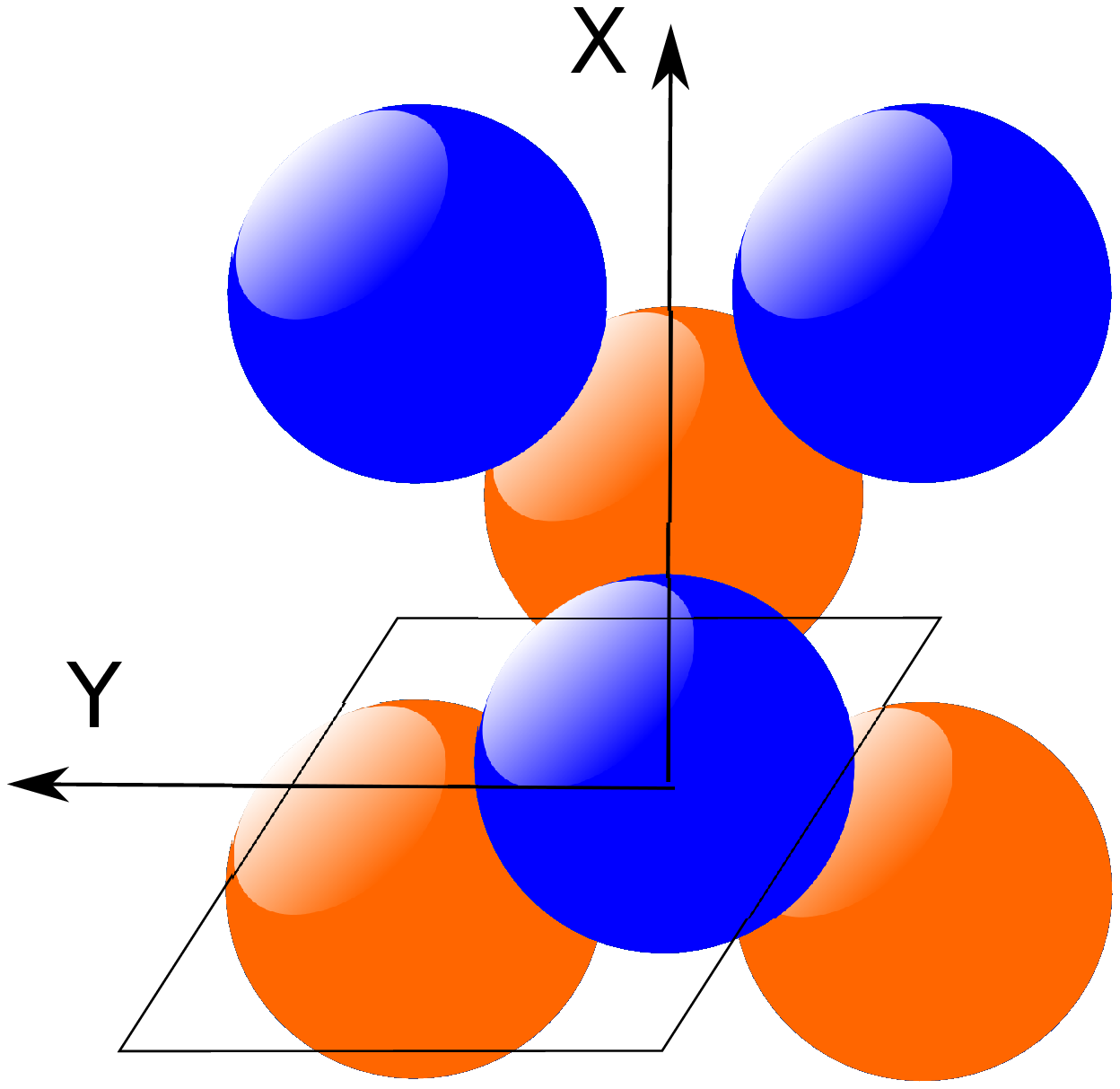}
\hspace{0.5cm}
\includegraphics[scale=0.3,clip]{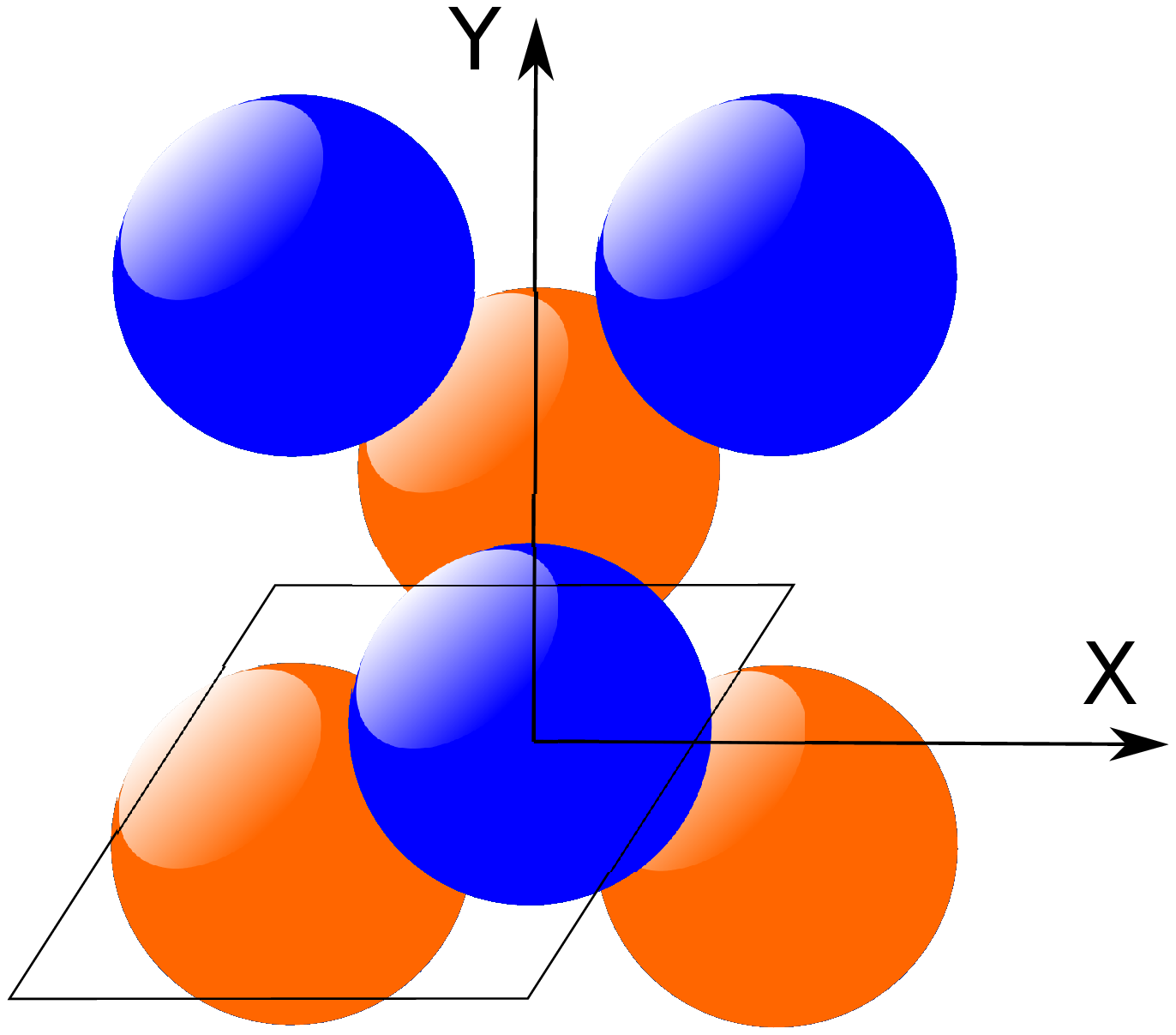}
\end{center}
   \caption{\LABEL{FIG-hcp-unit-cell}  Unit cell of a hcp solid 
 with two different
choices of the coordinate system. In both cases the 
z-axis points out of the plane.}
\end{figure}
%FFFFFFFFFFFFFFFFFFFFFFFFFFFFFFFFFFFFFFFFFFFFFFFFFFFFFFFFFFFFFFFFFFFFFFFFFFFFFFFFFFFFFFFFFFFFFFFFF
%
According to the picking
rules given in Table \ref{TAB-PICK-RULE-1}
there will be only positive $m$-values if the xz-plane is 
a mirror plane (left), while there will be 
positive, even and negative, odd 
 $m$-values if the yz-plane is 
a mirror plane (right). This is confirmed by Table  
\ref{FIG-L-table} that  gives the allowed $(l,m)$-values for an
expansion of the potential according to Eq.\ \eqref{EQ:EXPAND-VLL} 
for this two equivalent 
descriptions of the system.
%
%FFFFFFFFFFFFFFFFFFFFFFFFFFFFFFFFFFFFFFFFFFFFFFFFFFFFFFFFFFFFFFFFFFFFFFFFFFFFFFFFFFFFFFFFFFFFFFF
\begin{table}[hbt!]
  \begin{center}
  \begin{tabular}{|c||c||rrrrrrrrr|}
\hline
&      $l$  &~  $ 0 $&~  $ 2 $&  $ 3 $&~  $ 4 $&  $ 5 $&  $ 6 $&  $ 7 $&  $ 8 $& \\ \hline \hline
 $\sigma_{y}$
&   $m$     &~  $ 0 $&~  $ 0 $&  $+3 $&~  $ 0 $&  $ +3$&  $ 0 $&  $ +3$&  $ 0 $& \\[-1.5ex]
(xz) 
&         &~  $   $&~  $   $&  $   $&~  $   $&  $   $&  $+6 $&  $   $&  $+6 $& \\ \hline \hline
 $\sigma_{x}$
&   $m$     &~  $ 0 $&~  $ 0 $&  $-3 $&~  $ 0 $&  $ -3$&  $ 0 $&  $-3 $&  $ 0 $& \\[-1.5ex]
(yz) 
&         &~  $   $&~  $   $&  $   $&~  $   $&  $   $&  $+6 $&  $   $&  $+6 $& \\ \hline
  \end{tabular} 
  \end{center} 
   \caption{\LABEL{FIG-L-table}  
Non-vanishing elements $ V_{L}(r)$ for the 
expansion of the potential according to Eq.\ \eqref{EQ:EXPAND-VLL}
for the atom shown in Fig.\ \ref{FIG-hcp-unit-cell} at the origin  
 with the  magnetization along the $\hat z$ direction
given for a maximum angular momentum $l_{max}=8$.
The first and second rows give the allowed values for $m$
for the  xz- and yz-planes, respectively, being a mirror plane
($\sigma_{y}$ and  $\sigma_{x}$, respectively).
}
\end{table}
%FFFFFFFFFFFFFFFFFFFFFFFFFFFFFFFFFFFFFFFFFFFFFFFFFFFFFFFFFFFFFFFFFFFFFFFFFFFFFFFFFFFFFFFFFFFFFFFFF
%
Obviously, the non-vanishing potential terms
 $  V_{L}(r)$ determine which potential matrix element functions
 $V_{\Lambda \, \Lambda '}^{\pm}(r)$
 are non-zero. These in turn determine which 
 $\Lambda$-channels are coupled in
 the radial Dirac equations 
 \eqref{EQ:RAD-DEQU-RHS} and 
 \eqref{EQ:RAD-DEQU-LHS}, i.e.\
 which $\Lambda$-channels contribute to the expansion of the RHS and LHS 
 solutions in Eqs.\ \eqref{EQ:ANSATZ-RHS}
 and \eqref{EQ:ANSATZ-LHS}.
 This information is given by the wave function coupling scheme
 shown in Fig.\ \ref{FIG-coupling}.
%
%FFFFFFFFFFFFFFFFFFFFFFFFFFFFFFFFFFFFFFFFFFFFFFFFFFFFFFFFFFFFFFFFFFFFFFFFFFFFFFFFFFFFFFFFFFFFFFF
\begin{figure}[hbt]
  \begin{center}
  \begin{tabular}{|c||cc|cc|cccc|cccc|cccccc|}
\hline
          & \multicolumn{2}{c|}{$s_{1/2}$} & \multicolumn{2}{c|}{$p_{1/2}$} 
          & \multicolumn{4}{c|}{$p_{3/2}$} & \multicolumn{4}{c|}{$d_{3/2}$} 
          & \multicolumn{6}{c|}{$d_{5/2}$} \\
\hline
\hline
{$s_{1/2}$}
         &A&   & &   & & & &   & &A& &   & & &A& & &   \\[-1.5ex]
         & &B  & &   & & & &   & & &B&   & & & &B& &   \\ \hline
{$p_{1/2}$}
         & &   &C&   & &C& &   & & & &   & & & & & &C  \\[-1.5ex]
         & &   & &D  & & &D&   & & & &   &D& & & & &   \\ \hline
         & &   & &   &E& & &   & & & &E  & & & & &E&   \\[-1.5ex]
{$p_{3/2}$}
         & &   &C&   & &C& &   & & & &   & & & & & &C  \\[-1.5ex]
         & &   & &D  & & &D&   & & & &   &D& & & & &   \\[-1.5ex]
         & &   & &   & & & &F  &F& & &   & &F& & & &   \\ \hline
         & &   & &   & & & &F  &F& & &   & &F& & & &   \\[-1.5ex]
{$d_{3/2}$}
         &A&   & &   & & & &   & &A& &   & & &A& & &   \\[-1.5ex]
         & &B  & &   & & & &   & & &B&   & & & &B& &   \\[-1.5ex]
         & &   & &   &E& & &   & & & &E  & & & & &E&   \\ \hline
         & &   & &D  & & &D&   & & & &   &D& & & & &   \\[-1.5ex]
         & &   & &   & & & &F  &F& & &   & &F& & & &   \\[-1.5ex]
{$d_{5/2}$}
         &A&   & &   & & & &   & &A& &   & & &A& & &   \\[-1.5ex]
         & &B  & &   & & & &   & & &B&   & & & &B& &   \\[-1.5ex]
         & &   & &   &E& & &   & & & &E  & & & & &E&   \\[-1.5ex]
         & &   &C&   & &C& &   & & & &   & & & & & &C  \\ \hline
  \end{tabular} 
  \end{center} 
   \caption{\LABEL{FIG-coupling} Wave function coupling scheme 
for the atom shown in Fig.\ \ref{FIG-hcp-unit-cell} at the origin  
 with the  magnetization along the $\hat z$ direction
in the spin angular  representation,
i.e.\ the rows and columns are indexed with $\Lambda=(\kappa,\mu)$
given for a maximum angular momentum $l_{max}=2$.
The sub-blocks denoted by $s_{1/2}$ and so on correspond
to $\kappa= +1,\, -1,\, +2,\, -2$ and $+3$.
Within each sub-block the magnetic quantum number  $\mu$ runs
from $-j$ to $+j$ with $j=|\kappa|-1/2$.
For each column the occurring letter indicates 
the  non-vanishing contributions 
$\phi_{\Lambda ' \Lambda}(\vec r, z)$ to the sum 
$ \sum_{\Lambda '} \, \phi_{\Lambda ' \Lambda} (\vec r, z) $.
All columns having the same letter contribute to a corresponding
sub-block of the t-matrix. 
}
\end{figure}
%FFFFFFFFFFFFFFFFFFFFFFFFFFFFFFFFFFFFFFFFFFFFFFFFFFFFFFFFFFFFFFFFFFFFFFFFFFFFFFFFFFFFFFFFFFFFFFFFF
%
 Each letter within a column ($\Lambda$) indicates a contribution 
$\phi_{\Lambda ' \Lambda}(\vec r, z)$ to the wave function expansion.
Columns with the same letter have the same coupling sequence. In case
of regular functions these solutions contribute to a common sub-block
of the t-matrix.

%****************************************************************
%****************************************************************
\section{Summary} 
%****************************************************************
%****************************************************************

The full potential version of relativistic multiple scattering
theory has been reviewed in detail and an extension to the case of
a non-local but site-diagonal complex 
self-energy $\Sigma(\vec r , \vec r \,', z)$ has been discussed.
The properties of the  right- (RHS) and left-hand side (LHS)
solutions to the corresponding single-site problem have been worked
out.
It was demonstrated that the presence of a non-local self-energy
has  far reaching  consequences for the construction
of the associated single-site Green function. In particular, it turned
out that a simple product representation of the single-site
Green function in terms of the RHS and LHS solutions can be given 
only if the self-energy $\Sigma(\vec r , \vec r \,', z)$
can be written as a product of suitable basis functions
w.r.t.\ its dependence on the spatial variables $\vec r$
and  $\vec r \,'$. 
Furthermore, it was shown that in contrast to the single-site
Green function the back-scattering Green function 
representing the effect of the environment can
still be expressed as a product of RHS and LHS
solutions as in the case of a local potential.
Finally, some practical aspects of relativistic 
calculations for a general potential 
have been discussed. In particular the use of symmetry 
when dealing with the coupled radial equations
for the RHS and LHS solutions
has been demonstrated.

%****************************************************************
%****************************************************************
\appendix 
%****************************************************************
%****************************************************************

%****************************************************************
%****************************************************************
\section{Product representation of the single site Green function} 
%****************************************************************
%****************************************************************

\LABEL{appendix-A}

In the following, we consider first the single site Green function 
$G(\vec r, \vec r\, ',z)$ for the case of a general but local potential 
$V(\vec r)$, i.e. $\Sigma(\vec r, \vec r\,',z) = 0$ will be assumed.
Butler et al.\ \cite{BGZ92} showed for the corresponding non-relativistic 
case that the product representation for $G(\vec r, \vec r\, ',z)$ as given 
by Eq.\ \eqref{EQ:SS-GF-RH-OUT} is indeed a proper solution for the 
corresponding defining Eq.\ \eqref{EQ:RGF1-GFDEF-RHS} for any $\vec r$
and $\vec r\,'$. For the relativistic case, this proof
 can be given in an analogous way, as it is shown 
in the following. To simplify notation the energy argument $z$ will 
be omitted throughout and the frequent factor $-i \pbar$ will be combined
with the Hankel function (i.e., $-i \pbar h_{\Lambda}(\vec r) \rightarrow 
h_{\Lambda}(\vec r)$) and correspondingly with all other irregular
functions ($H_{\Lambda}(\vec r)$ and $F_{\Lambda}(\vec r)$) used below.
In contrast to the non-relativistic case, the RHS and LHS solutions
to the Dirac equations have to be distinguished properly, with all
quantities referring to the LHS solution indicated by ``$\times$''.\\

Following the scheme of Butler et al.\ \cite{BGZ92} we introduce 
an alternative RHS solution $\phi_{\Lambda}(\vec r)$ to the Dirac equation 
\eqref{EQ:DEQU-RHS} by imposing the boundary condition
\begin{eqnarray}
%----------------------------------------------------------------------------------------
\phi_{\Lambda}(\vec r) \rightarrow j_{\Lambda}(\vec r) \qquad \text{for} \qquad
\vec r \rightarrow 0 \nonumber \;.
%----------------------------------------------------------------------------------------
\end{eqnarray}
Setting up the Lippmann-Schwinger equation for  $\phi_{\Lambda}(\vec r)$ that
accounts for that asymptotic behavior one may express it in terms of the
Bessel and Hankel functions $j_{\Lambda}(\vec r)$ and $h_{\Lambda}(\vec r)$,
respectively, as:
\begin{eqnarray}
\LABEL{PHI-j-h-RHS}
%----------------------------------------------------------------------------------------
\phi_{\Lambda}(\vec r) = \sum\limits_{\Lambda '}\, j_{\Lambda '}(\vec r)\, C_{\Lambda ' \Lambda}
(r) + h_{\Lambda '}(\vec r)\, S_{\Lambda ' \Lambda}(r)
%----------------------------------------------------------------------------------------
\end{eqnarray}
with the expansion coefficients:
\begin{eqnarray}
%----------------------------------------------------------------------------------------
 C_{\Lambda ' \Lambda}(r) =& \delta_{\Lambda ' \Lambda}& - \int\limits_{0}^{r} d^{3} r\,'
 h_{\Lambda '}^{\times}(\vec r\,')\, V(\vec r\,')\, \phi_{\Lambda}(\vec r\,')\nonumber \\
%----------------------------------------------------------------------------------------
%----------------------------------------------------------------------------------------
  = &\,\, C_{\Lambda ' \Lambda}(r_{\rm crit})\,\,& + \int\limits_{r}^{r_{\rm crit}} d^{3} r\,'
 h_{\Lambda '}^{\times}(\vec r\,')\, V(\vec r\,')\, \phi_{\Lambda}(\vec r\,')\nonumber\\
%----------------------------------------------------------------------------------------
%----------------------------------------------------------------------------------------
 S_{\Lambda ' \Lambda}(r) =& & \int\limits_{0}^{r} d^{3} r\,'
 j_{\Lambda '}^{\times}(\vec r\,')\, V(\vec r\,')\, \phi_{\Lambda}(\vec r\,')\nonumber
%----------------------------------------------------------------------------------------
\end{eqnarray}
and a corresponding expansion for the LHS regular solution  $\phi_{\Lambda}^{\times}(\vec r)$:
\begin{eqnarray}
\LABEL{PHI-j-h-LHS}
%----------------------------------------------------------------------------------------
\phi_{\Lambda}^{\times}(\vec r) = \sum\limits_{\Lambda '}\, C_{\Lambda \Lambda '}^{\times}(r)\,
 j_{\Lambda '}^{\times}(\vec r) +  S_{\Lambda \Lambda '}^{\times}(r)\, h_{\Lambda '}^{\times}(\vec r)
%----------------------------------------------------------------------------------------
\end{eqnarray}
with 
\begin{eqnarray}
%----------------------------------------------------------------------------------------
 C_{\Lambda \Lambda '}^{\times}(r) =& \delta_{\Lambda \Lambda '}& - \int\limits_{0}^{r} d^{3} r\,'
 \phi_{\Lambda}^{\times}(\vec r\,')\, V(\vec r\,')\, h_{\Lambda '}(\vec r\,')\nonumber \\
%----------------------------------------------------------------------------------------
%----------------------------------------------------------------------------------------
  = &\,\, C_{\Lambda ' \Lambda}^{\times}(r_{\rm crit})\,\, & + \int\limits_{r}^{r_{\rm crit}} d^{3} r\,'
 \phi_{\Lambda}^{\times}(\vec r\,')\, V(\vec r\,')\, h_{\Lambda '}(\vec r\,')\nonumber\\
%----------------------------------------------------------------------------------------
%----------------------------------------------------------------------------------------
 S_{\Lambda \Lambda '}^{\times}(r) =& & \int\limits_{0}^{r} d^{3} r\,'
 \phi_{\Lambda}^{\times}(\vec r\,')\, V(\vec r\,')\, j_{\Lambda '}(\vec r\,')\nonumber
 \;.
%----------------------------------------------------------------------------------------
\end{eqnarray}
It should be mentioned that the expansion used here is completely equivalent to
that in terms of the Bessel and von Neumann functions
 used  in context of the  variable-phase approach of 
Calogero with the expansion 
coefficients $C_{\Lambda ' \Lambda}(r)$ and $ S_{\Lambda ' \Lambda}(r)$ redefined  (see section  \ref{SEC:variable-phase}). 

Considering the behavior of these functions for $r > r_{\rm crit}$ one finds
the connection to the regular functions $R_{\Lambda}(\vec r)$ and 
$R_{\Lambda}^{\times}(\vec r)$ introduced in Eqs.\eqref{EQ:RHS-ASYMPT-R}
and \eqref{EQ:LHS-ASYMPT-R}, respectively:
\begin{eqnarray}
\LABEL{R-PHI-RHS}
%----------------------------------------------------------------------------------------
R_{\Lambda}(\vec r) &=& \sum\limits_{\Lambda '}\, \phi_{\Lambda '}(\vec r)\, C_{\Lambda \Lambda '}^{-1}(r)\\
%----------------------------------------------------------------------------------------
\LABEL{R-PHI-LHS}
%----------------------------------------------------------------------------------------
R_{\Lambda}^{\times}(\vec r) &=& \sum\limits_{\Lambda '}\, C_{\Lambda \Lambda '}^{\times -1}(r)\, 
\phi_{\Lambda '}^{\times}(\vec r)
%----------------------------------------------------------------------------------------
\end{eqnarray}
as well as an expression for the $t$-matrix in terms of the expansion coefficients for 
$r > r_{\rm crit}$:
\begin{eqnarray}
%----------------------------------------------------------------------------------------
t_{\Lambda ' \Lambda ''} &=& \sum\limits_{\Lambda}\, S_{\Lambda ' \Lambda}(r_{\rm crit})\, C_{\Lambda \Lambda ''}^{-1}(r_{\rm crit})\\
%----------------------------------------------------------------------------------------
%----------------------------------------------------------------------------------------
t_{\Lambda '' \Lambda '} &=& \sum\limits_{\Lambda}\, C_{\Lambda '' \Lambda}^{\times -1}(r_{\rm crit})\, 
S_{\Lambda \Lambda '}^{\times}(r_{\rm crit}) \;.
%----------------------------------------------------------------------------------------
\end{eqnarray}
Replacing the regular function $R_{\Lambda}(\vec r)$ ($R_{\Lambda}^{\times}(\vec r)$) in
Eq.\ \eqref{EQ:SS-GF-RH-OUT} for the Green function by $ \phi_{\Lambda}(\vec r)$
($\phi_{\Lambda}^{\times}(\vec r)$) via Eqs.\  \eqref{R-PHI-RHS} and  \eqref{R-PHI-LHS}
implies a corresponding replacement of the irregular function  $H_{\Lambda}^{\times}(\vec r)$ 
($H_{\Lambda}(\vec r)$) by its counterpart  $F_{\Lambda}^{\times}(\vec r)$ 
($F_{\Lambda}(\vec r)$) defined by
\begin{eqnarray}
%----------------------------------------------------------------------------------------
F_{\Lambda '}(\vec r) &=& \sum\limits_{\Lambda}\, H_{\Lambda}(\vec r)\, C_{\Lambda \Lambda '}^{\times -1}(r)\\
%----------------------------------------------------------------------------------------
%----------------------------------------------------------------------------------------
F_{\Lambda '}^{\times}(\vec r) &=& \sum\limits_{\Lambda}\, C_{\Lambda ' \Lambda}^{-1}(r)\, 
H_{\Lambda}^{\times}(\vec r) \;.
%----------------------------------------------------------------------------------------
\end{eqnarray}
Again an expansion of these functions in terms of $j_{\Lambda}(\vec r)$ and
$h_{\Lambda}(\vec r)$ can be given starting from the corresponding Lippmann-Schwinger equation
and imposing the appropriate boundary conditions:
\begin{eqnarray}
\LABEL{F-j-h-RHS}
%----------------------------------------------------------------------------------------
F_{\Lambda}(\vec r) = \sum\limits_{\Lambda '}\, j_{\Lambda '}(\vec r)\,\bar C_{\Lambda ' \Lambda}
(r) + h_{\Lambda '}(\vec r)\, \bar S_{\Lambda ' \Lambda}(r)
%----------------------------------------------------------------------------------------
\end{eqnarray}
with
\begin{eqnarray}
%----------------------------------------------------------------------------------------
 \bar C_{\Lambda ' \Lambda}(r)  = &&  \int\limits_{r}^{r_{\rm crit}} d^{3} r\,'
 h_{\Lambda '}^{\times}(\vec r\,')\, V(\vec r\,')\, F_{\Lambda}(\vec r\,')\nonumber\\
%----------------------------------------------------------------------------------------
%----------------------------------------------------------------------------------------
 \bar S_{\Lambda ' \Lambda}(r) =&\,\,C_{\Lambda ' \Lambda}^{\times -1}(r_{\rm crit})\,\,& 
- \int\limits_{r}^{r_{\rm crit}} d^{3} r\,' j_{\Lambda '}^{\times}(\vec r\,')\, V(\vec r\,')\, F_{\Lambda}(\vec r\,')\nonumber
%----------------------------------------------------------------------------------------
\end{eqnarray}
\begin{eqnarray}
\LABEL{F-j-h-LHS}
%----------------------------------------------------------------------------------------
F_{\Lambda}^{\times}(\vec r) = \sum\limits_{\Lambda '}\,\bar C_{\Lambda \Lambda '}^{\times}
(r)\, j_{\Lambda '}^{\times}(\vec r) +  \bar S_{\Lambda \Lambda '}^{\times}(r)\, h_{\Lambda '}^{\times}(\vec r)
%----------------------------------------------------------------------------------------
\end{eqnarray}
with
\begin{eqnarray}
%----------------------------------------------------------------------------------------
 \bar C_{\Lambda \Lambda '}^{\times}(r)  = &&  \int\limits_{r}^{r_{\rm crit}} d^{3} r\,'
 F_{\Lambda}^{\times}(\vec r\,')\, V(\vec r\,')\, h_{\Lambda '}(\vec r\,')\nonumber\\
%----------------------------------------------------------------------------------------
%----------------------------------------------------------------------------------------
 \bar S_{\Lambda \Lambda '}^{\times}(r) =&\,\,C_{\Lambda \Lambda '}^{-1}(r_{\rm crit})\,\,& 
- \int\limits_{r}^{r_{\rm crit}} d^{3} r\,' F_{\Lambda}^{\times}(\vec r\,')\, V(\vec r\,')\, j_{\Lambda '}(\vec r\,')\nonumber  \;.
%----------------------------------------------------------------------------------------
\end{eqnarray}
Inserting now the expansion given in Eqs.\  \eqref{PHI-j-h-RHS},  \eqref{PHI-j-h-LHS},
 \eqref{F-j-h-RHS} and  \eqref{F-j-h-LHS} into the expression for the Green function
\begin{eqnarray}
\LABEL{GF-PHI-F}
%----------------------------------------------------------------------------------------
G(\vec r, \vec r\, ') =  \sum\limits_{\Lambda}&& \phi_{\Lambda}(\vec r)\,
 F_{\Lambda}^{\times}(\vec r\, ')\,\theta(r ' - r) \nonumber\\
%----------------------------------------------------------------------------------------
%----------------------------------------------------------------------------------------
 +&&   F_{\Lambda}(\vec r)
 \phi_{\Lambda}^{\times}(\vec r\,')\, \theta(r - r ')
%----------------------------------------------------------------------------------------
\end{eqnarray}
one is led to a corresponding expansion for the Green function in terms of 
$j_{\Lambda}(\vec r)$ and  $h_{\Lambda}(\vec r)$:
\begin{eqnarray}
\LABEL{GF-j-h-exp}
%----------------------------------------------------------------------------------------
G(\vec r, \vec r\, ') =  \sum\limits_{\Lambda '' \Lambda '''}& \Bigg\{ & j_{\Lambda ''}(\vec r)\,
\Bigg [ \sum\limits_{\Lambda} C_{\Lambda '' \Lambda}(r) \bar C_{\Lambda \Lambda '''}^{\times}(r ')\Bigg ]
j_{\Lambda '''}^{\times}(\vec r\, ') \nonumber\\
%----------------------------------------------------------------------------------------
&+& j_{\Lambda ''}(\vec r)\,
\Bigg [ \sum\limits_{\Lambda} C_{\Lambda '' \Lambda}(r) \bar S_{\Lambda \Lambda '''}^{\times}(r ')\Bigg ]
h_{\Lambda '''}^{\times}(\vec r\, ') \nonumber\\
%----------------------------------------------------------------------------------------
&+& h_{\Lambda ''}(\vec r)\,
\Bigg [ \sum\limits_{\Lambda} S_{\Lambda '' \Lambda}(r) \bar C_{\Lambda \Lambda '''}^{\times}(r ')\Bigg ]
j_{\Lambda '''}^{\times}(\vec r\, ') \nonumber\\
%----------------------------------------------------------------------------------------
&+& h_{\Lambda ''}(\vec r)\,
\Bigg [ \sum\limits_{\Lambda} S_{\Lambda '' \Lambda}(r) \bar S_{\Lambda \Lambda '''}^{\times}(r ')\Bigg ]
h_{\Lambda '''}^{\times}(\vec r\, ')\Bigg \} \nonumber\\
%----------------------------------------------------------------------------------------
&\times& \theta(r '\!\! -\! r) \nonumber\\
%----------------------------------------------------------------------------------------
+  \sum\limits_{\Lambda '' \Lambda '''}& \Bigg\{ &j_{\Lambda ''}(\vec r)\,
\Bigg [ \sum\limits_{\Lambda} \bar C_{\Lambda '' \Lambda}(r)\, C_{\Lambda \Lambda '''}^{\times}(r ')\Bigg ]
j_{\Lambda '''}^{\times}(\vec r\, ') \nonumber\\
%----------------------------------------------------------------------------------------
%----------------------------------------------------------------------------------------
&+& j_{\Lambda ''}(\vec r)\,
\Bigg [ \sum\limits_{\Lambda} \bar C_{\Lambda '' \Lambda}(r)\, S_{\Lambda \Lambda '''}^{\times}(r ')\Bigg ]
h_{\Lambda '''}^{\times}(\vec r\, ')\nonumber\\
%----------------------------------------------------------------------------------------
%----------------------------------------------------------------------------------------
&+& h_{\Lambda ''}(\vec r)\,
\Bigg [ \sum\limits_{\Lambda} \bar S_{\Lambda '' \Lambda}(r)\, C_{\Lambda \Lambda '''}^{\times}(r ')\Bigg ]
j_{\Lambda '''}^{\times}(\vec r\, ')\nonumber\\
%----------------------------------------------------------------------------------------
%----------------------------------------------------------------------------------------
&+& h_{\Lambda ''}(\vec r)\,
\Bigg [ \sum\limits_{\Lambda} \bar S_{\Lambda '' \Lambda}(r)\, S_{\Lambda \Lambda '''}^{\times}(r ')\Bigg ]
h_{\Lambda '''}^{\times}(\vec r\, ')\Bigg \}  \nonumber\\
%----------------------------------------------------------------------------------------
&\times& \theta(r\!\! -\! r ')  \;.
%---------------------------------------------------------------------------
%----------------------------------------------------------------------------------------
\end{eqnarray}
To evaluate these expressions it is helpful to use in addition the following 
Lippmann-Schwinger equations
\begin{eqnarray}
%----------------------------------------------------------------------------------------
R_{\Lambda}(\vec r) =  j_{\Lambda}(\vec r) +  \int\limits_{0}^{r_{\rm crit}}d^{3} r '\,
G(\vec r, \vec r\,')\, V(\vec r\,')\,j_{\Lambda}(\vec r\, ')\  \nonumber \\
%----------------------------------------------------------------------------------------
%----------------------------------------------------------------------------------------
F_{\Lambda}(\vec r) =  h_{\Lambda}(\vec r) +  \int\limits_{0}^{r_{\rm crit}}d^{3} r '\,
G(\vec r\,', \vec r)\, V(\vec r\,')\, h_{\Lambda}(\vec r\, ') \nonumber
%----------------------------------------------------------------------------------------
\end{eqnarray}
as well as their counterparts for the LHS solutions $R_{\Lambda}^{\times}(\vec r)$ and $F_{\Lambda}^{\times}(\vec r)$
together with the relation for the $t$-matrix operator
\begin{eqnarray}
%----------------------------------------------------------------------------------------
\hat t = \hat V + \hat V\, \hat G\, \hat V.
%----------------------------------------------------------------------------------------
\end{eqnarray}
The sums [...] in Eq.\ \eqref{GF-j-h-exp} can now be reformulated more or less 
the same way as sketched in the appendix of Ref.\ \onlinecite{BGZ92} for the non-relativistic case, 
however, taking the difference between RHS and LHS solutions into account. This way 
one can express the various sums [...] in Eq.\ \eqref{GF-j-h-exp} in terms of the 
real space representation of the $t$-matrix operator
\begin{widetext}
\begin{eqnarray}
\LABEL{CC-hth}
%----------------------------------------------------------------------------------------
 \sum\limits_{\Lambda ''} C_{\Lambda \Lambda ''}(r)\, 
\bar C_{\Lambda '' \Lambda '}^{\times}(r ') = \sum\limits_{\Lambda ''} \bar C_{\Lambda \Lambda ''}(r)\, 
C_{\Lambda '' \Lambda '}^{\times}(r ') &= \int\limits_{r}^{r_{\rm crit}} d^{3}r_{1} 
\int\limits_{r '}^{r_{\rm crit}} d^{3}r_{2}
\, h_{\Lambda}^{\times}(\vec r_{1})\, t(\vec r_{1},\vec r_{2})\, h_{\Lambda '}(\vec r_{2})
&  = G_{\Lambda \Lambda '}^{jj}(r,r')\\
\LABEL{CS-htj}
%----------------------------------------------------------------------------------------
 \sum\limits_{\Lambda ''} C_{\Lambda \Lambda ''}(r)\, 
\bar S_{\Lambda '' \Lambda '}^{\times}(r ') = \sum\limits_{\Lambda ''} \bar C_{\Lambda \Lambda ''}(r)\, 
S_{\Lambda '' \Lambda '}^{\times}(r ') &= \int\limits_{r}^{r_{\rm crit}} d^{3}r_{1} 
\int\limits_{0}^{r '} d^{3}r_{2}
\, h_{\Lambda}^{\times}(\vec r_{1})\, t(\vec r_{1},\vec r_{2})\, j_{\Lambda '}(\vec r_{2})\nonumber\\
%----------------------------------------------------------------------------------------
 & + \,\, \delta_{\Lambda \Lambda '}\,  \theta(r ' - r) 
&= G_{\Lambda \Lambda '}^{jh}(r,r') 
\\
%----------------------------------------------------------------------------------------
%
\LABEL{SC-jth}
%----------------------------------------------------------------------------------------
 \sum\limits_{\Lambda ''} S_{\Lambda \Lambda ''}(r)\, 
\bar C_{\Lambda '' \Lambda '}^{\times}(r ') = \sum\limits_{\Lambda ''} \bar S_{\Lambda \Lambda ''}(r)\, 
C_{\Lambda '' \Lambda '}^{\times}(r ') &= \int\limits_{0}^{r} d^{3}r_{1} 
\int\limits_{r '}^{r_{\rm crit}} d^{3}r_{2}
\, j_{\Lambda}^{\times}(\vec r_{1})\, t(\vec r_{1},\vec r_{2})\, h_{\Lambda '}(\vec r_{2})\nonumber\\
%----------------------------------------------------------------------------------------
 & + \,\, \delta_{\Lambda \Lambda '}\,  \theta(r - r ') 
&= G_{\Lambda \Lambda '}^{hj}(r,r') \\
%----------------------------------------------------------------------------------------
%
\LABEL{SS-jtj}
%----------------------------------------------------------------------------------------
\sum\limits_{\Lambda ''} S_{\Lambda \Lambda ''}(r)\, 
\bar S_{\Lambda '' \Lambda '}^{\times}(r ') = \sum\limits_{\Lambda ''} \bar S_{\Lambda \Lambda ''}(r)\, 
S_{\Lambda '' \Lambda '}^{\times}(r ') &= \int\limits_{0}^{r} d^{3}r_{1} 
\int\limits_{0}^{r '} d^{3}r_{2}
\, j_{\Lambda}^{\times}(\vec r_{1})\, t(\vec r_{1},\vec r_{2})\, j_{\Lambda '}(\vec r_{2})
 &= G_{\Lambda \Lambda '}^{hh}(r,r') 
%%----------------------------------------------------------------------------------------
\;.
\end{eqnarray}
\end{widetext}
This result is exactly what has to be expected on the basis of the representation 
of the Green function $G(\vec r, \vec r\,')$ in terms of the $t$-matrix and the 
free electron Green function $G^{0}(\vec r, \vec r\,')$ as given in 
Eq.\ \eqref{EQ:SS-DYSON-EQ-T}. Inserting the explicit expression for
 $G^{0}(\vec r, \vec r\,')$ in terms of the Bessel and Hankel functions
 (see Eq.\ \eqref{EQ:GF-FEG}) into Eq.\ \eqref{EQ:SS-DYSON-EQ-T} one is led
to the expression for the Green function $G(\vec r, \vec r\,')$ in terms
of $j_{\Lambda}(\vec r)$ and  $h_{\Lambda}(\vec r)$ with
 the expansion coefficients 
given by the integrals as listed in Eqs.\  \eqref{CC-hth} to  \eqref{SS-jtj}.
This direct derivation of the integral form of the 
expansion coefficients 
$G_{\Lambda \Lambda '}^{\alpha \beta}(r,r')$ ($\alpha,\, \beta = j,\, h$) does obviously not
depend on the specific form of the potential. For that reason these expressions
have to hold also for the more 
general situation of a non-local potential (see below).

The expansion of the Green function given by Eq.\ \eqref{GF-j-h-exp} allows now
straight forwardly to demonstrate that Eq.\ \eqref{EQ:RGF1-GFDEF-RHS} is solved
for any $\vec r$ and $\vec r\, '$. For  $\vec r \neq \vec r\, '$ this is 
ensured by constructing $G(\vec r, \vec r\,')$ in terms of 
RHS and LHS solutions to the 
corresponding Dirac equations. For  $\vec r = \vec r\, '$ the inhomogeneity
represented by the $\delta$-function has to be recovered in addition.
To demonstrate this one integrates Eq.\ \eqref{EQ:RGF1-GFDEF-RHS} w.r.t.\ $r$
over the interval $[r ' - \epsilon, r ' + \epsilon]$ and takes the limit 
$\epsilon \rightarrow 0$ afterwards. For this purpose it is advantageous
to split the Dirac Hamiltonian in Eq.\ \eqref{EQ:DEQU-RHS} into the term
involving the radial derivative $\hat {\cal H}_{\rm rad}(\vec r) = -ic\,\alpha_{r}
\frac{\partial}{\partial r}$ and the remaining rest $\hat {\cal H}_{\rm rest}(\vec r) = 
\hat {\cal H}(\vec r) - \hat {\cal H}_{\rm rad}(\vec r)$. Here we combined 
$\gamma_{s}\, \sigma_{r}$ used in Eq.\ \eqref{EQ:RGF1-DEQ} to 
$- \alpha_{r} = - \vec \alpha \cdot \vec r/r$ \cite{Ros61}. With this one has:
\begin{widetext}
\begin{eqnarray}
%----------------------------------------------------------------------------------------
-\int\limits_{r ' - \epsilon}^{r ' + \epsilon} dr\,  \hat {\cal H}_{\rm rad}(\vec r)\, 
G(\vec r, \vec r\,') + \int\limits_{r ' - \epsilon}^{r ' + \epsilon} dr\, \Big( z - 
\hat {\cal H}_{\rm rest}(\vec r)\Big )G(\vec r, \vec r\,') = 
\int\limits_{r ' - \epsilon}^{r ' + \epsilon} dr\, \delta(\vec r - \vec r\, ')
 {\mathbbm{1}}_{4}   \; . \nonumber
%----------------------------------------------------------------------------------------
\end{eqnarray}
\end{widetext}
As $G(\vec r, \vec r\,')$ is continuous at $r = r '$ w.r.t.\ $r$ the second 
integral will vanish for $\epsilon \rightarrow 0$. Using the relation
$\delta(\vec r - \vec r\, ') = \frac{1}{r^{2}}\, \delta(r - r')
\, \delta(\hat r, \hat r')$ for the $\delta$-function one gets
\begin{eqnarray}
%----------------------------------------------------------------------------------------
-\!\!\int\limits_{r ' - \epsilon}^{r ' + \epsilon}\!\! dr \Big( -ic\,\alpha_{r}
\frac{\partial}{\partial r} \Big)\, 
G(\vec r, \vec r\,') = 
\!\!\int\limits_{r ' - \epsilon}^{r ' + \epsilon}\!\! dr\, \frac{1}{r^{2}}\delta(r - r\, ')
\delta(\hat r, \hat r')
 {\mathbbm{1}}_{4} \nonumber\\
%----------------------------------------------------------------------------------------
%----------------------------------------------------------------------------------------
 -ic\,\alpha_{r}
\Big [ G(\vec r\,' + \epsilon, \vec r\,') - G(\vec r\,' - \epsilon, \vec r\,')\Big ]  = 
 \frac{1}{r'^{2}}\, \delta(\hat r, \hat r')  {\mathbbm{1}}_{4} \nonumber
\;.
%----------------------------------------------------------------------------------------
\end{eqnarray}
Inserting now the expansion of $G(\vec r, \vec r\,')$ in terms of $j_{\Lambda}$ and
$h_{\Lambda}$ as given by Eq.\ \eqref{GF-j-h-exp} and making use of Eqs.\  
\eqref{CC-hth} to \eqref{SS-jtj} one is led    in the limit $\epsilon \rightarrow 0$  to:
\begin{eqnarray}
\LABEL{jh-hj-delta}
%----------------------------------------------------------------------------------------
 -ic\,\alpha_{r}
\sum\limits_{\Lambda} \Big(j_{\Lambda}(\vec r) h_{\Lambda}^{\times}(\vec r\,') - 
h_{\Lambda}(\vec r) j_{\Lambda}^{\times}(\vec r\,') \Big )  = 
 \frac{1}{r'^{2}}\, \delta(\hat r, \hat r')  {\mathbbm{1}}_{4} \nonumber
\; .
\\
%----------------------------------------------------------------------------------------
\end{eqnarray}
The expression on the left hand side can be evaluated straight forwardly using 
the definition of $\alpha_{r}$ together with 
Eq.\ \eqref{EQ:SO-K-AIG-VAL}. Finally, 
making use of the Wronskian of the Bessel and 
Hankel functions (with the later ones including in this section the factor $-i \pbar$)
given by 
Eq.\ \eqref{EQ:WRONSKIAN-jh-hj}  together
with the relations
\begin{eqnarray}
%----------------------------------------------------------------------------------------
\sum\limits_{\Lambda}  \chi_{\Lambda}(\hat r) \,  \chi_{\Lambda}^{\dagger}(\hat r') =
\sum\limits_{\Lambda}  \chi_{-\Lambda}(\hat r) \,  \chi_{-\Lambda}^{\dagger}(\hat r') = 
\delta(\hat r, \hat r')  {\mathbbm{1}}_{2} \nonumber
%----------------------------------------------------------------------------------------
\end{eqnarray}
the equivalency of the left and right hand side of Eq.\ \eqref{jh-hj-delta}
can be demonstrated completing the proof this way.

 \medskip

To investigate whether the product representation of  $G(\vec r, \vec r\,')$
given by  Eq.\ \eqref{EQ:SS-GF-RH-OUT} is also a proper solution for the defining 
Eq.\ \eqref{EQ:RGF1-GFDEF-RHS} for a non-local potential ${\cal V}(\vec r, \vec r\,') = 
V(\vec r)\,\delta(\vec r - \vec r\,') + \Sigma(\vec r, \vec r\,')$ one may 
generalize the expansion of the various wave functions in terms of $j_{\Lambda}(\vec r)$ and
$h_{\Lambda}(\vec r)$ in an appropriate way. For example for the expansion coefficient 
$C_{\Lambda ' \Lambda}(r)$ of $\phi_{\Lambda}(\vec r)$ in Eq.\ \eqref{PHI-j-h-RHS} one gets:
\begin{eqnarray}
%----------------------------------------------------------------------------------------
C_{\Lambda ' \Lambda}(r) = \delta_{\Lambda \Lambda '} - \int\limits_{0}^{r} d^{3} r ' 
h_{\Lambda '}^{\times}(\vec r\,') \int\limits_{0}^{r_{\rm crit}} d^{3} r '' {\cal V}(\vec r\,',\vec r\,'')\,
\phi_{\Lambda}(\vec r\,'') \nonumber
\; .
%----------------------------------------------------------------------------------------
\end{eqnarray}
Considering now for example the case $r ' > r$, this together with a corresponding 
expression for $\bar C_{\Lambda ' \Lambda}^{\times}(r)$ leads to:
\begin{widetext}
\begin{eqnarray}
%----------------------------------------------------------------------------------------
\sum\limits_{\Lambda ''}C_{\Lambda \Lambda ''}(r)\, \bar C_{\Lambda '' \Lambda '}^{\times}(r ')& =&
\sum\limits_{\Lambda ''} \Bigg [  \delta_{\Lambda '' \Lambda} - \int\limits_{0}^{r}  d^{3} r_{1}\,
h_{\Lambda}^{\times}(\vec r_{1})  \int\limits_{0}^{r_{\rm crit}} d^{3} r_{2} {\cal V}(\vec r_{1},\vec r_{2})
\phi_{\Lambda ''}(\vec r_{2})\Bigg ]\,
\nonumber \\
&& \qquad\qquad \Bigg[ \int\limits_{r '}^{r_{\rm crit}} d^{3} r_{3},
F_{\Lambda ''}^{\times}(\vec r_{3}) \!\! \int\limits_{0}^{r_{\rm crit}} d^{3} r_{4} {\cal V}(\vec r_{3},\vec r_{4})
h_{\Lambda '}(\vec r_{4}) \Bigg]\nonumber
\; .
%----------------------------------------------------------------------------------------
\end{eqnarray}
\end{widetext}
In contrast to the case of a local potential $V(\vec r)$ the restriction
$r_{3} \geq r ' > r \geq r_{2}$ does not hold anymore.
As a consequence the combined term $\sum\limits_{\Lambda ''}\phi_{\Lambda ''}(\vec r_{2})
F_{\Lambda ''}^{\times}(\vec r_{3})$ cannot be identified with the Green function 
$G(\vec r_{2}, \vec r_{3})$ anymore. Due to this, the sum $\sum\limits_{\Lambda ''}
C_{\Lambda \Lambda ''}(r)\, \bar C_{\Lambda '' \Lambda '}^{\times}(r ')$ is not identical to the 
integral $\int\limits_{r}^{r_{\rm crit}}  d^{3} r_{1} \int\limits_{r '}^{r_{\rm crit}} d^{3} r_{2}
\,h_{\Lambda}^{\times}(\vec r_{1})\, t(\vec r_{1},\vec r_{2})\, h_{\Lambda '}(\vec r_{2})$.
As this is required also for the case of a non-local potential (see above) the
product representation for the Green function given in Eq.\ \eqref{GF-PHI-F}
cannot be valid in case of a non-local potential.
This obviously also applies if an expansion 
of  the self-energy
in terms of basis functions 
 as given by Eq.\ \eqref{EQ:SELF-EXPANSION} is assumed.
In this case the double integral over $ \vec r_1$ and
$ \vec r_2$ and also that over 
$ \vec r_3$  and$ \vec r_4$ factorize, but the above mentioned necessary restriction concerning
$ r_2$ and 
$ r_3$ still does not hold.

%****************************************************************
%****************************************************************
%                         ACKNOWLEDGMENTS    
%****************************************************************
%****************************************************************
\begin{acknowledgments}
  This    work    was   supported    financially    by   the    Deutsche
  Forschungsgemeinschaft (DFG) 
within the research group  FOR 1346.
Helpful discussions with Rudi Zeller and
Rino Natoli 
and 
within the COST Action MP1306 EUSpec
are gratefully acknowledged.
\end{acknowledgments}

%\bibliography{akhelit,RKKR-add}

\begin{thebibliography}{10}
\bibitem{Tam82-WRONSKIAN}
\bibinfo{note}{It should be noted that the definition for the relativistic
  Wronskian given here deviates from that of Tamura \cite{Tam92} by the factor
  $r^2$ that is not included in Eq.\ \eqref{EQ:WRON-DEF-gf}. This definition
  for the relativistic Wronskian leads in the non-relativistic limit ($ c
  \rightarrow \infty$) to its conventional counterpart $ [ u(r),v(r)]= u (r)
  \,\frac{d}{dr}v(r) - \frac{d}{dr}u(r)\,v(r)$ if the small component is
  eliminated by means of the radial Dirac equation.}
\end{thebibliography}

\bibliographystyle{apsrev}

\end{document}